\documentclass[twocolumn]{aastex63}

\usepackage{subfigure,graphicx,amsmath}

\graphicspath{{./}{figures/}}

\submitjournal{ApJ}

\shorttitle{ICRF3 Imaging}
\shortauthors{Hunt et al.}
\begin{document}

\title{Imaging Sources in the Third Realization of the International Celestial Reference Frame}

\email{lucas.hunt.ctr@navy.mil}

\author[0000-0001-8587-9285]{Lucas Hunt}
\affiliation{United States Naval Observatory \\
3450 Massachusetts Ave NW \\
Washington, DC 20392, USA}
\affiliation{Computational Physics, Inc. \\
8001 Braddock Road, Suite 210 \\
Springfield, VA 22151-2110}

\author[0000-0002-4146-1618]{Megan C. Johnson}
\affiliation{United States Naval Observatory \\
3450 Massachusetts Ave NW \\
Washington, DC 20392, USA}

\author[0000-0002-8736-2463]{Phillip J. Cigan}
\affiliation{United States Naval Observatory \\
3450 Massachusetts Ave NW \\
Washington, DC 20392, USA}
\affiliation{George Mason University \\
4400 University Dr \\
Fairfax, VA 22030}

\author[0000-0001-8009-995X]{David Gordon}
\affiliation{United States Naval Observatory \\
3450 Massachusetts Ave NW \\
Washington, DC 20392, USA}
\affiliation{George Mason University \\
4400 University Dr \\
Fairfax, VA 22030}

\author[0000-0001-8184-3672]{John Spitzak}
\affiliation{Computational Physics, Inc. \\
8001 Braddock Road, Suite 210 \\
Springfield, VA 22151-2110}



\begin{abstract}

The third iteration of the International Celestial Reference Frame (ICRF3) is made up of 4536 quasars observed at S/X bands using Very Long baseline interferometry (VLBI). These sources are high redshift quasars, typically between $1<z<2$, that are believed to host active galactic nuclei (AGN) at their centers. The position of compact radio sources can be determined better than sources with large amounts of extended radio structure.\deleted{The compactness, or core dominance, of a source}\deleted{ can be determined with imaging and because the structure of radio sources is dynamic,}\deleted{these sources need to be regularly monitored.} Here we report information on a series of 20 observations from January 2017 through December 2017 which were designed for precise astrometry and to monitor the structure of sources included in the ICRF3\deleted{, which was approved at the International Astronomical Union General Assembly} \deleted{meeting in Vienna, Austria in August of 2018, and formerly}\deleted{adopted on January 1, 2019}.\added{We targeted 3627 sources over the one year campaign and found the median flux density of 2697 detected sources at S-band is 0.13 Jy, and the flux density of 3209 sources detected at X-band is 0.09 Jy. We find that $70\%$ of detected sources in our campaign are considered compact at X-band and ideal for use in the ICRF and $89\%$ of the 2615 sources detected at both frequencies have a flat spectral index, $\alpha>0.5$  }

\end{abstract}



\section{Introduction} \label{sec:intro}

 The International Celestial Reference Frame (ICRF) is the definitive standard framework for precise astronomical positions at radio wavelengths, and underpins a wide range of applications including navigation, the GPS satellite network, and defines right ascension and declination coordinates in astronomy.  The ICRF is determined from positions of compact quasars, observed using the Very Long Baseline Interferometry (VLBI) technique, and has undergone three realizations (hereafter ICRF1, ICRF2, and ICRF3) (\citet{Ma1998,Fey2015,charlot2020}, respectively). The third realization \citep[ICRF3;][]{charlot2020} was adopted in January 2019 by the International Astronomical Union (IAU) as the international standard reference frame. 

 ICRF1 \citep{Ma1998} contained  positions of 608 sources, 212 of which were ``defining sources,'' so called because they serve to define the axis of the frame. Observations for ICRF1 were carried out between 1979 and 1995. ICRF2 \citep{Fey2015} built upon ICRF1, having incorporated sources from the Very Long Baseline Array (VLBA) Calibrator Survey \citep[VCS;][]{Beasley2002,Gordon2016}, and ultimately included positions of 3,414 total sources, 295 of which were defining sources.  
 The list of ICRF2 defining sources was formed by first selecting quasars with the most stable and well-determined positions and then, second, by selecting sources that formed an isotropic distribution across the sky.   
The ICRF3 improves upon previous iterations and contains 4536 sources at S/X bands, with improved astrometry over the ICRF2 catalog primarily due to incorporation of VLBA 24-hour astrometric and geodetic sessions that make up $\sim$68\% of the ICRF3 data \citep{charlot2020}.  The ICRF3 contains 303 defining sources that were selected based on the following criteria (in order of priority) (i) isotropic over the celestial sphere, (ii) contain positional stability, and (iii) are compact. Notably, ICRF3 includes for the first time observations at multiple radio frequencies including K-band and X/Ka-band reference frames in addition to the legacy S/X-band reference frame.

As described in \citet{charlot2020}, the ideal ICRF sources are compact, point-like objects, with astrometrically stable positions, which allow for high positional accuracy.  Selected sources are  distant, radio-loud quasars; the great distances to these sources translates to very small proper motions, which satisfies the requirement of a quasi-inertial reference frame.  They are bright enough in the radio to require only short integration times; $\sim$300 sources can be observed during a single 24 hour observing epoch.  ICRF imaging campaigns have been carried out since 1995 and with the advent of the VLBA in 1993, it has been used to conduct almost all ICRF imaging campaigns.  These campaigns are critical to refining, monitoring, maintaining, and improving ICRF source selection.

Though the VLBI technique allows for precisely measured positions of radio loud quasars, its unmatched resolution also means that some of these quasars are resolved and this effect increases as a function of increasing frequency. Furthermore, the spatial scales at which we observe these sources, and the turbulent nature of the active galactic nuclei (AGN) means that the sources are variable on timescales of hours, days, months, and years (e.g., 3C48). Changes in source structure can affect astrometric source positions, reducing the precision and adversely affecting the accuracy of the ICRF. To mitigate the effects of variability and maintain the integrity of the ICRF, it is important to image these sources and to continuously monitor them for changes in source structure.   

 In January 2017, the United States Naval Observatory (USNO) entered into an agreement with the National Science Foundation to contribute 50\% of the operating costs for the NRAO's VLBA in exchange for 50\% of the observing time. 
 With this time, USNO and the National Aeronautics and Space Administration Goddard Space Flight Center (NASA GSFC) have carried out a joint observing campaign  of ICRF3 sources in order to improve positions for those that had a limited number of past observations, and to image those sources to establish a snapshot set of observations in order to begin monitoring and understanding their positional and intensity variability at radio frequencies as well as their spatial structure. This is an ongoing USNO effort to continue monitoring these sources to contribute improved astrometric positions to the ICRF and look for changes in source structure.  \added{This imaging monitoring campaign will aid in source selection for projects such as the next generation geodetic VLBI array known as the VLBI Global Observing System (VGOS).  VGOS will observe ICRF sources across multiple frequency bands from $\sim$2.5 GHz to 14 GHz and our image data archive can be used to inform users on which targets have source structure that may require a more sophisticated model in the group delay measurements, for example.  Also, if VGOS sessions are designed for imaging, then our pipeline may be useful for automating images from the VGOS databases.} We are providing the image FITS files, calibrated $uv$-data, and other image products through a web-based interface for use by the astronomical and geodetic community.

This paper is laid out as follows: In \S~\ref{sec:Data} we describe the observations, calibration, and imaging of the data.  In \S~\ref{sec:FRIDA} we introduce the database of images and other data products, Fundamental Reference Image Data Archive (FRIDA).  In \S~\ref{sec:properties} we explore global properties of the ICRF sources, including fluxes and band-to-band spectral indices.  Finally, we provide a summary in \S~\ref{sec:Summary}.

\section{Data} \label{sec:Data}
\subsection{Observations and Scheduling} \label{sec:ObsSched}

The observations were made using the VLBA S/X dual frequency system, providing compatibility with earlier VLBA astrometry sessions and with nearly 40 years of astrometric/geodetic VLBI. A description of the VLBA system can be found in \citet{Napier1994}. The simultaneous recording of data at both S and X-band, often used for a more accurate ionosphere calibration that is important for astrometric and geodetic experiments, is enabled by a dichroic mirror that is deployed over the S-band receiver and reflects the higher frequency radiation to a deflector that then leads to the X-band receiver.

The primary goal of this VLBA observing campaign is to improve the ICRF3 \citep{charlot2020} with improved source positions and with images to help in the selection of defining sources. The setup was very similar to that of the VCS-II campaign \citep{Gordon2016}. Frequencies and bandwidths were identical to VCS-II, with 12, 32-MHz channel windows at X-band,  and 4 at S-band, using 2-bit sampling for a total recording rate of 2 Gbps. Table \ref{tab:obs} gives the observing parameters and list of sessions observed during 2017.  The target list was created with most sources being  previously observed at S/X bands in only three or fewer sessions, and some being sources not previously detected. Schedules were made using the NRAO SCHED\footnote{http://www.aoc.nrao.edu/software/sched/index.html} program. 

\begin{figure}
\centering
\includegraphics[width=0.48\textwidth]{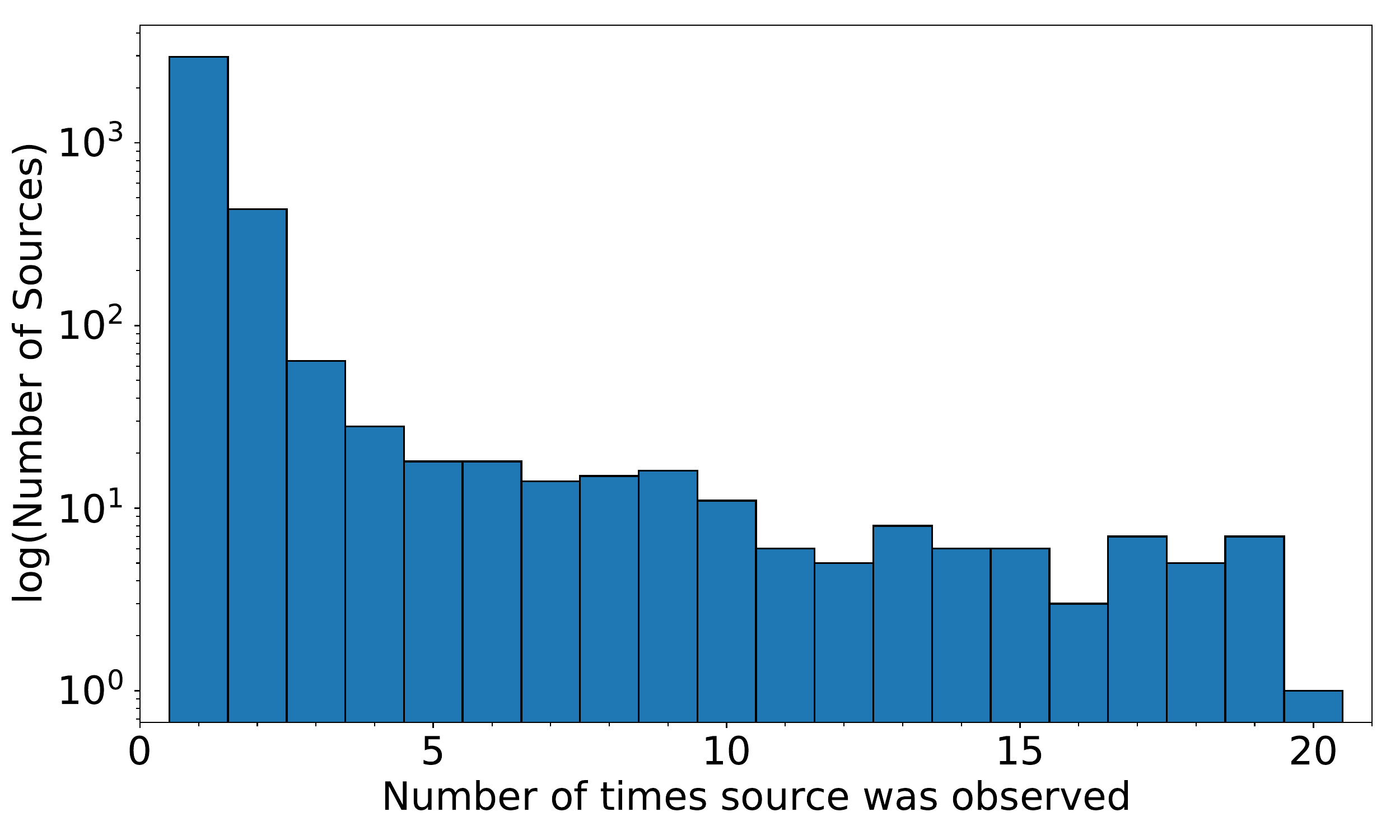}
\caption{Histogram of the number of times a given source was observed. Most sources were only observed one time. Stronger sources which were used to tie the other objects to the ICRF were observed in multiple sessions.}
\label{fig:ObsHist}
\end{figure}

\begin{figure*}
\centering
\includegraphics[width=\textwidth]{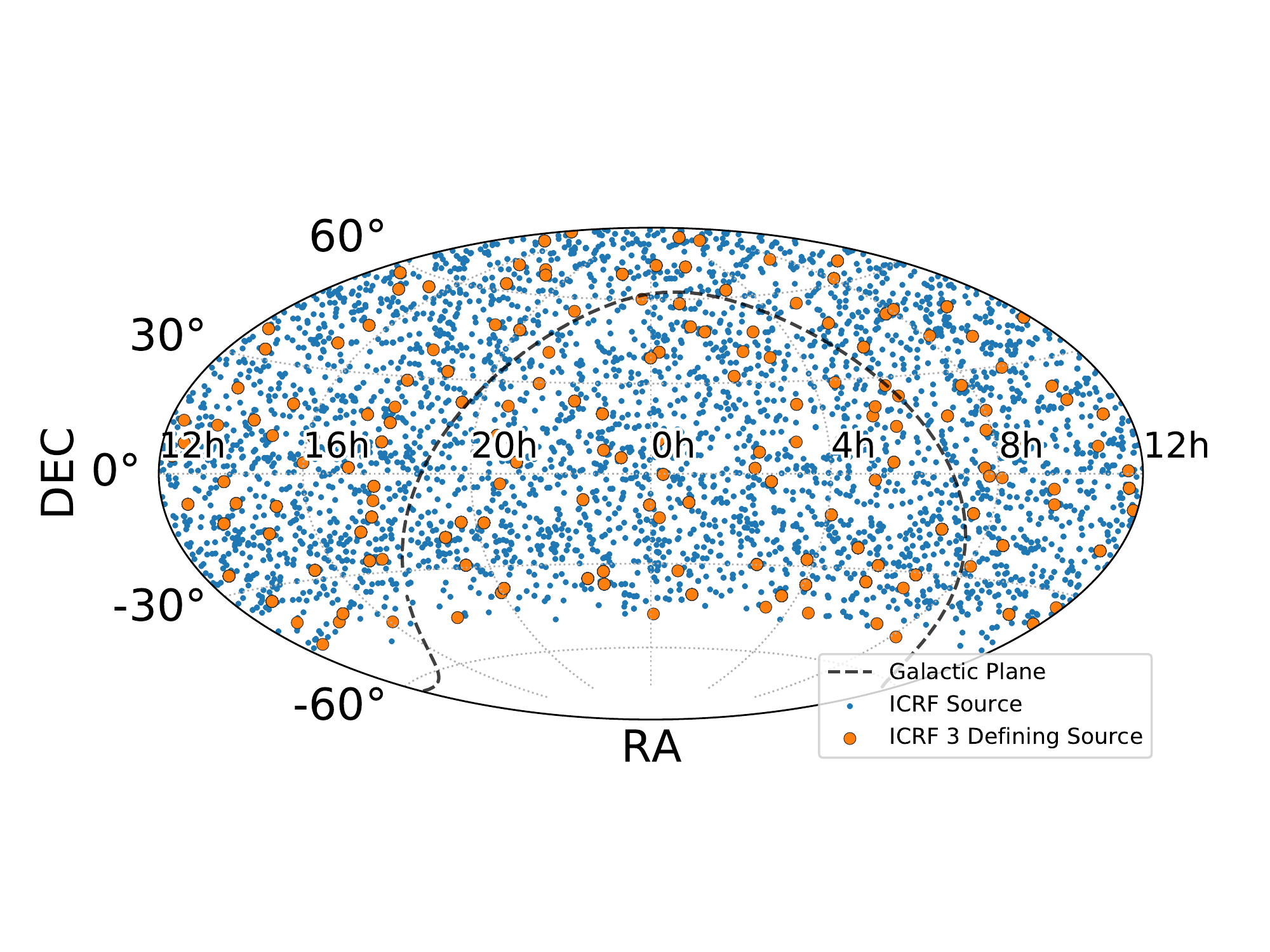}
\caption{Distribution of imaged sources on the sky. Objects highlighted in orange are ICRF3 defining sources.}
\label{fig:SkyMap}
\end{figure*}

Schedules were written using the SCHED dynamic mode, with each taking approximately one sidereal day, allowing them to be run at any day and starting time. To minimize slewing time, the sources were split into groups of 4 nearby sources, to be observed sequentially along with an ICRF2 defining source for troposphere calibration and for ties into the ICRF. We scheduled $\sim$300 target sources in each session, with integration times between 60 and 160 seconds. Table \ref{tab:obs} lists the observing properties of all 20 sessions.  Most sources north of $-20^{\circ}$ declination were comprised of three scans, with those south of $-20^{\circ}$ having two scans. The declination limit was approximately $-45^{\circ}$. There were two sessions per month during the first ten months of 2017, for a total of 20 sessions. The distribution of the number of visits to individual sources over the 20 included observing sessions and their positions on the sky are shown in Figure~\ref{fig:ObsHist} and Figure~\ref{fig:SkyMap}, respectively.

\added{For four sessions at least one antenna that was supposed to be included in the observations was completely flagged in the calibration stage. These four observations are listed in Table \ref{tab:obs}, and denoted below. In session UF001B, the antenna at Fort Davis was not operating properly and was removed from most of the observation. In session UF001C the antenna at Pie Town was having focusing issues and was completely flagged. In session UF001Q the antennas at both the Brewster station and the Mauna Kea station had no usable data. In session UF001S, the antenna at the Owens Valley station was taken out due to focusing issues.     

The final four sessions of this observing campaign, from September 18th through October 21st, were scheduled without the St. Croix antenna, which was badly damaged from Hurricane Maria, and was not available. This antenna is part of the longest baseline in the VLBA, and therefore resolution for these four sessions was $\approx15\%$ larger than sessions that included the St. Croix station. The schedules for these observations were made knowing that we would not have the St. Croix antenna available to us.}

\begin{table*}
\centering
\caption{Observation Parameters}
\begin{tabular}{lcccc}
\hline
\hline
\multicolumn{5}{c}{All Sessions}\\
\hline
\hline
\multicolumn{3}{l}{Parameter}& \multicolumn{2}{l}{Value}\\
\hline
\multicolumn{3}{l}{Backend System} &
\multicolumn{2}{l}{Polyphase Filterbank (PFB)}\\
\multicolumn{3}{l}{Total channel windows} &
\multicolumn{2}{l}{16}\\
\multicolumn{3}{l}{Single channel window bandwidth (MHz)} &
\multicolumn{2}{l}{32}\\
\multicolumn{3}{l}{No. of spectral channels per window} &
\multicolumn{2}{l}{64}\\
\multicolumn{3}{l}{Total bandwidth at X-band (MHz)} &
\multicolumn{2}{l}{384}\\
\multicolumn{3}{l}{Total bandwidth at S-band (MHz)} &
\multicolumn{2}{l}{128}\\
\multicolumn{3}{l}{Frequency resolution (MHz)} &
\multicolumn{2}{l}{0.5} \\ 
\multicolumn{3}{l}{Polarization} &
\multicolumn{2}{l}{Right-hand circular}\\
\multicolumn{3}{l}{Data rate (Gbps)} &
\multicolumn{2}{l}{2}\\
\multicolumn{3}{l}{Sampling rate (bits)} &
\multicolumn{2}{l}{2}\\
\multicolumn{3}{l}{X-band channel frequencies (MHz)} &
\multicolumn{2}{l}{8460.0, 8492.0, 8524.0, 8556.0, 8620.0, 8652.0,} \\ 
\multicolumn{3}{l}{} & 
\multicolumn{2}{l}{8716.0, 8748.0, 8812.0, 8844.0, 8876.0, 8908.0}\\
\multicolumn{3}{l}{S-band channel frequencies (MHz)} &
\multicolumn{2}{l}{2220.0, 2252.0, 2284.0, 2348.0}\\
\hline
\hline
\multicolumn{5}{c}{Individual Sessions}\\
\hline
\hline
Session  & \bf{Antennas in array$^a$} & \# of Sources & \# of Scans  & Obs. Date Range (2017)\\
\hline
UF001A &\bf{BR, FD, HN, KP, LA, MK, NL, OV, PT, SC}& 297 & 798 &  Jan-16 23:26$-$Jan-17 23:21\\
UF001B &\bf{BR, FD$^b$, HN, LA, MK, NL, OV, PT, SC}& 309 & 780 &  Jan-21 23:06$-$Jan-22 23:00\\
UF001C &\bf{BR, FD, HN, KP, LA, MK, NL, OV, PT$^b$, SC}& 292 & 769 &  Feb-19 21:12$-$Feb-20 21:06\\
UF001D &\bf{BR, FD, HN, KP, LA, MK, NL, OV, PT, SC}& 281 & 773 &  Feb-24 20:52$-$Feb-25 20:47\\
UF001E &\bf{BR, FD, HN, KP, LA, MK, NL, OV, PT, SC}& 284 & 785 &  Mar-23 19:06$-$Mar-24 19:01\\
UF001F &\bf{BR, FD, HN, KP, LA, MK, NL, OV, PT, SC}& 281 & 785 &  Mar-27 07:02$-$Mar-28 06:57\\
UF001G &\bf{BR, FD, HN, KP, LA, MK, NL, OV, PT, SC}& 289 & 765 &  Apr-28 16:45$-$Apr-29 16:40\\
UF001H &\bf{BR, FD, HN, KP, LA, MK, NL, OV, PT, SC}& 281 & 752 &  May-01 16:33$-$May-02 16:28\\
UF001I &\bf{BR, FD, HN, KP, LA, MK, NL, OV, PT, SC}& 287 & 765 &  May-27 22:02$-$May-28 21:57\\
UF001J &\bf{BR, FD, HN, KP, LA, MK, NL, OV, PT, SC}& 297 & 777 &  May-31 04:34$-$Jun-01 04:28\\
UF001K &\bf{BR, FD, HN, KP, LA, MK, NL, OV, PT, SC}& 291 & 772 &  Jun-10 03:08$-$Jun-11 03:03\\
UF001L &\bf{BR, FD, HN, KP, LA, MK, NL, OV, PT, SC}& 280 & 760 &  Jun-15 19:15$-$Jun-16 19:11\\
UF001M  &\bf{BR, FD, HN, KP, LA, MK, NL, OV, PT, SC}& 293 & 776  & Jul-09 05:08$-$Jul-10 05:01\\
UF001N &\bf{BR, FD, HN, KP, LA, MK, NL, OV, PT, SC}& 297 & 763 &  Jul-16 10:00$-$Jul-17 09:54\\
UF001O &\bf{BR, FD, HN, KP, LA, MK, NL, OV, PT, SC}& 281 & 742 &  Aug-05 04:40$-$Aug-06 04:34\\
UF001P &\bf{BR, FD, HN, KP, LA, MK, NL, OV, PT, SC}& 286 & 759 &  Aug-12 15:11$-$Aug-13 15:05\\
UF001Q &\bf{BR$^b$, FD, HN, KP, LA, MK$^b$, NL, OV, PT}& 285 & 751 &  Sep-18 19:50$-$Sep-19 19:44\\
UF001R &\bf{BR, FD, HN, KP, LA, MK, NL, OV, PT}& 175 & 352 &  Sep-26 07:11$-$Sep-27 07:06\\
UF001S &\bf{BR, FD, HN, KP, LA, MK, NL, OV$^b$, PT}& 251 & 523 &  Oct-09 12:03$-$Oct-10 11:59\\
UF001T &\bf{BR, FD, HN, KP, LA, MK, NL, OV, PT}& 253 & 525 &  Oct-21 13:56$-$Oct-22 13:48\\
\hline
Notes: & \multicolumn{4}{l}{$^a$ Antennas listed were present in at least one scan, \deleted{$^b$ Used for comparison}} \\
&\multicolumn{4}{l}{\added{$^b$ Antenna was removed in calibration stage and not included in any imaging}|}
\label{tab:obs}
\end{tabular}
\end{table*}

\subsection{Calibration} \label{sec:CalIm}

\begin{figure*}
\centering
\includegraphics[width=\textwidth]{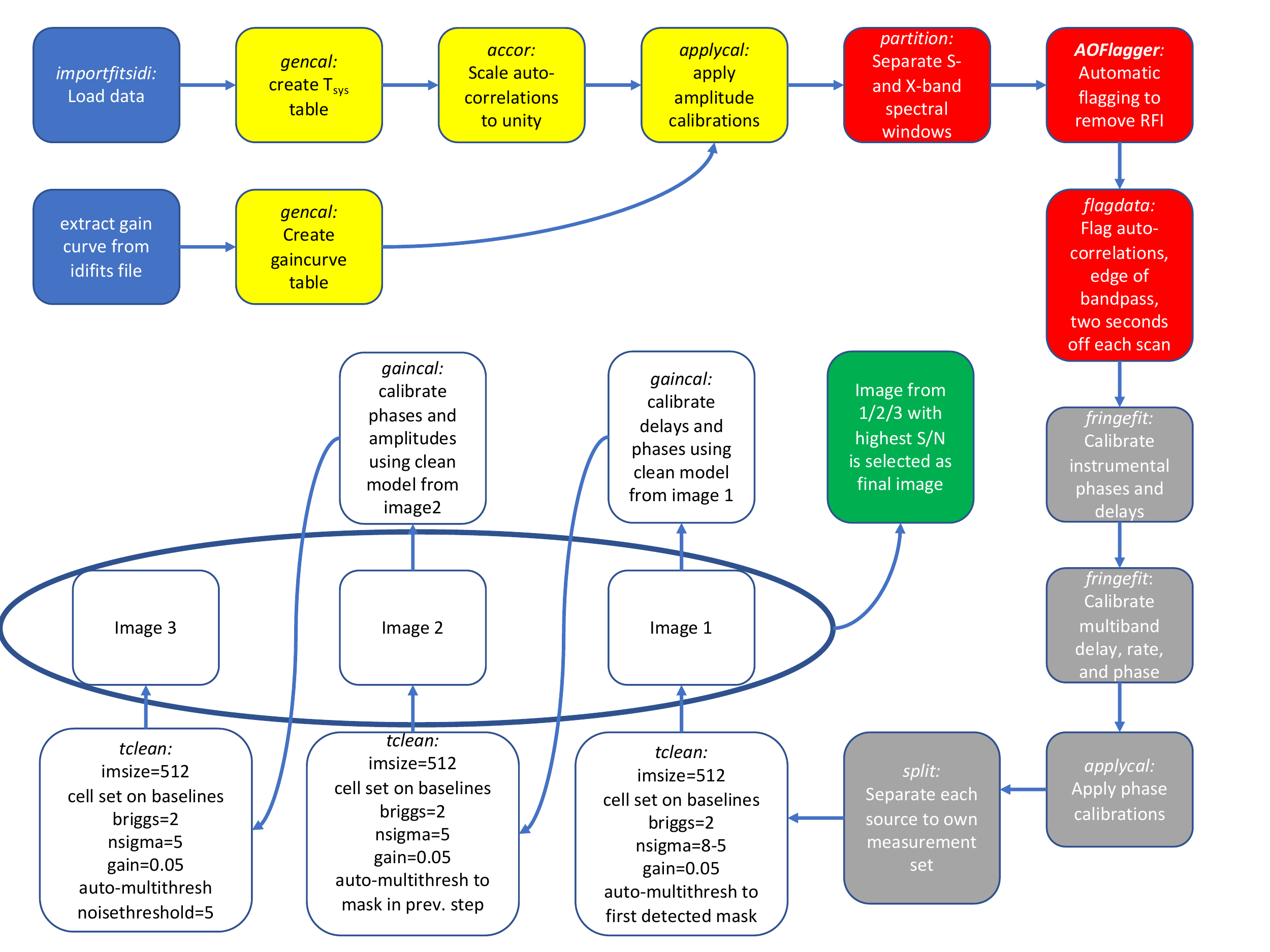}
\caption{A flowchart outlining our calibration routine. The CASA task used for each step is written in italics. Programs run within the script but outside of CASA are written in bold italics.  The colors of the squares correspond to categories for the steps: blue for importing data; yellow for amplitude calibration; red for flagging; gray for phase calibration; and white for imaging and self-calibration. The final image is selected (green) as the highest S/N image created in the process described in Section~\ref{sec:imaging}.}
\label{fig:Pipeline_flowchart}
\end{figure*}

The raw data were correlated at the Array Operations Center in Socorro, New Mexico with the Socorro-DiFX correlator \citep{Deller2011}. Initial amplitude and phase calibration was performed using Common Astronomy Software Applications \citep[CASA;][]{McMullin2007}. The calibration and imaging steps used in our CASA pipeline are briefly described in the following sections. Section \ref{sec:amplitude_calibration} covers how the amplitude calibration was carried out. Section \ref{sec:flagging} outlines some of the flags that were used to ensure data quality. Section \ref{sec:phasecal} outlines the phase calibration steps, and Section \ref{sec:imaging} outlines the imaging and self-calibration steps. A flowchart of our pipeline is shown in Figure \ref{fig:Pipeline_flowchart}, with the importing tasks in blue squares, the amplitude calibration steps in yellow squares, the flagging steps in red squares, the phase calibration steps in gray squares, and the imaging and self-calibration steps in white squares.  For a given step, CASA tasks are italicized while separate programs are listed in bold text.

\subsubsection{Amplitude Calibration}
\label{sec:amplitude_calibration}

The first step carried out by our pipeline is to ingest the data from a standard format to a format that can be modified within the CASA environment. \deleted{We use a customized script to import the gain curve} \deleted{information required for the amplitude calibration.}\added{We use a customized script to extract the VLBA gain curve information from the idifits file, and convert it into a CASA readable calibration table. The gain curve and system temperature of the telescope are used to calculate the flux density of sources in our observation in the process further described below.} We use the CASA task {\sc importfitsidi} to convert the idifits files to the CASA measurement set format. We do not use phase referencing so our observations do not require corrected earth orientation parameters, and we currently do not make corrections for the ionosphere. 

The steps required to calibrate the amplitudes with our pipeline are highlighted in yellow in Figure \ref{fig:Pipeline_flowchart}. The first step in amplitude calibration is to correct for errors in sampler thresholds when the analog signal is converted to a digital signal. This correction can be calculated by determining how much the autocorrelation spectrum deviates from unity and applying that scaling factor at a later time. 

The second step in amplitude calibration is to determine amplitudes from the antenna information. The VLBI technique allows us to probe some of the smallest physical scales in all of observational astronomy, which means that source amplitudes can, and do, appear to vary on short timescales. We do not have calibrators of constant, known brightness at these resolutions, and so we use the system equivalent flux density (SEFD) for each antenna to calibrate amplitudes. The SEFD is defined as the flux of a radio source which doubles the antenna's system temperature and can be written as

\begin{equation}
{\rm SEFD} = \frac{T_{sys} }{{\rm DPFU}\cdot{\rm gc}}
\end{equation}
where $T_{sys}$ is the system temperature measured at the telescope in Jy. DPFU is the degrees per flux unit, or the gain in units Jy K$^{-1}$ which relates the measured system temperature to a flux value at a specific elevation. gc is the aforementioned gain curve which describes how the telescope's gain changes as a function of elevation. The correlated flux density of a source on a given baseline can then be calculated using

\begin{equation}
S_{c,ij}=C^{acc}_{i}C^{acc}_{j}\sqrt{SEFD_{i}SEFD_{j}} 
\end{equation}
where $S_{c,ij}$ is the flux density of the source on the baseline $ij$ in units of Jy, $C^{acc}_{n}$ is the correction for correlator offsets for antenna $n$, and SEFD is defined above. 

\subsubsection{Separating Observing Bands and Flagging Data}
\label{sec:flagging}

This section covers the tasks marked by red boxes in Figure \ref{fig:Pipeline_flowchart}, and starts with the separation of the S-band (4 spectral windows) and X-band (12 spectral windows) into different measurement set files. All of the following steps, through separating the individual sources into their own files, are applied to each band.
We flag the data using the automated flagging program AOFlagger \footnote{\software{AOFlagger}} \citep{Offringa2012}. We then flag the autocorrelations because they are not needed. Next we flag 7 channels from each edge of each spectral window where the sensitivity drops off dramatically. These channels do not calibrate well, and contribute little to the overall signal. Finally, we flag one second at the beginning and end of each scan. The task that calibrates the phases, {\sc fringefit}, uses a Fourier transform code that
crashes the script when encountering a stray one second integration. In these cases, the phases do not properly calibrate, resulting in corrupted initial images. We found that flagging one second at the beginning and end of each scan removed these errant one second integrations and enable the {\sc fringefit} task to run without error. 

\subsubsection{Phase and Delay Calibration}
\label{sec:phasecal}

Accurate correlation of interferometric data requires a good estimate of the difference in time it takes for the same wave front to reach each telescope, based on numerous factors including the distance between the telescopes and the rotation of the Earth. Even with careful estimation, it is often not possible to accurately calculate all of the factors that contribute to that time offset: for example, the atmosphere above each telescope can be notably different, or the signal may have to travel through different hardware paths to get to the correlator. These differences manifest as changes in phase with time and frequency and can cause loss of coherence when the data are averaged in time and frequency during the imaging steps. We use a calibration technique known as fringe fitting to remove these errors after correlation. 

Phase calibration is done in two steps. The first is to remove the change in phase over frequency due to different path lengths through telescope electronics for each spectral window. This is done by finding a strong source and correcting the slope as a function of frequency. Our code does this by running the CASA task {\sc fringefit} on scans that include all antennas, and selecting the scan where the {\sc fringefit} task finds solutions to all antennas. This calibration step is only applied to the change in phase over frequency, and is applied to all scans. 

The second phase calibration step corrects the phase, change in phase as a function of frequency, and change in phase as a function of time, for each source individually. Since the previous calibration step removed the change in phase with frequency, we can average spectral windows together, over the whole scan, in order to increase the chance of finding a solution. 

\subsubsection{Imaging and Self-calibration}
\label{sec:imaging}

\deleted{Imaging was performed using the CASA task {\sc tclean}. For each source, we created} \deleted{an initial image with 512x512 pixels and estimated the pixel size to fit five pixels across} \deleted{the synthesized beam, which we estimated using}
\begin{equation}
    \Theta_{{\rm hpbw}} = \frac{\lambda}{D}
\end{equation}
\deleted{where $\lambda$ is the central wavelength of the observation and D is the distance between two} \deleted{antennas on the longest baseline, both in meters. For sources that had at least one scan} \deleted{on the longest baseline, St. Croix to Mauna Kea, the pixel size was set to 0.165 mas} \deleted{(usually 0.165 mas, but dependent on the longest baseline).} \added{The antennas included for observations of each source varied over the entire observing period, so we calculated the size of the synthesized beam for each source using} 
\begin{equation}
    \Theta_{{\rm hpbw}} = \frac{\lambda}{D}
\end{equation}
\added{where $\lambda$ is the central wavelength of the observation and D is the distance between two antennas on the longest baseline, both in meters. We then chose the binning cell size, or pixel size of the image, that would place five pixels across the minor axis of the snythesized beam. All images were 512$\times$512 pixels.} A model of clean components was built using the auto-masking feature of {\sc tclean} to generate a mask that selected only the brightest emission in the field.  This was done using the H\"ogbom  algorithm \citep{Hogbom1974} and setting the Briggs robust weighting equal to 2, traditionally defined as natural weighting in the {\sc tclean}. The masks created using the automatic masking feature in {\sc tclean} were controlled by the {\texttt auto-multithresh} parameter \citep{Kepley2020}. Masking is mainly controlled by the {\texttt noisethreshold} parameter which masks peaks in the residual image that are some multiple of the noise. The sources in our survey have varying flux densities so it is impossible to predict what value for {\texttt noisethreshold} will create a conservative clean mask. We therefore start with the {\texttt noisethreshold} value set to 15, and continue to check if a clean mask has been made. If it has not, the code will try again until it reaches a {\texttt noisethreshold} of 5. If it doesn't make a mask at that point it, will end with the dirty image. 

Self-calibration was attempted next --- a process of successively using models produced from previous images to improve the calibration for the phases, delays, and amplitudes, which can often yield images of higher fidelity and improved S/N.  The clean component model from the initial image was fed back into the calibration task {\sc gaincal}, to calibrate the delays and phases.  An image was produced from this "phase-only" self-calibration, using the same parameters as the initial image.  This in turn was used to repeat the process, but this time calibrating for phase and amplitude for each antenna.  \added{The antenna gains were recalculated and applied per scan, and re-normalised to maintain the flux density scale.} The self-calibration process was repeated until the brightest residual emission in {\sc tclean} reached a threshold of three times the noise \added{ of the residual image}.  The image with the highest S/N from among the initial images, phase-only self-calibration, and phase+amplitude self-calibration was then selected as the final image to include in the sample.

\subsection{Comparing CASA to other software}

\subsubsection{Calibration Comparison to AIPS}\label{sec:CompareCasaAips}

Previous large VLBI campaigns including the VLBA Imaging and Polarimetry Survey \citep[VIPS;][]{Helmboldt2007}, VCS \citep{Beasley2002,Petrov2006,Gordon2016}, and Monitoring Of Jets in Active galactic nuclei with VLBA Experiments \citep[MOJAVE;][]{Lister2005} have used the Astronomical Image Processing System \citep[AIPS;][]{Greisen2003} to apply the initial calibrations to their data, and Difmap \citep{Shepherd1994} for imaging and self-calibration. 
We considered using AIPS and Difmap to carry out calibration and imaging in this campaign as well, but due to having thousands of sources and plans for continued observations, we needed to ensure that the software package we used was easy to script, continued to receive new features and updates, and allowed simple use of outside packages. We chose CASA as a better fit to our requirements than AIPS and Difmap, but we still needed to compare CASA to AIPS and Difmap to ensure that the results were consistent.

The calibration steps carried out using CASA were outlined above in Section \ref{sec:CalIm}. We carried out the same calibration tasks using the equivalent AIPS tasks. We did not carry out the initial fringe fitting, where we corrected the phase for instrumental errors, nor did we run AOFlagger on the data.

To carry out a detailed calibration and imaging comparison, we selected three sources from the data which represent a stronger source (0955+476), a source with complex structure (0733+261), and a weak source (1257+839). Figure \ref{fig:CASA_AIPS_DIFMAP_COMPARE} shows images of each source, and Table \ref{tab:CASA_AIPS_DIFMAP_COMPARE} shows the rms and peak values in Jy bm$^{-1}$ extracted from each image. The main focus of this section is the initial calibration comparison, so we focus on the statistics where the sources were both imaged in CASA. \deleted{The noise and peak flux of sources calibrated in CASA are both lower than sources that were} \deleted{calibrated in AIPS. This difference is visibile in panel (d) of Figure \ref{fig:CASA_AIPS_DIFMAP_COMPARE}, which shows a plot} \deleted{of amplitude vs baseline length for each source. The data calibrated with AIPS (blue) for} \deleted{each source has a higher amplitude than the data calibrated with CASA (green). We are} \deleted{working with NRAO to investigate the origin of this scaling difference, and plan to release} \deleted{a memo to discuss those results.} \added{All three sources show that the peak flux density are approximately the same, independent of the software used for calibration.}  

\subsubsection{Imaging Comparison to \textit{Difmap}}\label{sec:comparedifmap}

\begin{figure*}
    \centering
    \begin{subfigure}[]{}
        \includegraphics[width=0.48\linewidth]{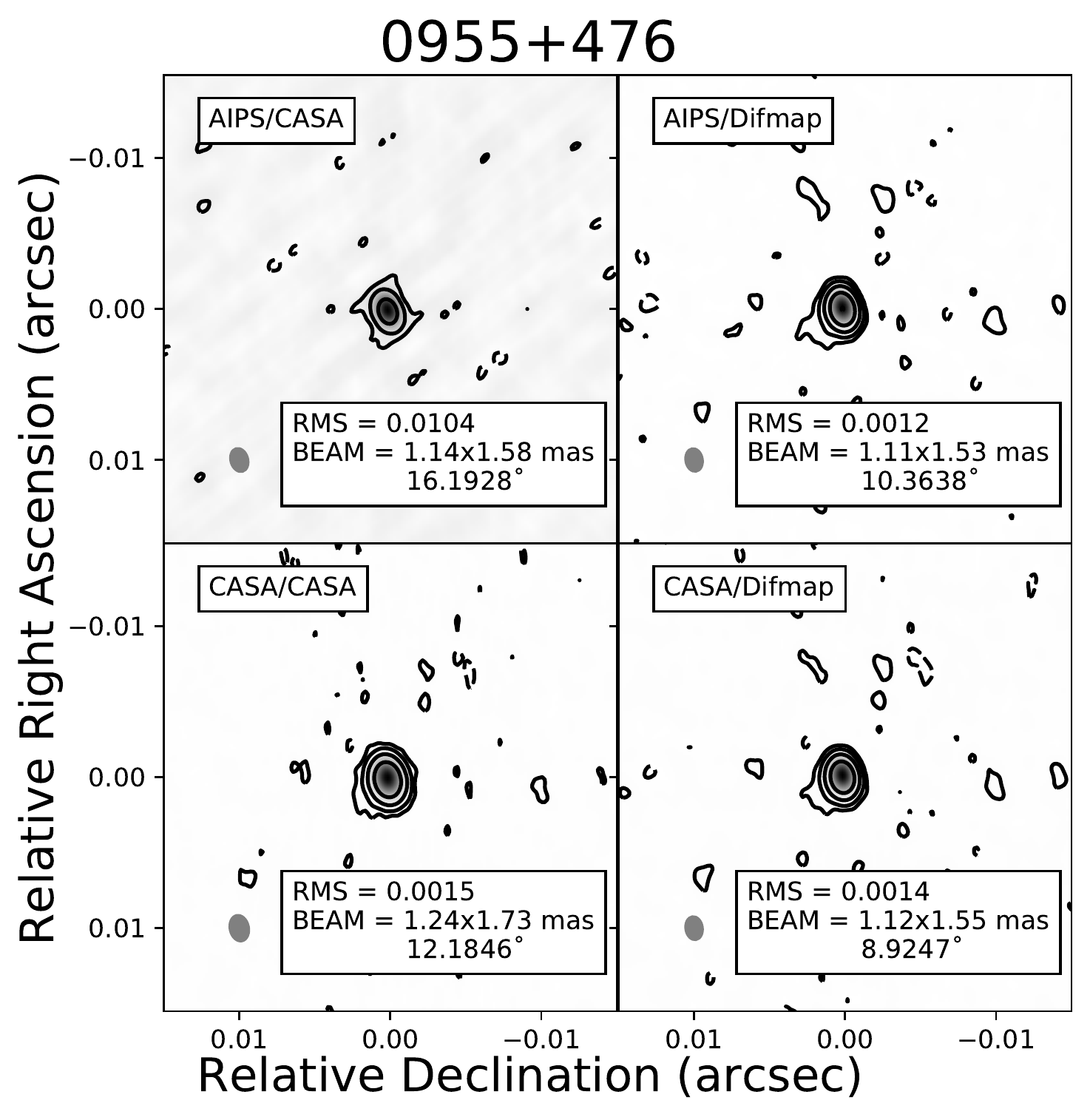}
    \end{subfigure}
    \begin{subfigure}[]{}
        \includegraphics[width=0.48\linewidth]{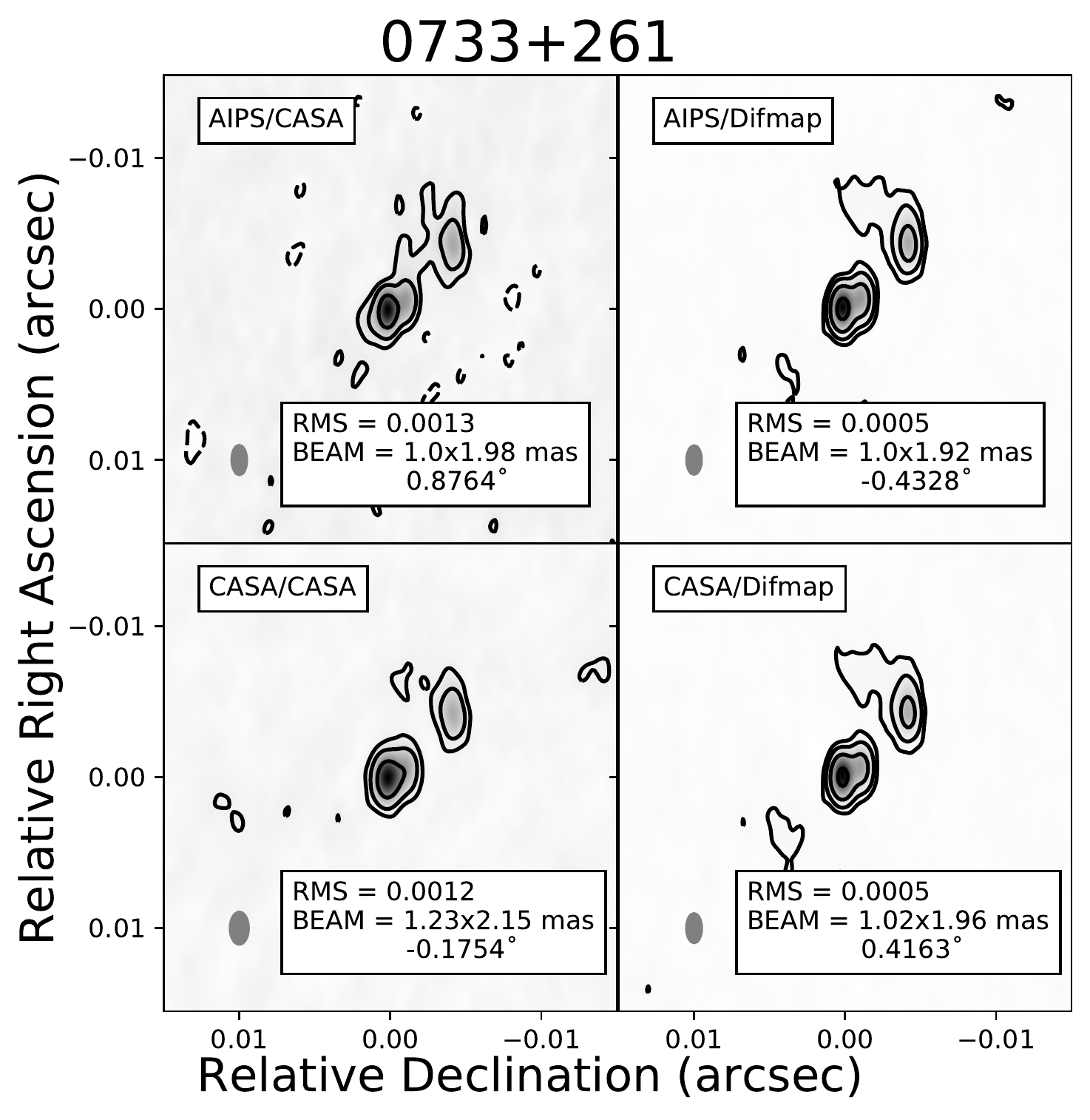}
    \end{subfigure}
    \\
    \begin{subfigure}[]{}
        \includegraphics[width=0.48\linewidth]{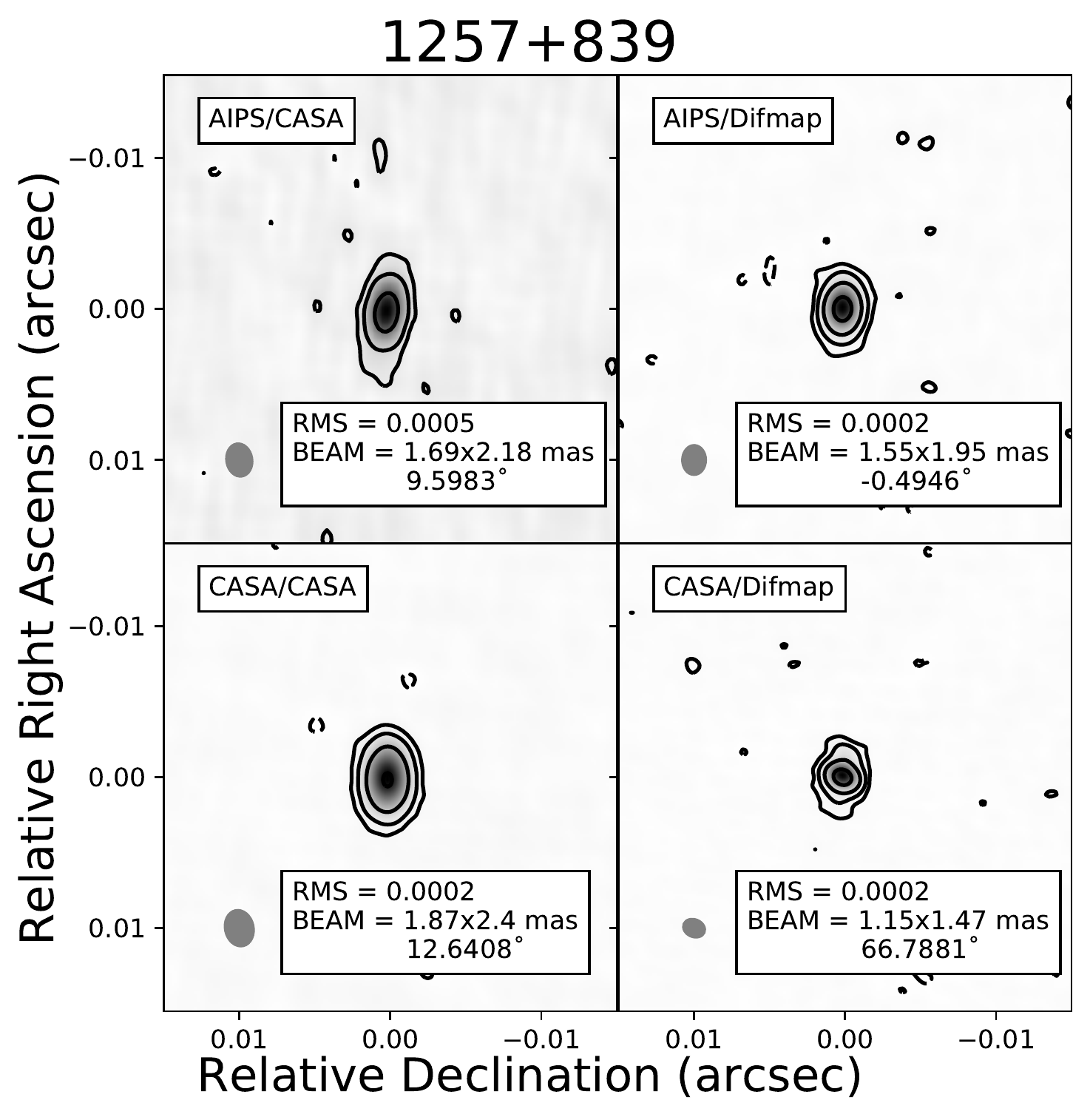}
    \end{subfigure}
    \begin{subfigure}[]{}
        \includegraphics[width=0.48\linewidth]{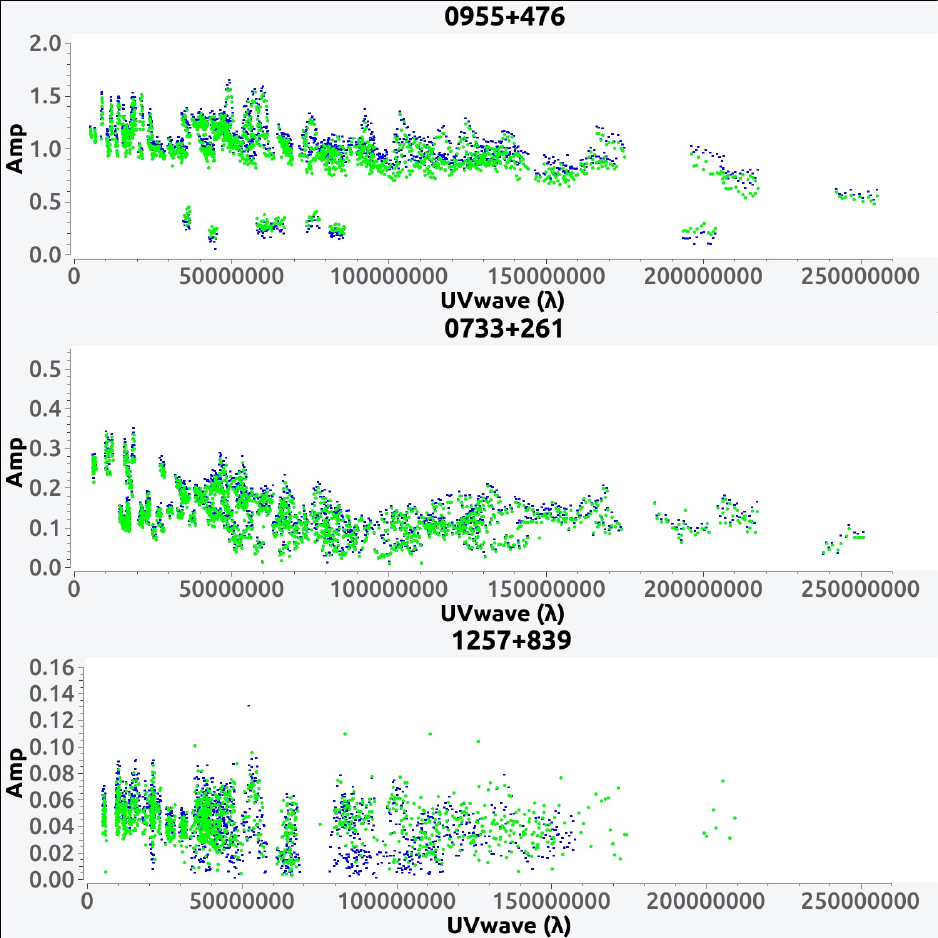}
    \end{subfigure}
    \caption{Comparison of RR polarization images made using different combinations of intial calibration and imaging/self-calibration steps. These images were made from the UF001K dataset observed on 10 June 2017.  We show the software used for calibration and imaging in the upper left hand corner of each image, and the rms measured in the image away from the source at the bottom of the image. The restoring beam is shown in gray in the lower left hand corner of the image. The contours in each image are $3\times~4^{n}\sigma$. Figures \textbf{(a)}, \textbf{(b)}, and \textbf{(c)} show that data imaged with CASA and Difmap produce very similar images for stronger sources (0955+476), complex sources (0733+261), and weak sources (1257+839). Panel \textbf{(d)} shows the amplitude vs UV-distance plot for sources that have been calibrated in AIPS (blue) and CASA \replaced{(yellow)}{(green)} both imaged in CASA. The plots show that the amplitude calibration between AIPS and CASA are \replaced{scaled differently}{nearly identical}.  
    }
    \label{fig:CASA_AIPS_DIFMAP_COMPARE}
\end{figure*}

Difmap \footnote{\software{Difmap} https://sites.astro.caltech.edu/~tjp/citvlb/} \citep{Shepherd1994} is part of the Caltech VLBI software package, and was designed to quickly make images from VLBI data. Difmap does not contain calibration packages like those described in \ref{sec:CalIm} and \ref{sec:CompareCasaAips}, but rather it is strictly a software package for imaging, and is optimized for VLBI data.  It has been used in conjuction with calibrated data from AIPS to make the images for the aforementioned large VLBI survey campaigns and is considered an industry standard. 
In Figure \ref{fig:CASA_AIPS_DIFMAP_COMPARE} we show images of sources mentioned in \ref{sec:CompareCasaAips}. The layout for each source image set is, from upper-left to lower-right: AIPS calibration, CASA imaging; AIPS Calibration, Difmap imaging; CASA Calibration, CASA imaging; and CASA calibration, Difmap imaging. 

The self-calibration steps using CASA are outlined above in section \ref{sec:imaging}. The Difmap script starts with a uniform weighted image. It runs 50 iterations of CLEAN, then calibrates the phases in a loop, until the peak in the residual image is 8 times the noise. It then creates a naturally weighted image in the same way. Finally it carries out a similar process but calibrates the amplitude per antenna and channel window and calibrates the phases.

\added{We tried to reduce the differences in parameters for both the Difmap and CASA imaging scripts. You can see the parameters set for each program in Table \ref{tab:CASA_AIPS_DIFMAP_COMPARE_IMAGING_PARAM}. One of the main ways the programs differ is how the clean iterations and self-calibration steps are carried out. The CASA self-calibration steps make a full clean component model until it reaches the stopping criteria, in our case three times the noise in the residual image. CASA then uses that model to calibrate the amplitudes and phases. Difmap carries out a set number of iterations, adding clean components to a model, and then calibrates the amplitudes and phases. Then Difmap continues to add clean components to the same model, and calibrates the data with the updated model.}

Though the steps are different, the results from imaging and self-calibration in CASA and Difmap are the same for 0955+476\replaced{. In 0723+261 the peak flux from CASA imaging is higher, and in}{ and 0723+261. In the weaker source,} 1257+839 the peak flux from Difmap is higher. This is likely due to the differences in the self-calibration procedure \added{mentioned above}.

\begin{deluxetable*}{|l|c|c|}
\tabletypesize{\footnotesize}
\tablecaption{Outlining imaging parameters to compare between AIPS and CASA \label{tab:CASA_AIPS_DIFMAP_COMPARE_IMAGING_PARAM}}
\tablehead{&CASA&Difmap}
\startdata
Final map size & 512 & 512\textsuperscript{a}\\
Number of iterations per cycle\textsuperscript{b} & 1000 & 100 \\
Stopping criteria & 5$\sigma$ in residual & 5$\sigma$ in residual\\
Cell size & 0.165 mas & 0.165 mas\\
Clean gain & 0.05 & 0.05\\
Solution Interval & Scan & Scan\\
\hline
\multicolumn{3}{|l|}{\textsuperscript{a} \footnotesize{Difmap map size starts at 1024}}\\
\multicolumn{3}{|l|}{\textsuperscript{b} \footnotesize{always reach stopping criteria before all clean iterations}}
\enddata
\end{deluxetable*}

%\begin{deluxetable*}{|l|cc|cc|cc|cc|}
%\tablecaption{Comparison of sample source RMS and peak flux density in Jy bm${-1}$ when processed with CASA, AIPS, and Difmap \label{tab:CASA_AIPS_DIFMAP_COMPARE}}
%\tablehead{&\multicolumn{2}{|c|}{AIPS/CASA}& \multicolumn{2}{c|}{AIPS/Difmap} & \multicolumn{2}{c|}{CASA/CASA} & \multicolumn{2}{c|}{CASA/Difmap}\\
%Source & $\sigma_{\rm obs}$ & Peak & $\sigma_{\rm obs}$ & Peak &$\sigma_{\rm obs}$ & Peak &$\sigma_{\rm obs}$ & Peak}
%\startdata
%0955+476 (Compact) & $0.01$ & \replaced{$1.2$}{$1.01$} & $0.0012$ & \replaced{$1.21$}{$0.97$} &  $0.0015$ & $1.02$&$0.0014$ & \replaced{$1.0$}{$0.96$} \\
%0733+261 (Complex) & $0.0013$ & \replaced{$0.161$}{$0.135$} & $0.0005$ & \replaced{$0.149$}{$0.127$} & $0.0012$ & $0.133$& $0.0005$ & $0.123$ \\
%1257+839 (Weak) &$0.0005$ & \replaced{$0.064$}{$0.040$} & $0.0002$ & \replaced{$0.074$}{$0.058$} & $0.0002$ & $0.045$ & $0.0002$ & $0.057$\\ 
%\enddata
%\end{deluxetable*}

\begin{deluxetable*}{|l|cc|cc|cc|cc|}
\tabletypesize{\footnotesize}
\tablecaption{Comparison of sample source RMS and peak flux density in Jy bm${-1}$ when processed with CASA, AIPS, and Difmap \label{tab:CASA_AIPS_DIFMAP_COMPARE}}
\tablehead{&\multicolumn{2}{|c|}{AIPS/CASA}& \multicolumn{2}{c|}{AIPS/Difmap} & \multicolumn{2}{c|}{CASA/CASA} & \multicolumn{2}{c|}{CASA/Difmap}\\
Source & $\sigma_{\rm obs}$ & Peak & $\sigma_{\rm obs}$ & Peak &$\sigma_{\rm obs}$ & Peak &$\sigma_{\rm obs}$ & Peak}
\startdata
0955+476 (Compact) & $0.01$ & $1.01$ & $0.0012$ & $0.97$ &  $0.0015$ & $1.02$&$0.0014$ & $0.96$ \\
0733+261 (Complex) & $0.0013$ & $0.135$ & $0.0005$ & $0.127$ & $0.0012$ & $0.133$& $0.0005$ & $0.123$ \\
1257+839 (Weak) &$0.0005$ & $0.040$ & $0.0002$ & $0.058$ & $0.0002$ & $0.045$ & $0.0002$ & $0.057$\\ 
\enddata
\end{deluxetable*}

\section{Fundamental Reference Image Data Archive (FRIDA)} \label{sec:FRIDA}

As part of USNO's membership in the IVS, USNO is an official IVS Analysis Center and an Analysis Center for Source Structure.  In support of these international arrangements, USNO has historically been responsible for providing images of ICRF sources to the community through what was previously known as the Radio Reference Frame Image Database (RRFID).  USNO has been undergoing many updates and changes to our networks and computer systems, and thus, access to RRFID has been unavailable for the past few years.  However, USNO has taken this opportunity to develop a new interactive web-based interface called the Fundamental Reference Image Data Archive (FRIDA).  FRIDA will debut in 2021 and it will contain all archival images from the RRFID along with the images presented in this work.  Currently, the USNO images of ICRF sources span frequencies from 2.3 to 42 GHz with the majority of images at 2.3 and 8.6 GHz.  FRIDA will host FITS files for all images as well as calibrated $uv$-data files and ancillary image quality diagnostic files such as amplitude versus $uv$-distance and $uv$ sky coverage plots for individual sources.  Users will be able to download all data available through the interactive website.

With the calibration and imaging pipeline developed in CASA, we aim to have new images populate FRIDA in an automated or semi-automated manner after each 24-hour VLBA session is correlated.  FRIDA is planned to grow by including the Research and Development with VLBA (RDV) sessions, an IVS-sponsored series which combine the VLBA with other IVS stations.  The goal of the RDV sessions is to monitor and maintain information on faint or non-detected sources, and this series is a collaborative effort between USNO and Bordeaux Observatory.  In fact, the Bordeaux VLBI Image Database (BVID\footnote{http://bvid.astrophy.u-bordeaux.fr/database.html}) contains images from half of all the RDV sessions and it is USNO's goal to host the remaining half on FRIDA. \added{We also acknowledge the Astrogeo\footnote{http://astrogeo.org/} Center as another source of images of sources that comprise the ICRF} In addition to the VLBA-only sessions, the RDV sessions, and archival images from RRFID, USNO plans to host K-band images from the USNO-sponsored UD001 series for multi-wavelength radio images of ICRF sources.  

\begin{figure*}
    \centering
    \begin{subfigure}[]{}
        \includegraphics[width=0.48\linewidth]{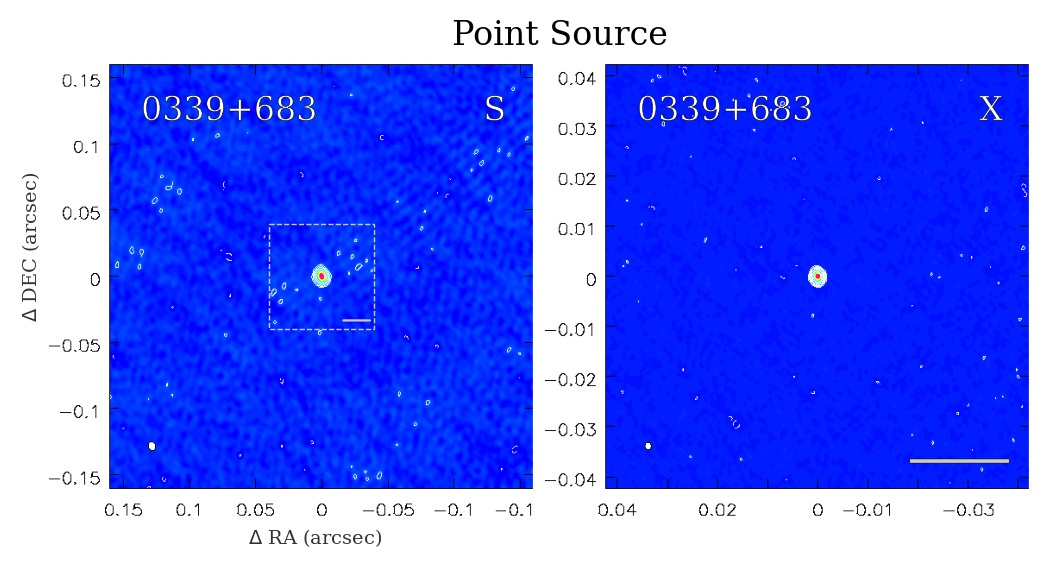}
    \end{subfigure}
    \begin{subfigure}[]{}
        \includegraphics[width=0.48\linewidth]{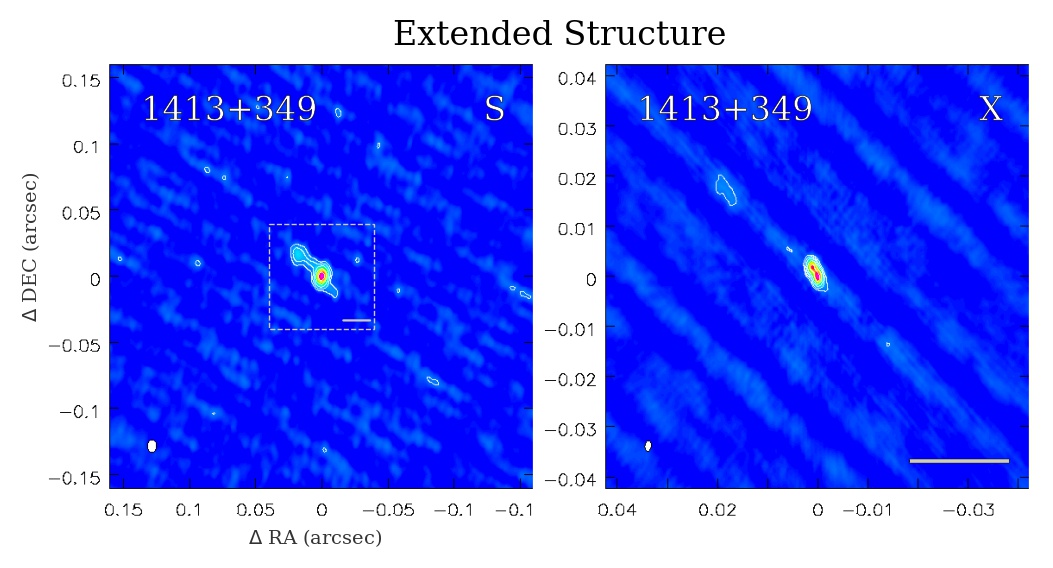}
    \end{subfigure}
    \\
    \begin{subfigure}[]{}
        \includegraphics[width=0.48\linewidth]{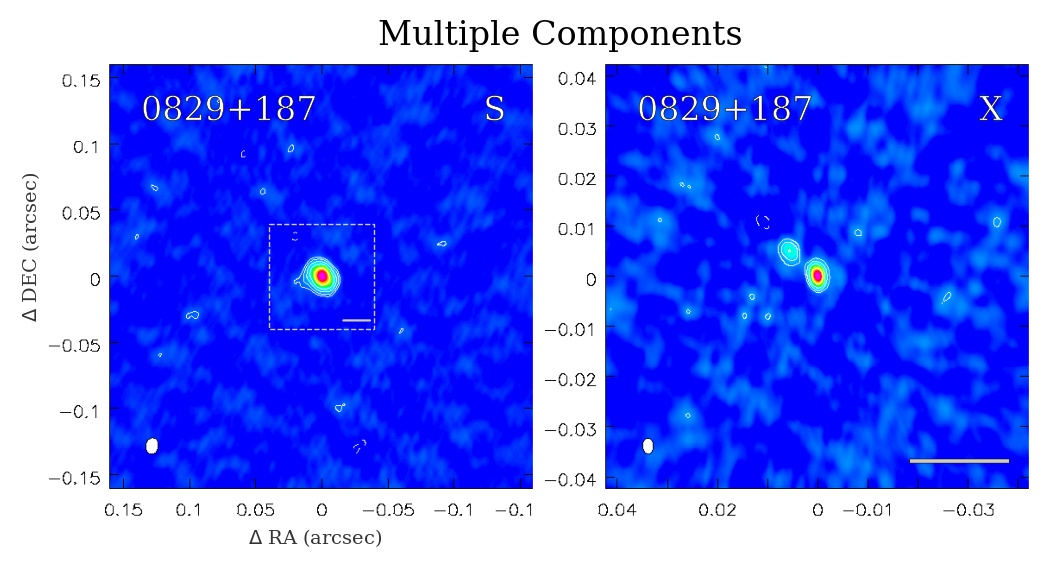}
    \end{subfigure}
    \begin{subfigure}[]{}
        \includegraphics[width=0.48\linewidth]{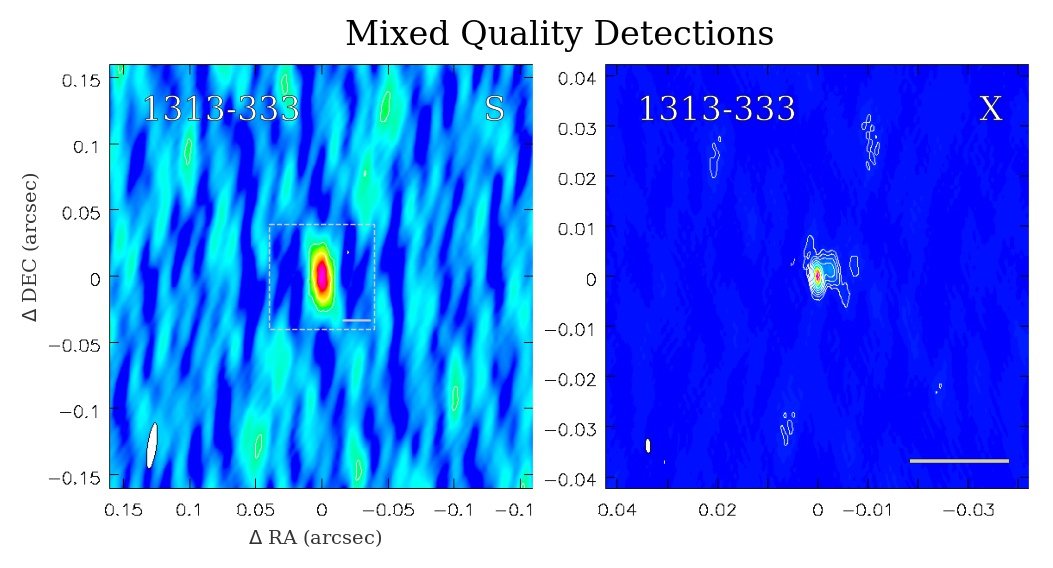}
    \end{subfigure}
    \caption{Examples highlighting the variety of source features in the ICRF catalog, in image pairs at S and X bands for selected sources. The field of view is narrower for the X band images, and the equivalent extent is denoted by the dashed boxes in the S band images, along with a 0\farcs2 size bar in each for reference.
    \textbf{(a)} 0339-683, a point-like source in both bands.  \textbf{(b)} 1413+349, with extended emission. \textbf{(c)} 0829+187, showing multiple components.  \textbf{(d)} 1313-333, with a clear detection and extended structure at X band but a low S/N detection at S band.    }
    \label{fig:source_examples}
\end{figure*}

Imaging ICRF sources is paramount for monitoring the physical properties intrinsic to the quasars --- these may lead to astrometric uncertainties, which in turn may contribute to uncertainties in geodetic measurements.  Characteristics such as source structure, variability in flux and position, core shift, and other physical phenomena make quasars problematic over long temporal ranges for maintaining precise astrometry for each target.  Therefore, it is vital to image ICRF sources regularly in order to monitor any changes that might lead to problems in astrometric or geodetic measurements.
Figure~\ref{fig:source_examples} shows example images of four sources in S and X bands, demonstrating the variety of source features within the sample.

\section{Global Properties of ICRF Sources} \label{sec:properties}
\subsection{Sources included in analysis}
We have observed 3,627 sources between one and 20 times, at two frequencies, for a possible 11,220 images. For sources that were observed in more than one 24-hour observation session, we select the image that has the highest dynamic range at X-Band, leaving us with 7,254 images. Our imaging pipeline automatically creates clean masks for residual images with emission brighter than fifteen times the noise level.  Sources with S/N lower than this level will not have any cleaning attempted, and only a `dirty image' will be created. We exclude the low S/N sources where only a dirty image has been made in this global property analysis. \added{We also exclude 10 sources at S-band and 11 sources at X-band that have a S/N greater than 15, but for which modelling the source fails.}  The figures below include \replaced{3381}{3371} sources for plots made at X-band, \replaced{2,669}{2,659} sources at S-band, and \replaced{2587}{2576} sources where information from both bands are included. 

\subsection{Flux properties} \label{sec:Flux_Properties}

\begin{figure}
\centering
\includegraphics[width=0.45\textwidth]{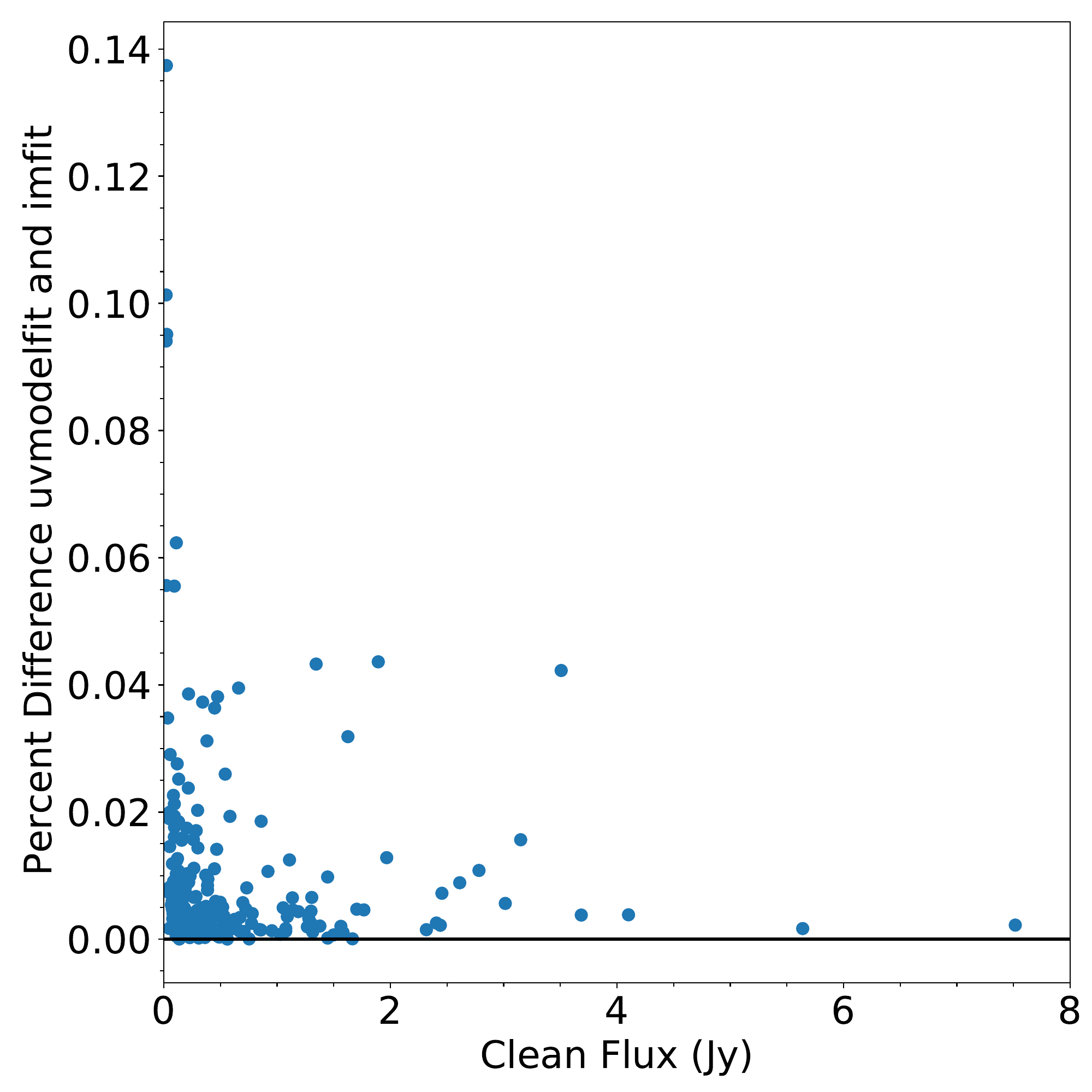}
\caption{The percent difference between the flux estimated for the gaussian fit to the brightest component in an image using the casa tasks {\sc uvmodelfit} and {\sc imfit}. For most sources the percent difference is less than 1\%, and only a small number of weak sources have a percent difference greater than 5\%}

\label{fig:percent_dif}
\end{figure}

We have used the CASA task {\sc imstat} to determine the noise in a region of each image that is free of emission, and the peak flux density of the image both in units Jy bm$^{-1}$. We calculated the theoretical RMS noise ($\sigma_{\rm theor}$), in units Jy bm$^{-1}$, for each image using the following equation from \citet{Wrobel1999}\footnote{ sensitivity calculation also found from NRAO at https://science.nrao.edu/facilities/vlba/docs/manuals/oss/imag-sens}

\begin{equation}
\sigma_{\rm theor} = \frac{\rm SEFD}{\eta_{c}(N(N-1)\delta\nu~t_{\rm int})^{1/2}}~{\rm Jy~bm^{-1}}
\end{equation}
where SEFD is the system equivalent flux density in Jy, the overall system noise defined as the flux density of a source that doubles the system temperature, $\eta_{c}$ is the correlator efficiency (0.75 for the VLBA \footnote{value for $\eta_{c}$ comes from https://science.nrao.edu/facilities/vlba /docs/manuals/oss/bsln-sens}), $N$ is the number of antennas, $\delta\nu$ is the bandwidth in Hz and $t_{\rm int}$ is the total on source integration time. We calculated the signal-to-noise ratio (S/N) using the peak flux and observed noise $\sigma_{\rm obs}$. We used the CASA task {\sc imfit} to fit a 2-D Gaussian to the center-most point source of each image \added{which is approximately equivalent to fitting a Gaussian model to the UV data. You can see in Figure \ref{fig:percent_dif} that for most sources with a signal to noise ratio of greater than 15, fitting a Gaussian to the image plane or the UV plane yields the same total flux for the brightest component. The handful of sources that have a percent difference larger than 5\% are weaker sources where no self-calibration was completed.} We label the flux density for the gaussian  $S_{Gauss}$.
Past studies have calculated the total flux density in an image by fitting a Gaussian to each component \citep[ex.][]{Pushkarev2012,Fey2002} and summing the flux density from all components. Fitting a Gaussian to each component by hand for over 10,000 images is prohibitively time consuming, but \citet{Pushkarev2012} show that the total flux density estimated from fitting 2-D Gaussians to components in an image is approximately equal to the sum of the flux density in the clean components in an image. Therefore we estimate the total flux density, $S_{\nu}$ by summing the flux density of clean components in an image.  All of these source properties calculated for each observation are included in Table \ref{tab:Glorious_Table} along with the source name, the ICRF3 Right Ascension and Declination to the median uncertainty of ICRF3 of 0.1 mas \citep{charlot2020}, and the date of each observation. The first 15 sources are included as an example and the full table will be available as a supplement in a machine-readable format.  

\begin{figure}
\centering
\includegraphics[width=0.45\textwidth]{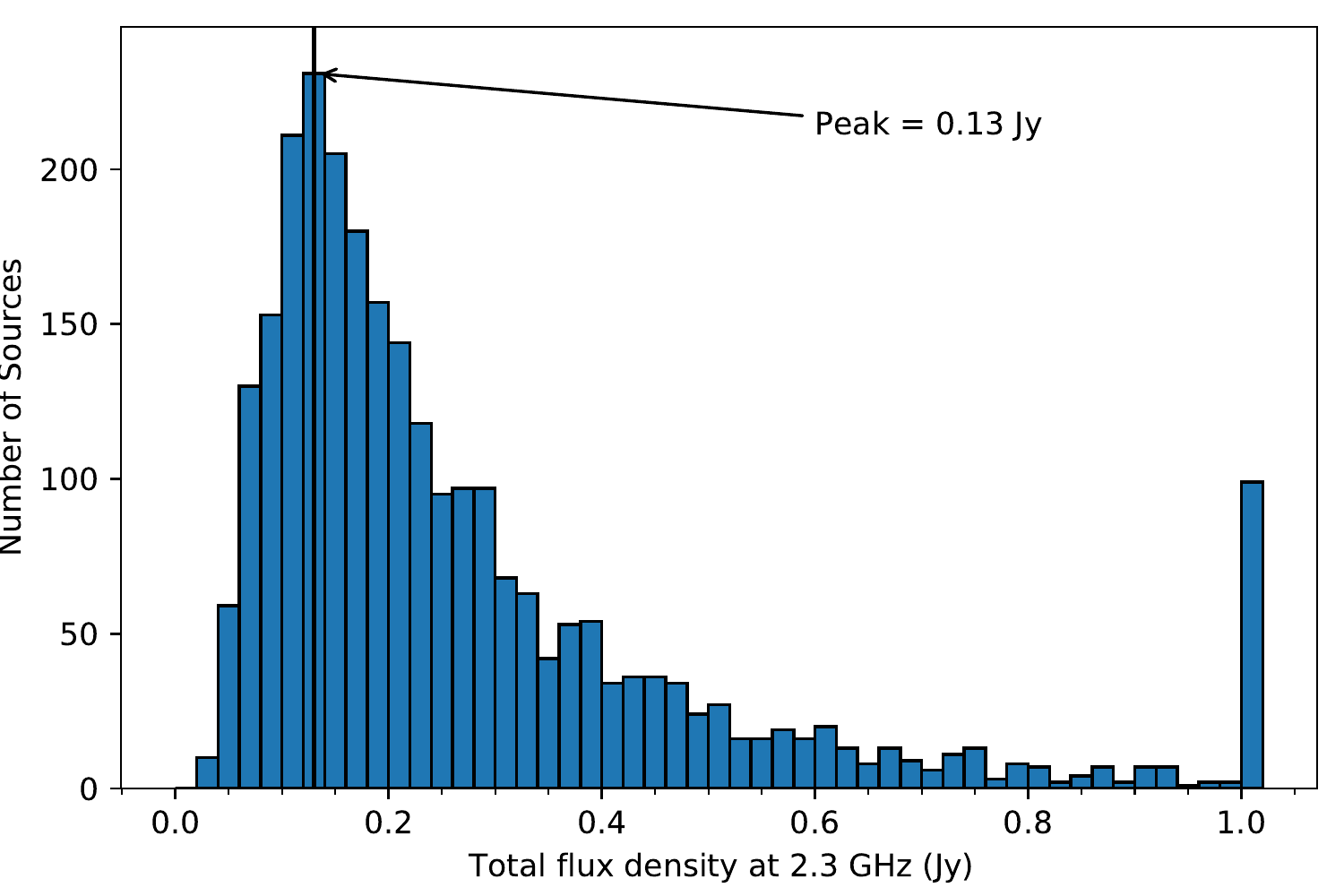}
\includegraphics[width=0.45\textwidth]{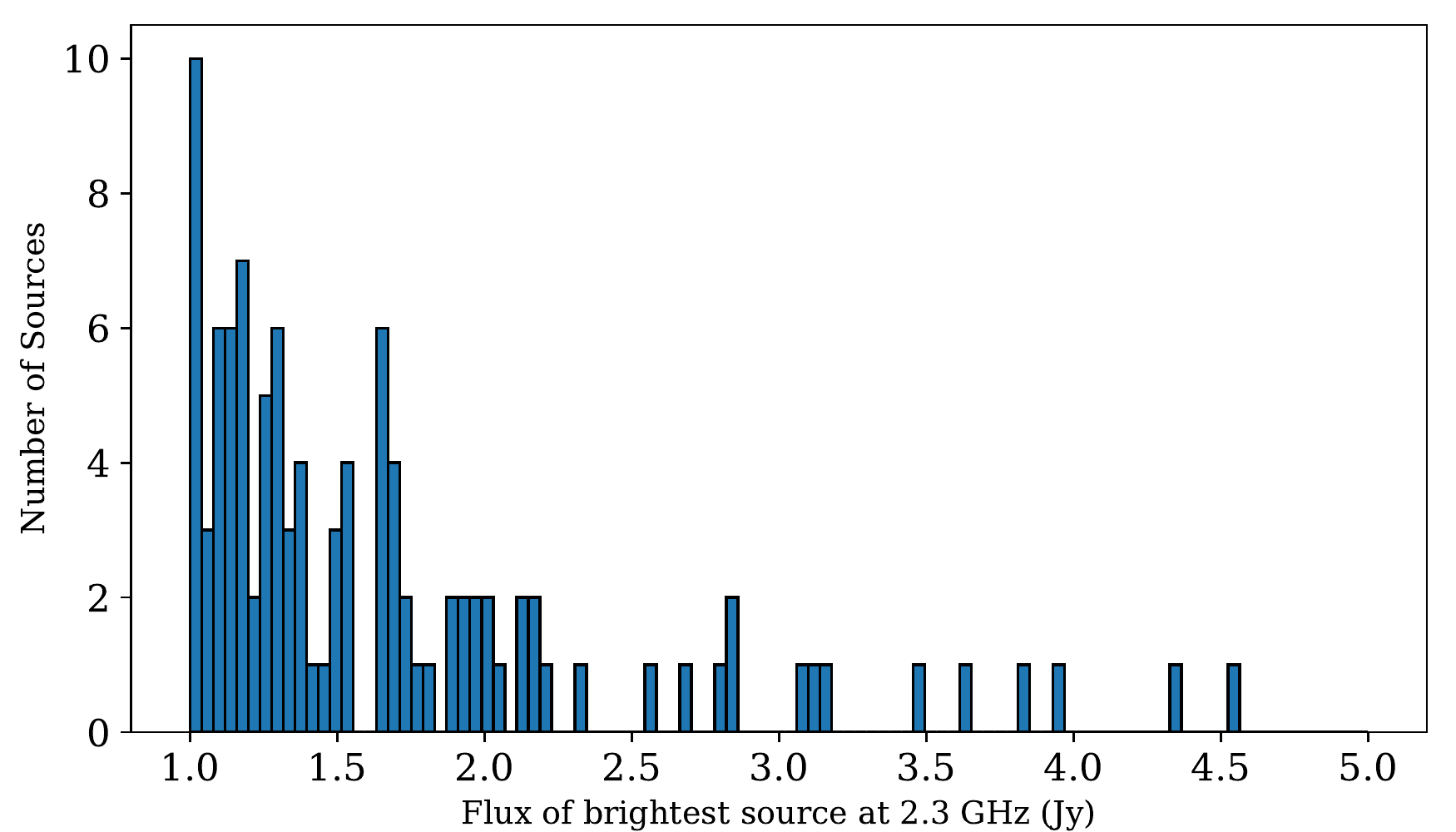}
\caption{Left: Distribution of total flux density, $S_{\nu}$, for each source at S-Band. Sources with flux density larger than 1 Jy are counted in the final bin on the right.  Right: the distribution of sources with flux density greater than 1 Jy. These histograms show the distribution of flux densities measured from a single observation for each source.  For sources that were observed and imaged more than once, the flux density from the image with the highest dynamic range was used. These histograms do not include sources that were not imaged or sources whose only image had a dynamic range of less than 15.}

\label{fig:S_Band_flux}
\end{figure}

\begin{figure}
\centering
\includegraphics[width=0.45\textwidth]{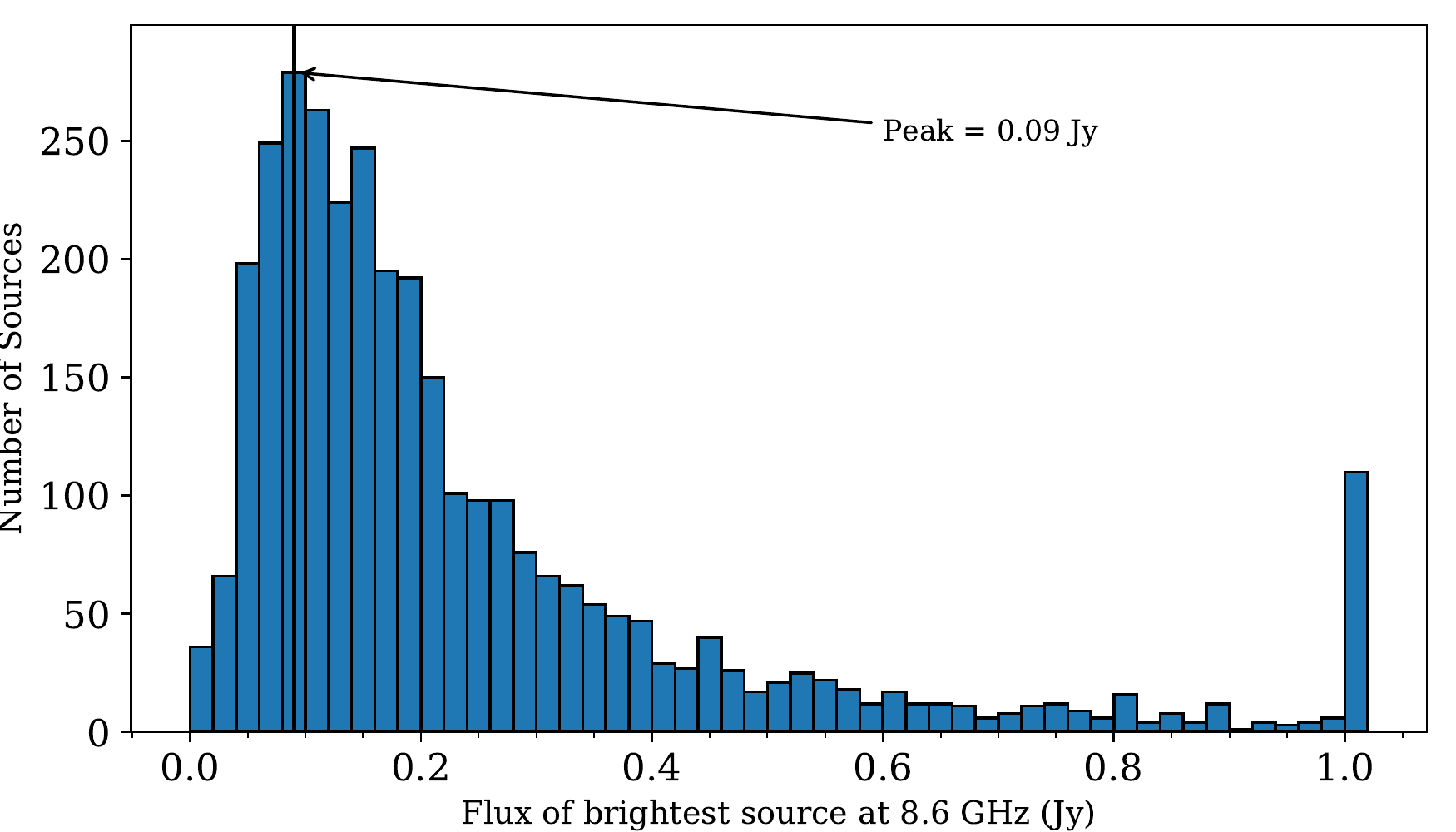}
\includegraphics[width=0.45\textwidth]{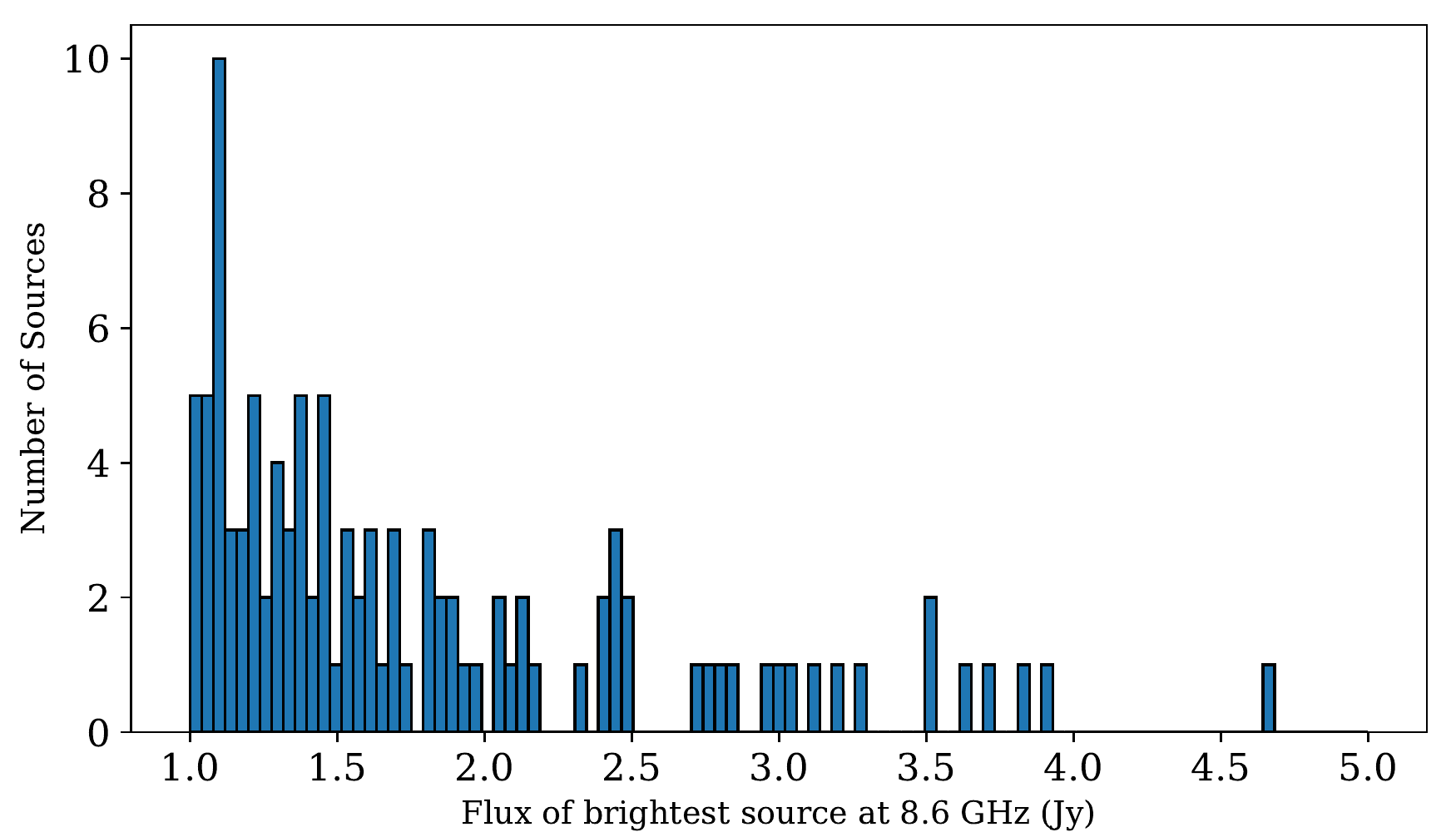}
\caption{Same as Figure \ref{fig:S_Band_flux} for all sources at X-band}
\label{fig:X_Band_flux}
\end{figure}

We show the distribution of $S_{\nu}$ at both S and X-bands in Figures \ref{fig:S_Band_flux} and \ref{fig:X_Band_flux}, respectively.  Each of these figures have two histograms: the left histogram shows the total flux density for all sources up to 1 Jy while the right histogram shows the distribution of the sources whose total flux density is greater than 1 Jy. For sources that were imaged more than once over the full duration of the \replaced{19}{20} observing sessions, the value of $S_{\nu}$ comes from the image with the highest dynamic range. \added{We find that the median flux density is 0.13 Jy at S-band and 0.09 Jy at X-band. This is fainter than the median flux of 1.01 and 0.7 Jy at S-band and X-band respectively from \citet{Fey2002} and the median flux of 0.76 and 0.69 Jy at S-band and X-band respectively. The median flux of sources in this campaign is about 6 times fainter than previous campaigns. This shows that while all three campaigns are targeting compact quasars, this campaign has targeted many more objects and therefore fainter objects. }  

\begin{figure}
\centering
\includegraphics[width=0.48\textwidth]{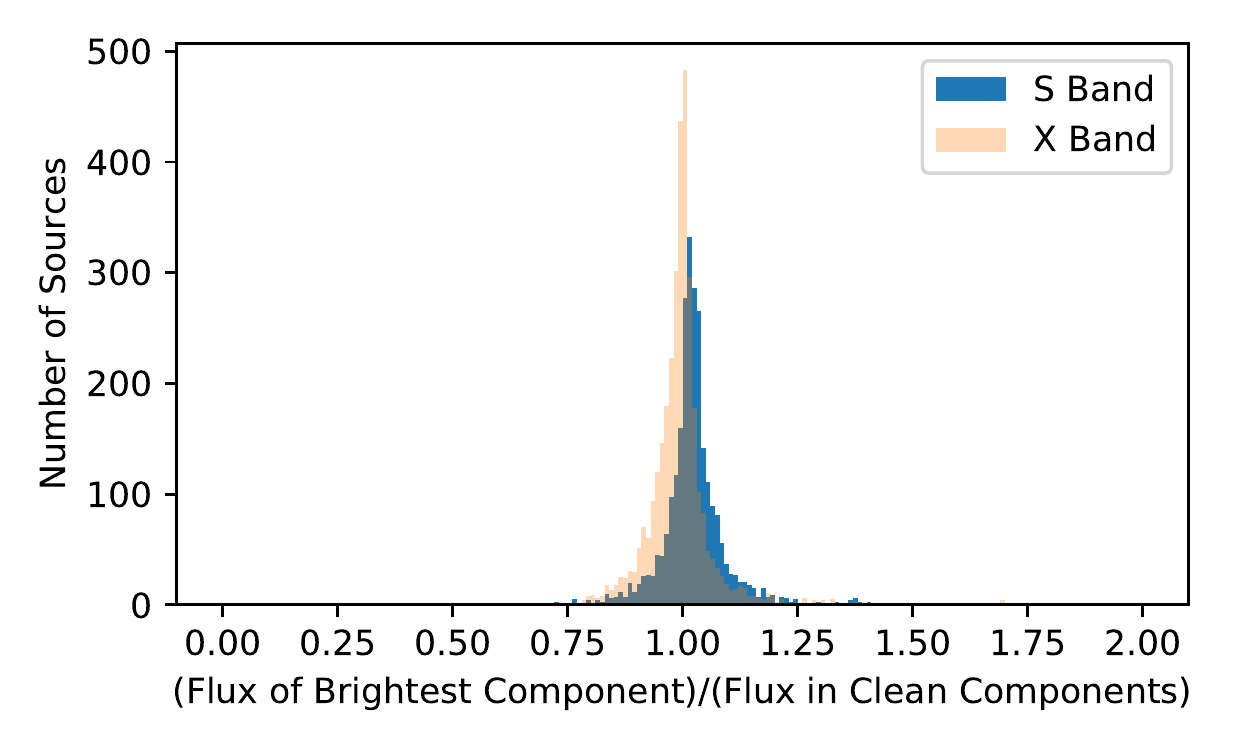}
\caption{The compactness ratios of observed sources. This serves as an estimation of how point-like an object is, where a source with a Gaussian model flux ratio of 1 has all of the flux contained within the beam (unresolved). The total flux is measured by adding the flux from the clean component map.}
\label{fig:compact_to_total}
\end{figure}

As mentioned in Section \ref{sec:intro}, unresolved, compact point sources typically provide better astrometric precision than sources with structure. Any extended emission can cause the astrometry solution to degrade in accuracy, especially when aggregated over long temporal timelines.  
As such, we aim to include as many compact, point-like sources as possible in the ICRF, and imaging campaigns such as this one provide a great way to study the nature of source structure in large numbers of quasars at high spatial resolutions. We estimate the compactness, or core dominance of sources in our survey as the ratio of the Gaussian model flux density, $S_{Gauss}$, to total flux density, $S_{\nu}$. We show this distribution in Figure \ref{fig:compact_to_total}. This ratio has been referred to previously as the core dominance \citep{Pushkarev2012, Fey2002} but because we don't want to draw any conclusions about the source of the emission for any given source we will refer to the ratio as the compactness ratio.

The calculated compactness ratio is sometimes greater than one because the {\sc imfit} task in CASA can produce a flux density value that is slightly larger than the sum of the clean components for that source if the wings of the Gaussian model fit to the central component are below the noise limit in the image. Since we used the sum of the clean components to estimate $S_{\nu}$, and the model flux density, $S_{Gauss}$ is estimated using {\sc imfit}, a point source could have a compactness ratio greater than one.  After visually inspecting $\sim$200 sources with a compactness ratio between 0.95 and 1.0 we find that sources with a compactness ratio greater than 0.975 are usually compact and less than 0.975 have some extended emission. 

As mentioned above, core dominance (referred to here as compactness ratio) was measured by \citet{Fey2002} and shown to be correlated with Source Structure Index, a measurement of how much time the source structure adds to the group delay measurement \citep{Charlot1990}.  
Though our measurement of the compactness ratio does not directly match the value from \citet{Fey2002} due to our use of the clean flux as opposed to their method of model fitting for every component in an image, we have compared sources that are in both samples and found that sources in our study that have a compactness greater than 0.975 typically have a source structure index of 1 or 2 in the \citet{Fey2002} sample, a good indication that these sources are reliable for use in the ICRF. \added{We find that, at X-band, 2351 sources, $\approx65\%$, have a compactness ratio of greater than 0.975, 1020, $\approx28\%$, have a compactness ratio less than 0.975, and 256 sources, $\approx7\%$ are not detected. At S-band we find 2262 soruces, $\approx62\%$, have a compactness ratio of greater than 0.975, 407, $\approx11\%$, have a compactness ratio less than 0.975, and 958 sources, $\approx26\%$ are not detected. At X-band, This is slightly higher than the $60\%$ of sources in \citet{Fey2002} that have a Source Structure Index of 1 or 2, which roughly matches the compactness ratio of greater than 0.975. These sources were determined by \citet{Fey2002} to be ideal for inclusion in the ICRF.}

\subsection{Spectral Index}

\begin{figure}
\centering
\includegraphics[width=0.48\textwidth]{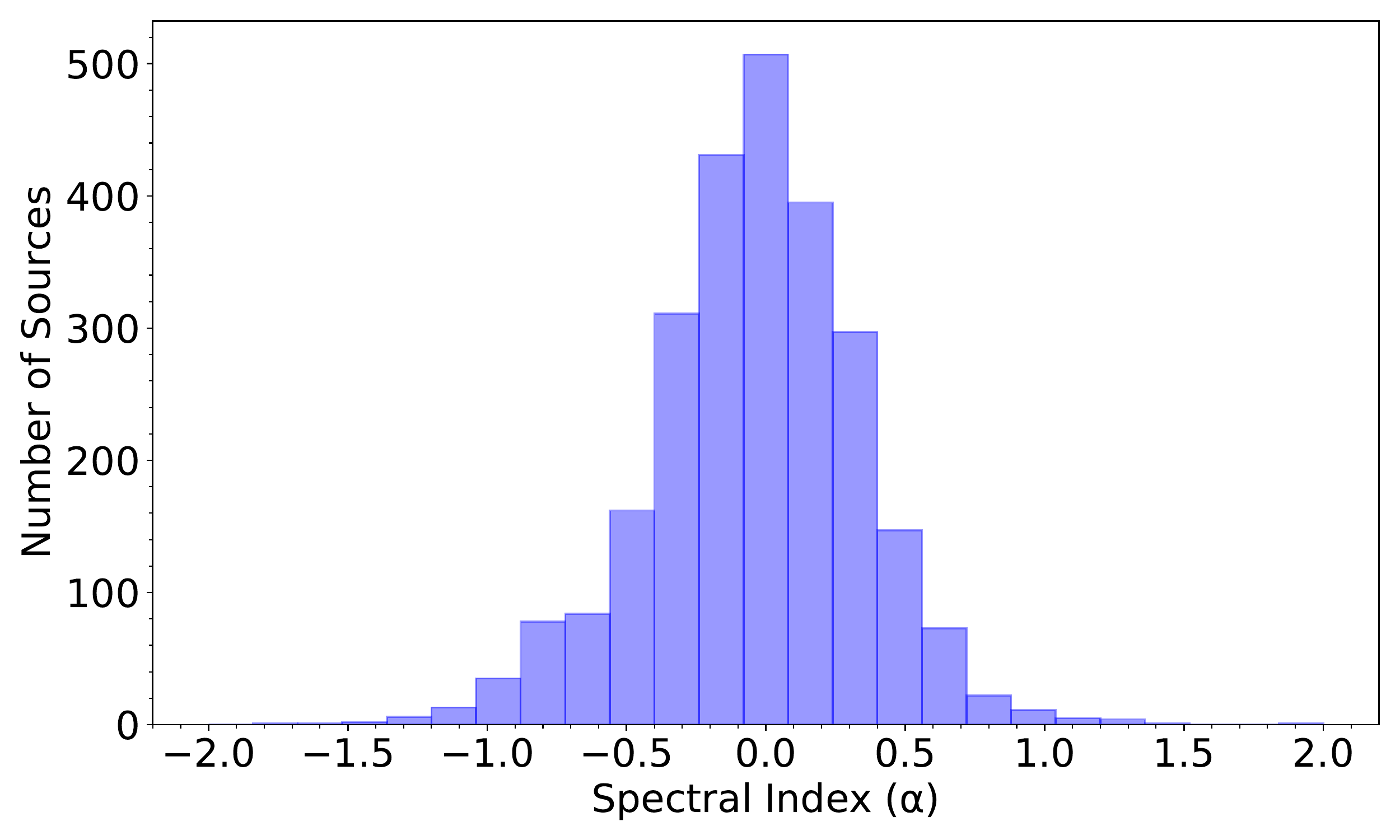}
\caption{Distribution of band-to-band spectral indices.}  
\label{fig:Flux_Spectral_Index}
\end{figure}

We measured the spectral index, $\alpha$,  using the total flux density, $S_{\nu}$ from each band and assuming the flux density changes as S$_{\nu} \propto \nu^{\alpha}$ where S$_{\nu}$ is the flux density at a given frequency $\nu$. With simultaneous S/X band observations, our measurements of the spectral index of sources are free from errors that might otherwise arise from variable fluxes between different epochs. We do acknowledge that the spectral index calculated is biased due to the different spatial resolution of the images at different frequencies. The images generated at 8.7 GHz are not sensitive to some of the extended emission that may be detected at 2.3 GHz, therefore the flux density at 2.3 GHz may include emission that would not be detectable at 8.7 GHz. The spectral index in such a case would be steeper than the actual spectral index of the source. For a point-like source, which we expect most of these sources to be, \deleted{the} all of the emission would \added{be} detected at both frequencies and any difference is due to the spectral index of the source. There will, however, be some number of sources in this catalog for which the spectral index we've measured is steeper than the actual spectral index of the source.  The distribution of spectral index values measured across our sample is shown in Figure \ref{fig:Flux_Spectral_Index}. Spectral indices vary from $-1.82$ to $1.85$ with a median value of $-0.02$. \replaced{This is an indication that most sources targeted in this campaign are near flat spectrum sources.}{We find that 2315 of the 2587, or $89\%$ of sources detected in both bands have  spectral index greater than -0.5, and are defined as flat spectrum sources. The other 272 sources have a spectral index less than -0.5.}

Previous imaging campaigns such as \citet{Fey2002,Pushkarev2012} that used a similar frequency setup and observed sources used in different versions of the ICRF found that most of their sources were also flat spectrum sources. The median spectral index cited by \citet{Fey2002} is $-0.28$.  While the total spectral index was calculated and a histogram was presented in \citet{Pushkarev2012} (see, e.g., their Figure 13), no median value was given, though it lies somewhere in the range between $-0.2$ and $0$. We find similar spectral index trends in our sample which contains roughly eight times the number of sources. The spectral index distribution found here is also similar to the 153 sources detected in the VLBA survey of a complete north polar cap sample at 2.3 and 8.6 GHz \citep{Popkov2020}.

\section{Summary} \label{sec:Summary}

We have presented results from our imaging campaign targeting 3,627 sources in ICRF3 at 2.3 and 8.7 GHz. We have used a CASA pipeline to successfully image 2697 sources at 2.3 GHz and 3209 sources at 8.7 GHz. We imaged 2615 of those sources simultaneously replaced{as}{at} both frequencies. \added{We have shown that data calibration steps using both the AIPS and CASA packages yield the same results, and CASA is ready for calibration and imaging of VLBA data.}

\added{Observations of the fainter sources in this observing campaign helped to improve the accuracy of ICRF3 and the imaging here will help with studies of how source structure affects the accuracy of the ICRF.} We found that the median flux density of our sample is 0.13 Jy at 2.3 GHz and 0.09 Jy as 8.7 GHz \added{which is approximately 6 times fainter than previous campaigns}. We found that \replaced{most sources in our sample are compact, with most of the flux density for any given source coming from a bright central component}{$70\%$ of the sources have a compactness ratio greater than 0.975, indicating that there is little or no emission coming from outside the central, bright component.} . Finally we found that the spectral index of sources in our sample ranges from -1.8 to 1.8 with a median value of -0.02. \replaced{indicating that most sources included in ICRF3 are flat spectrum sources at VLBI resolutions.}{Approximately $90\%$ of the sources in our campaign that were detected at both frequencies have a spectral index greater than -0.5, the cutoff for flat spectrum sources.} 

\begin{deluxetable*}{lccccccccccc}
\tabletypesize{\footnotesize}
\tablecaption{Radio Properties of ICRF Sources \label{tab:Glorious_Table}}
\tablehead{\colhead{Source}  & \colhead{RA} & \colhead{Dec} &  \colhead{date} & \colhead{$\alpha$} & \colhead{$\nu$}&  \colhead{$\sigma_{\rm theor}$} & \colhead{$\sigma_{\rm obs}$} & \colhead{Peak} & \colhead{$S_{\rm \nu}$}  & \colhead{$S_{\rm Gauss}$}  & \colhead{S/N} \\
& \colhead{{\scriptsize (deg)}}&\colhead{{\scriptsize (deg)}} &&& {\scriptsize (GHz)} & \colhead{{\scriptsize(mJy bm$^{-1}$)}}& \colhead{{\scriptsize (mJy bm$^{-1}$)}}& \colhead{{\scriptsize (Jy bm$^{-1}$)}} & \colhead{{\scriptsize (Jy)}} & \colhead{{\scriptsize (Jy)}}}
\startdata
0000-160 & $0.86360065$ & $-15.78484871$ & 2017 Feb 19 & 0.2       & 2.3 GHz & $0.27$  & $0.839$  & $0.062291$ & $0.06085$  & $0.065444$ & $74.2$\\
         &              &                &             &           & 8.6 GHz & $0.191$ & $0.289$  & $0.067377$ & $0.074851$ & $0.075682$ & $233.3$\\
0000-197 & $0.82781262$ & $-19.45620993$ & 2017 Feb 19 & -0.3      & 2.3 GHz & $0.27$  & $0.866$  & $0.082143$ & $0.095579$ & $0.103313$ & $94.9$\\
         &              &                &             &           & 8.6 GHz & $0.191$ & $0.648$  & $0.057467$ & $0.067454$ & $0.06853$  & $88.6$\\
0000-199 & $0.81645585$ & $-19.69733383$ & 2017 Feb 19 & -0.3      & 2.3 GHz & $0.304$ & $2.376$  & $0.196167$ & $0.205944$ & $0.209789$ & $82.6$\\
         &              &                &             &           & 8.6 GHz & $0.215$ & $0.456$  & $0.089882$ & $0.140468$ & $0.122478$ & $197.0$\\
0001+459 & $1.06719853$ & $46.25499185$  & 2017 Mar 23 & -0.0      & 2.3 GHz & $0.352$ & $0.818$  & $0.15733$  & $ 0.157872$& $0.163444$ & $192.3$\\
         &              &                &             &           & 8.6 GHz & $0.249$ & $0.365$  & $0.150068$ & $0.155065$ & $0.15553$  & $411.1$\\
0001+478 & $0.94183991$ & $48.11781535$  & 2017 Mar 23 & -1.2      & 2.3 GHz & $0.269$ & $1.933$  & $0.157052$ & $0.238967$ & $0.259524$ & $81.3$\\
         &              &                &             &           & 8.6 GHz & $0.19$  & $1.582$  & $ 0.05662$ & $0.051817$ & $0.056724$ & $35.8$\\
0001-120 & $1.02047917$ & $-11.81621839$ & 2017 Feb 24 & -0.0      & 2.3 GHz & $0.238$ & $1.223$  & $0.540535$ & $0.624813$ & $0.60506$  & $442.1$\\
         &              &                &             &           & 8.6 GHz & $0.168$ & $0.817$  & $0.462684$ & $0.60403$  & $0.588762$ & $566.1$\\
0002+051 & $1.33423129$ & $5.403001$     & 2017 Jul 16 & -0.5      & 2.3 GHz & $0.313$ & $2.031$  & $0.17243$  & $0.175952$ & $0.173318$ & $84.9$\\
         &              &                &             &           & 8.6 GHz & $0.221$ & $1.211$  & $0.083285$ & $0.093682$ & $0.086632$ & $68.8$\\
0002+200 & $1.14899286$ & $20.32842159$  & 2017 May 27 & 0.1       & 2.3 GHz & $0.238$ & $1.815$  & $0.286337$ & $0.33006$  & $0.322637$ & $157.8$\\
         &              &                &             &           & 8.6 GHz & $0.168$ & $0.541$  & $0.252224$ & $0.362979$ & $0.365341$ & $466.4$\\
0002+541 & $1.26818061$ & $54.47359014$  & 2017 Mar 27 & 0.3       & 2.3 GHz & $0.27$  & $2.362$  & $0.263699$ & $0.256789$ & $0.264302$ & $111.7$\\
         &              &                &             &           & 8.6 GHz & $0.191$ & $0.671$  & $0.343928$ & $0.390912$ & $0.388459$ & $512.8$\\
0002-170 & $1.32472412$ & $-16.8012996$  & 2017 Feb 24 & -0.2      & 2.3 GHz & $0.338$ & $0.865$  & $0.1509$   & $0.169368$ & $0.172039$ & $174.4$\\
         &              &                &             &           & 8.6 GHz & $0.239$ & $0.373$  & $0.091568$ & $0.130629$ & $0.11989$  & $245.7$ \\
0002-350 & $1.27468791$ & $-34.76379337$ & 2017 Jan 16 & \nodata   & 2.3 GHz & $0.292$ & $30.877$ & $0.081618$ & $0.0$      & \nodata    & $2.6$ \\
         &              &                &             &           & 8.6 GHz & $0.206$ & $0.425$  & $0.098578$ & $0.102574$ & $0.103216$ & $232.0$\\
         &              &                & 2017 Jan 22 & \nodata   & 2.3 GHz & $0.577$ & \nodata  & \nodata    & \nodata    & \nodata    & \nodata \\
         &              &                &             &           & 8.6 GHz & $0.407$ & $1.517$  & $0.094508$ & $0.089684$ & $0.098981$ & $62.3$\\
\hline
\multicolumn{4}{l}{Notes: $^a$ indicates ICRF3 Defining Source}\\
\enddata
\caption{The Source is the source name in the observation. RA and Dec are the Right Ascension and Declination of the source taken from the ICRF3 catalog \citep{charlot2020} with precision to the median uncertainty of the frame to 0.1 mas. Date is the observation date for a given image. $\alpha$ is the 2.3 GHz to 8.7.GHz spectral index of a source, $\nu$ is the frequency at which the values in the next colums are measured. $\sigma_{\rm theor}$ and $\sigma_{\rm obs}$ are the theoretical and measured RMS in the image. Peak is the peak flux density of the brightest pixel in an image in Jy bm$^{-1}$. $S_{\rm \nu}$ is the total flux density of the image determined by summing the clean components.  $S_{\rm Gauss}$ is the flux density calculated from fitting a Gaussian to the brightest component in an image.  S/N is the signal-to-noise ratio of an image calculated as the peak flux density divided by $\sigma_{\rm obs}$. The full, machine readable table will be available through the journal. } 
\end{deluxetable*}

\section{Acknowledgements}
We sincerely thank Justin Linford for help with our CASA/AIPS/Difmap calibration comparison. We thank Bob Zavala, Brian Luzum, Mike Dutka, and Bryan Hemingway for helpful feedback. 

The National Radio Astronomy Observatory is a facility of the National Science Foundation operated under cooperative agreement by Associated Universities, Inc.  The authors acknowledge use of the Very Long Baseline Array under the US Naval Observatory's time allocation. This work supports USNO's ongoing research into the celestial reference frame and geodesy.

\begin{figure*}[h]
\figurenum{11}
    \centering
    \gridline{\fig{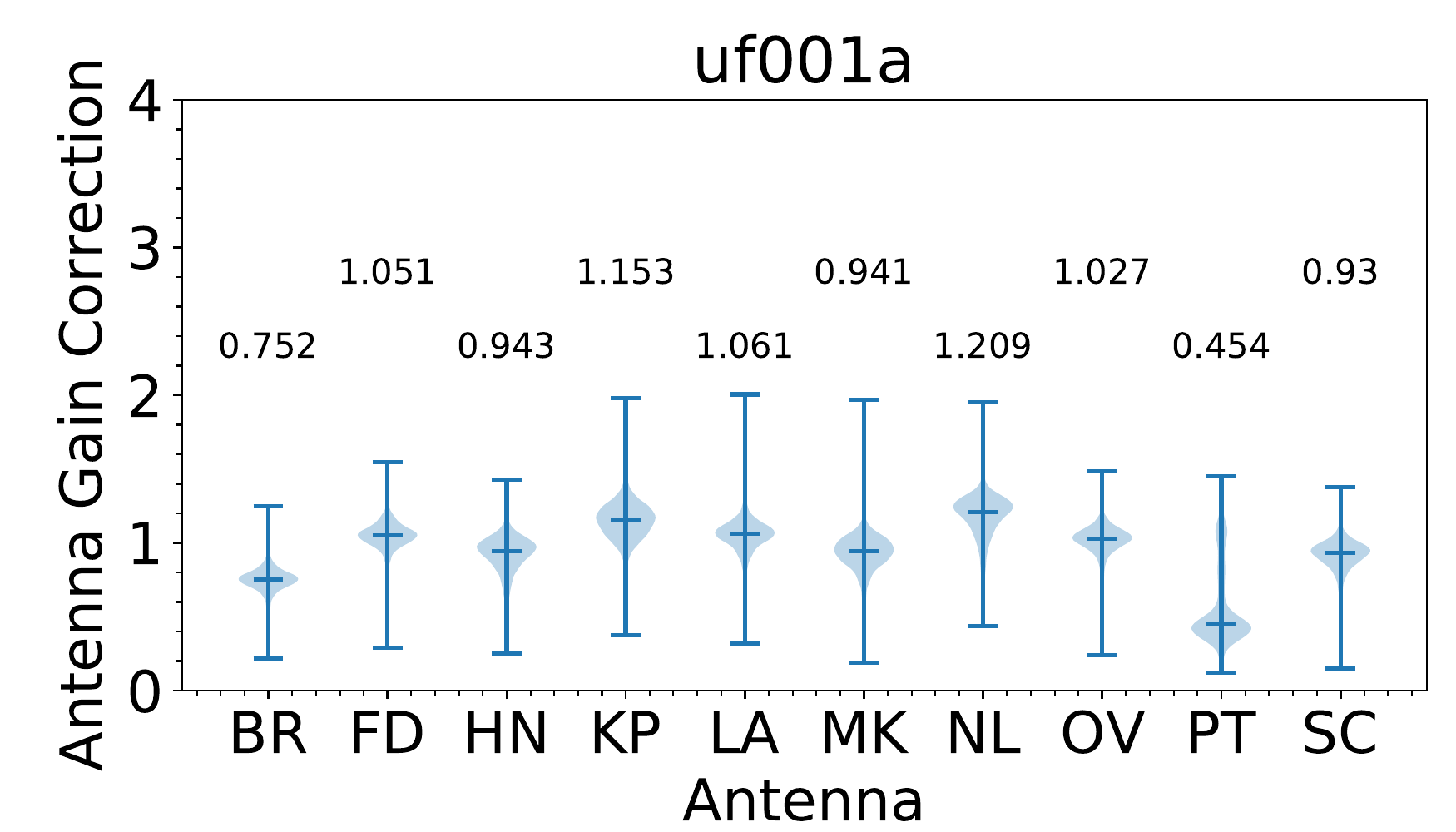}{0.48\linewidth}{(a)}
              \fig{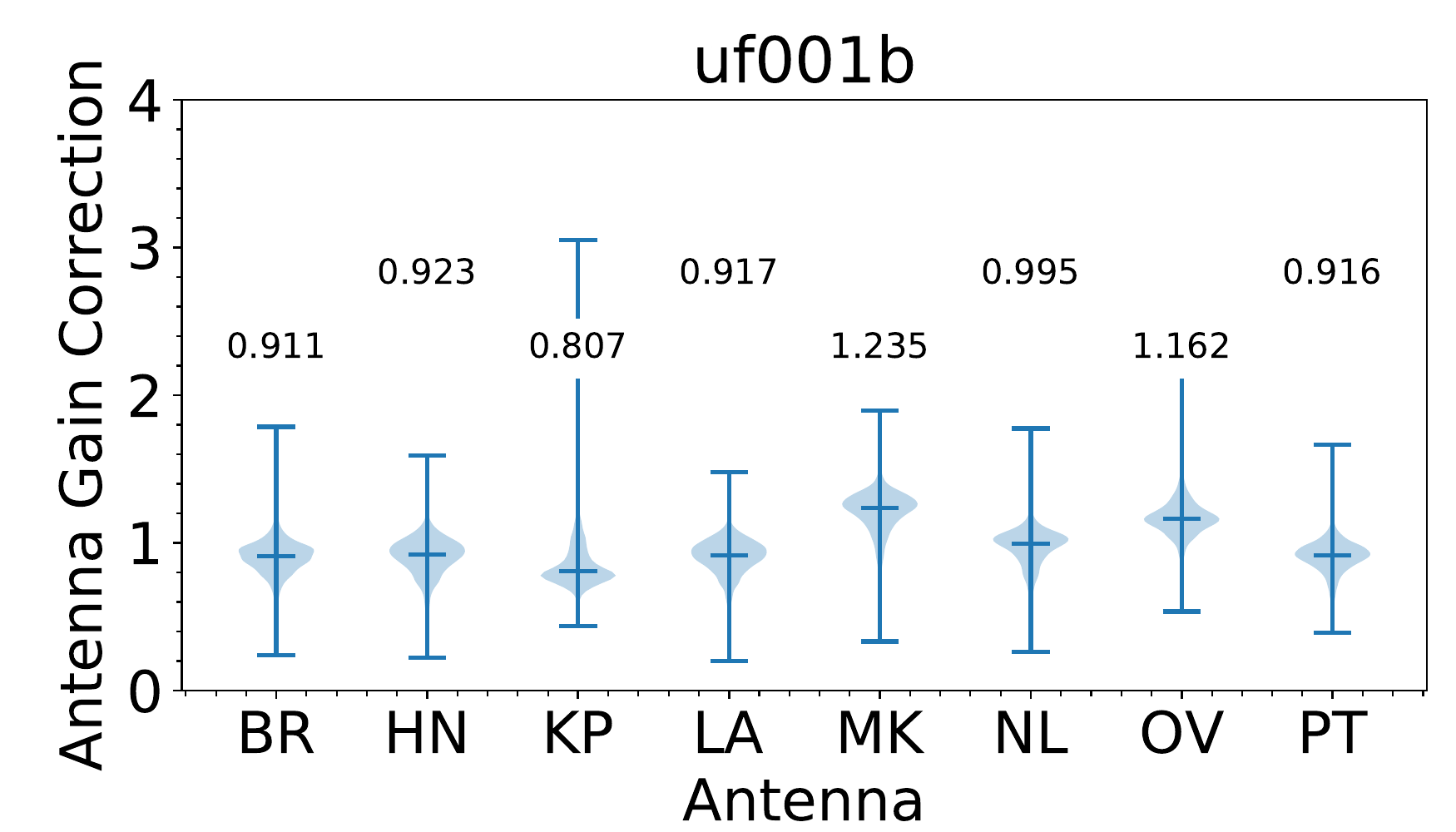}{0.48\linewidth}{(b)}}
    \gridline{\fig{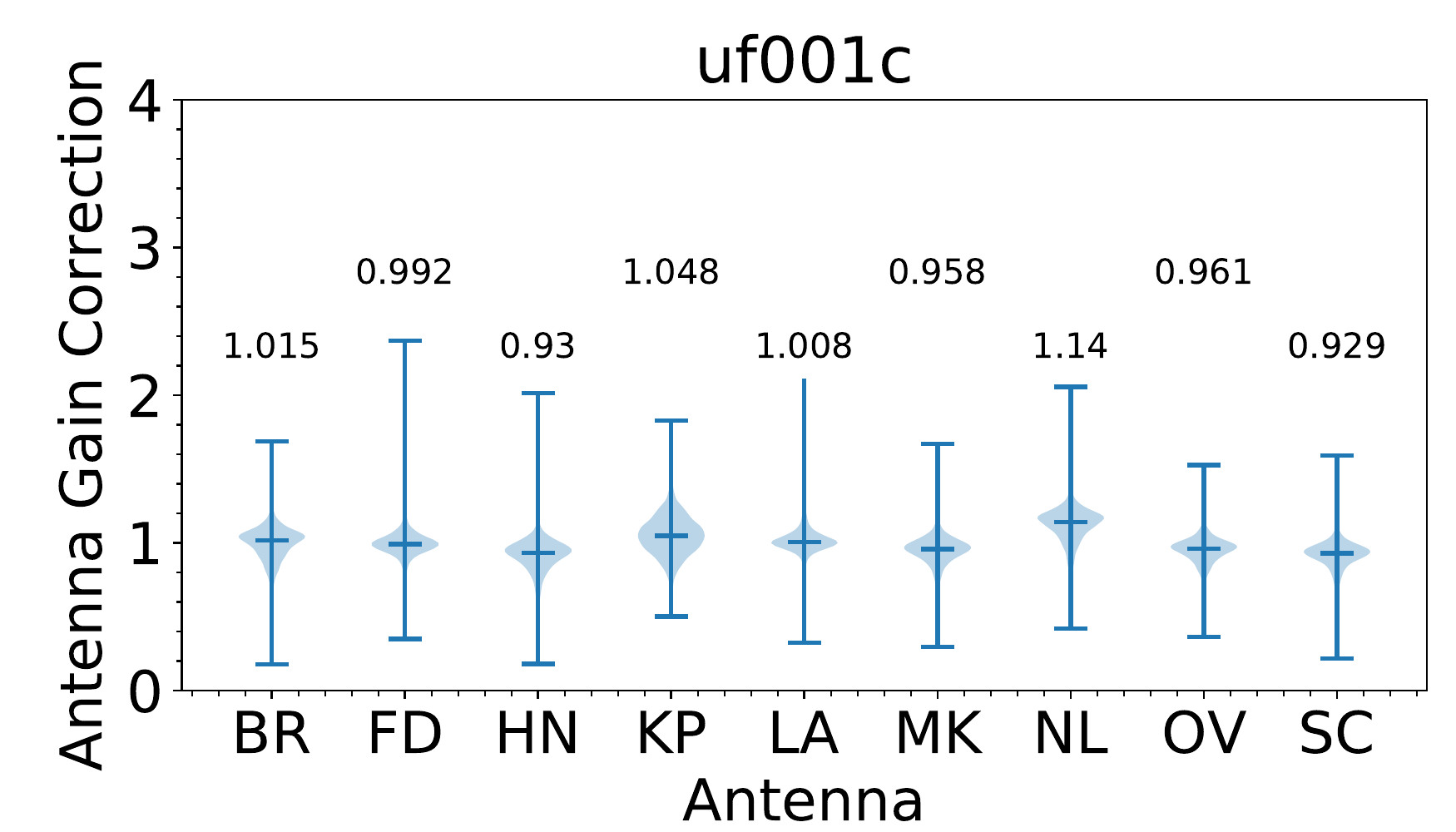}{0.48\linewidth}{(c)}
              \fig{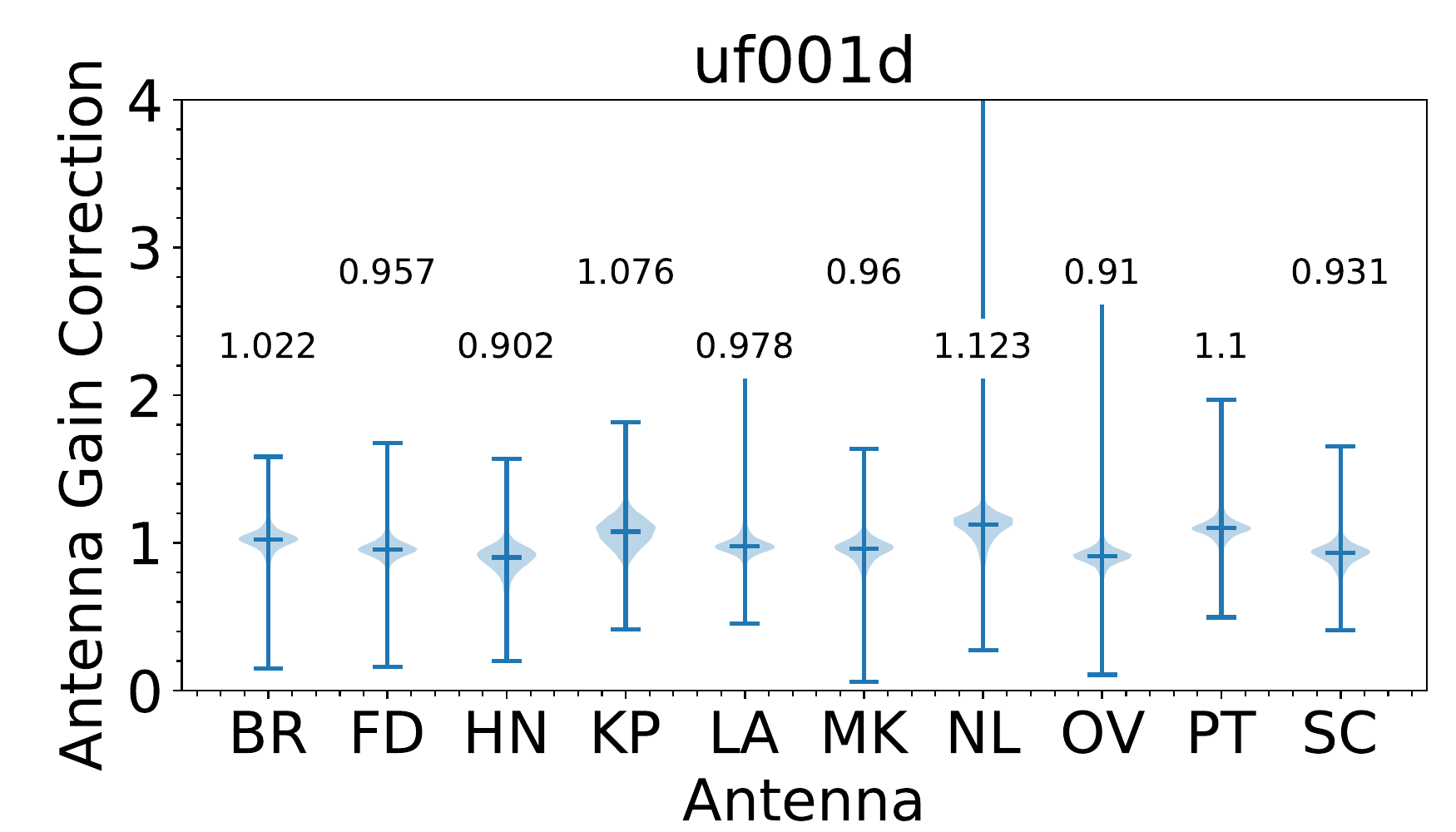}{0.48\linewidth}{(d)}}
    \gridline{\fig{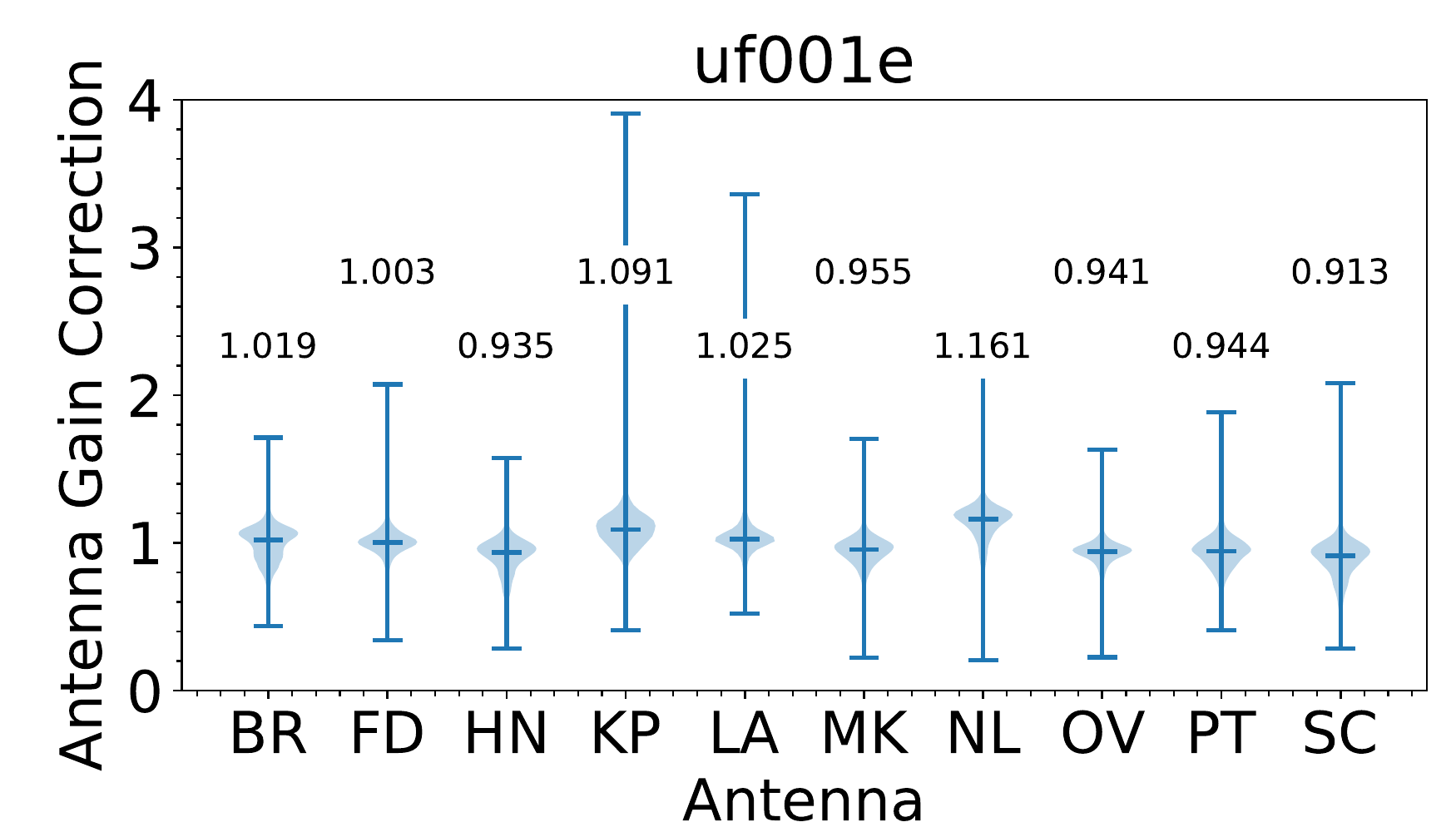}{0.48\linewidth}{(e)}
              \fig{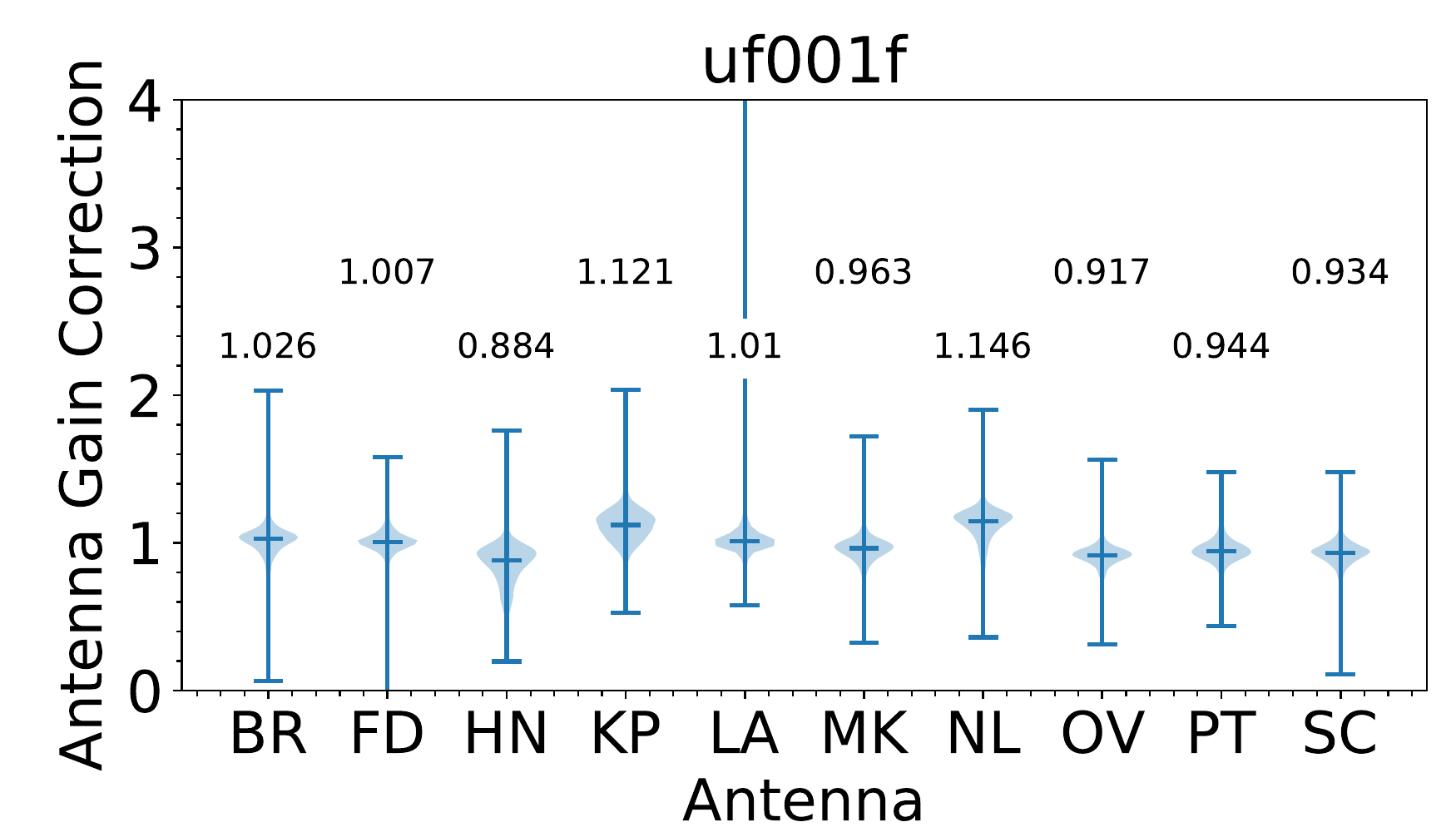}{0.48\linewidth}{(f)}}
    \caption{Violin plots showing the distribution and span of the gain corrections calculated from self-calibration for each antenna for each scan at 8.7 GHz. The median value of the gain corrections is indicated by the horizontal line in the center, and shown as text in each plot. }
    \label{fig:X_BAND_GAINCAL_VIOLIN_PLOTS}
\end{figure*}

\appendix
\added{\section{Gain Correction Factors}\label{app:gain_corr}

As mentioned in Section \ref{sec:imaging}, the final self-calibration step adjusts the gains per antenna, per scan to re-normalize the flux density scale. We include plots that show the median, range, and density of corrections to the antenna gain for each experiment and each imaging band in Figures \ref{fig:X_BAND_GAINCAL_VIOLIN_PLOTS} and \ref{fig:S_BAND_GAINCAL_VIOLIN_PLOTS}.  We also include a table listing the median gain for each antenna. }

\begin{figure*}[h]
\figurenum{11}
    \gridline{\fig{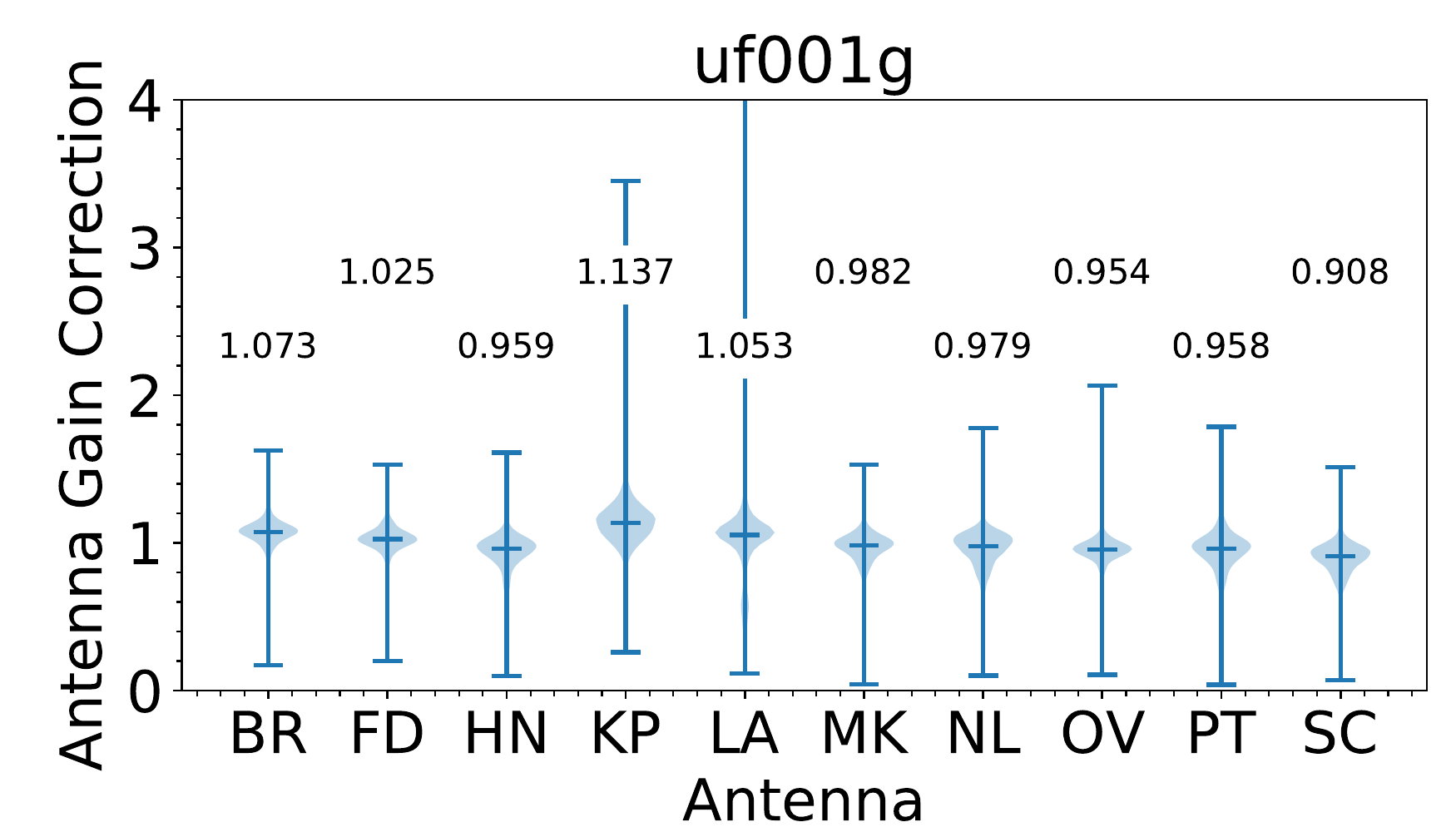}{0.48\linewidth}{(g)}
              \fig{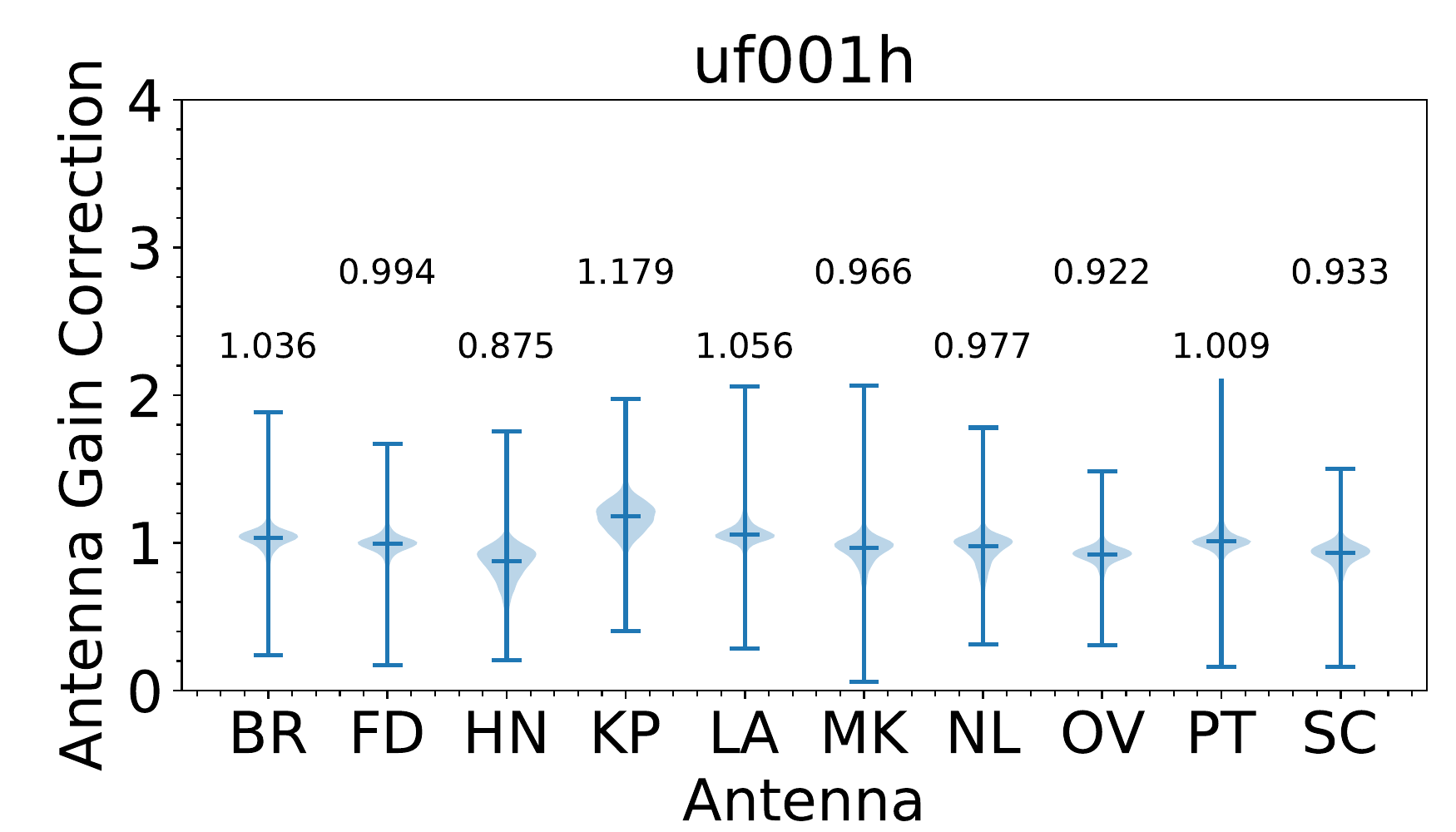}{0.48\linewidth}{(h)}}
    \gridline{\fig{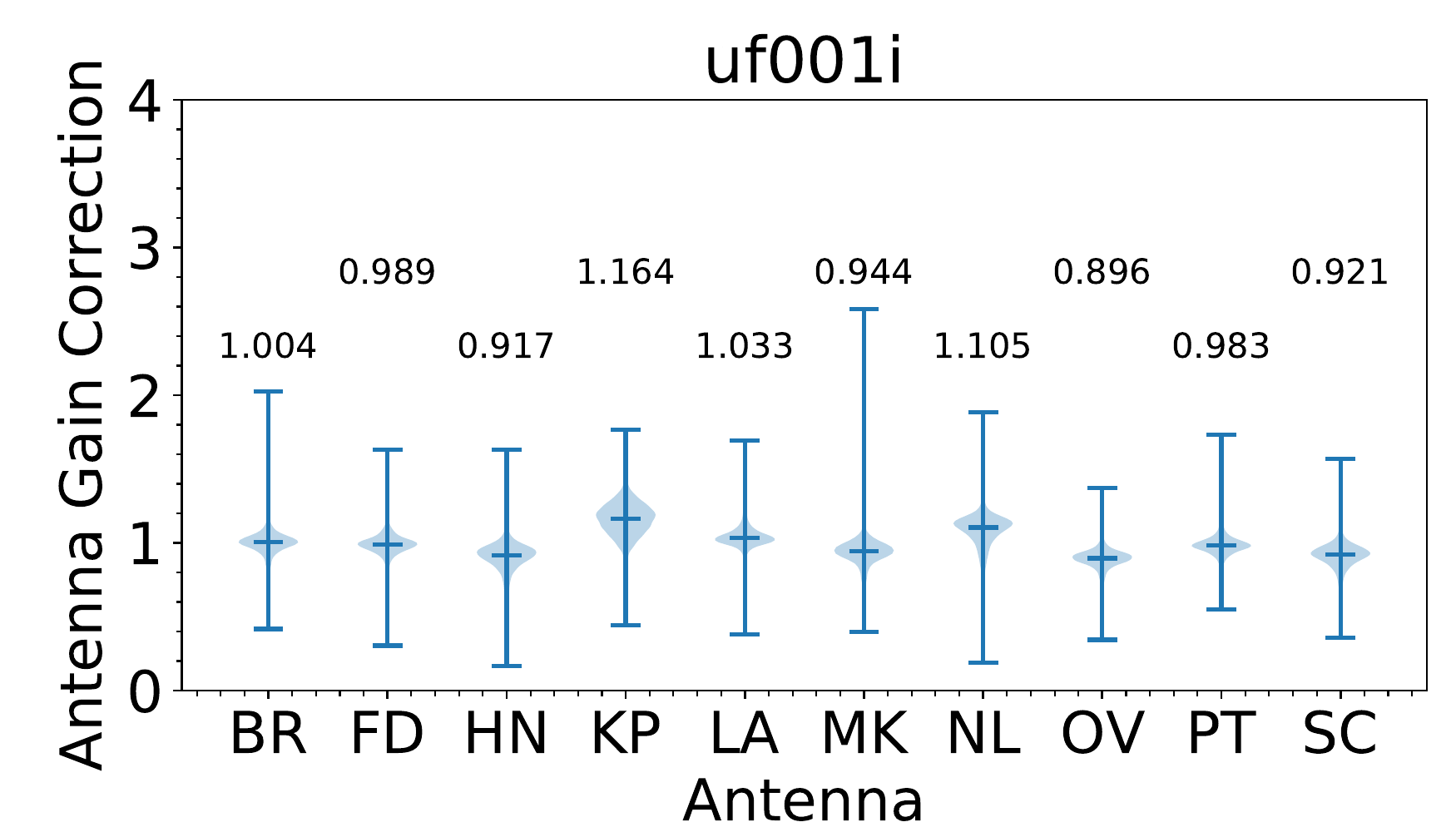}{0.48\linewidth}{(i)}
              \fig{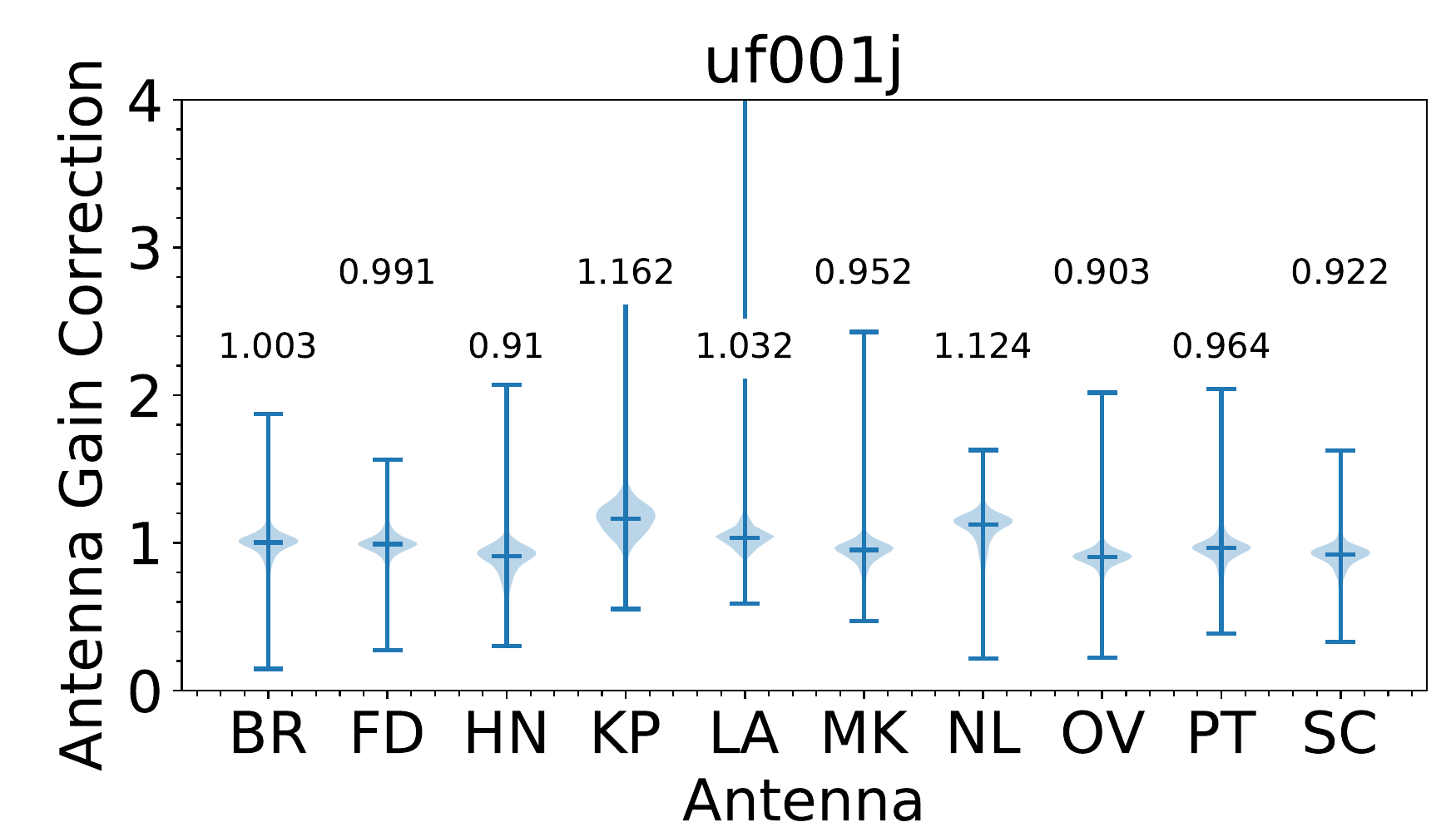}{0.48\linewidth}{(j)}}
    \gridline{\fig{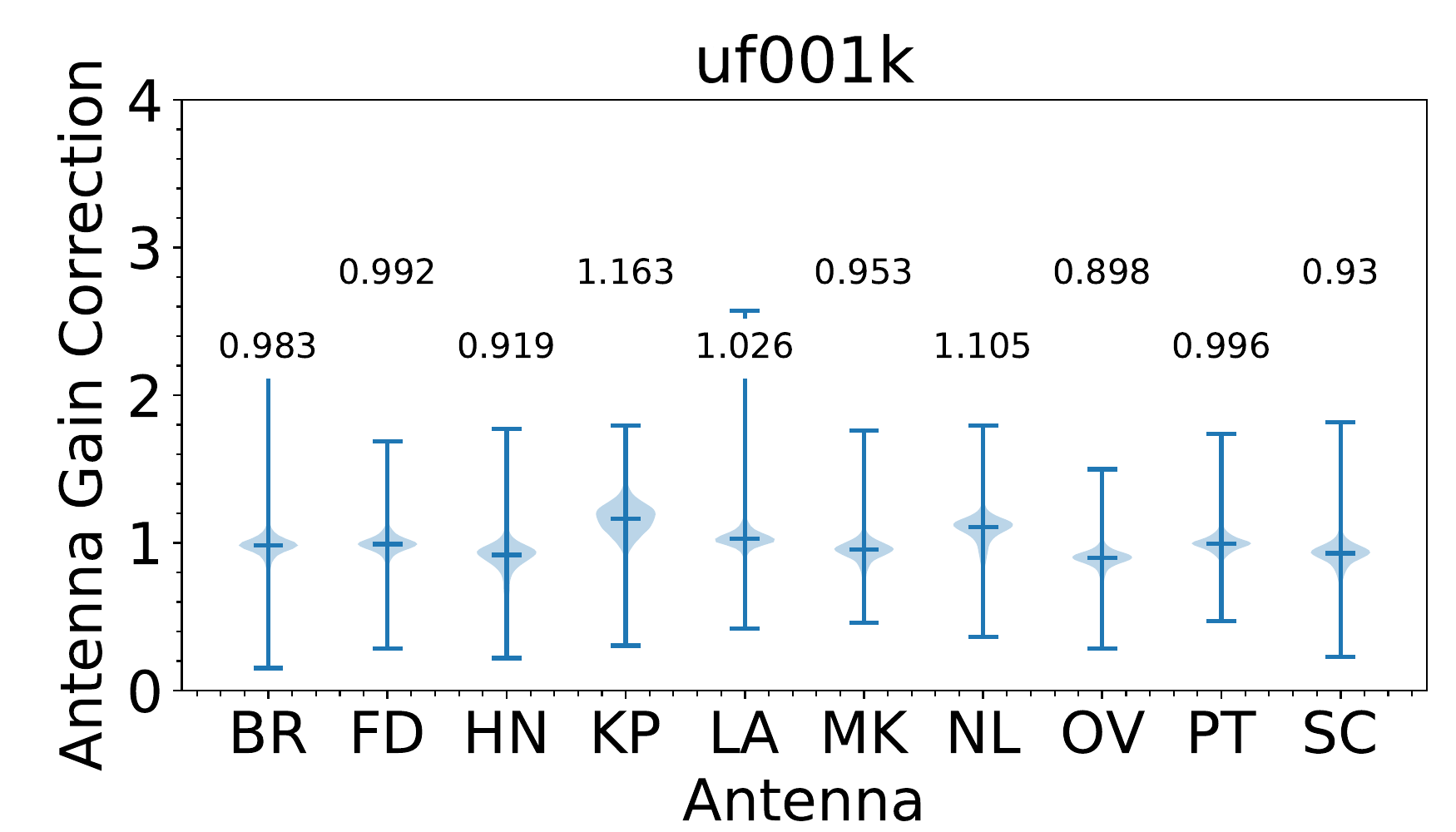}{0.48\linewidth}{(k)}
              \fig{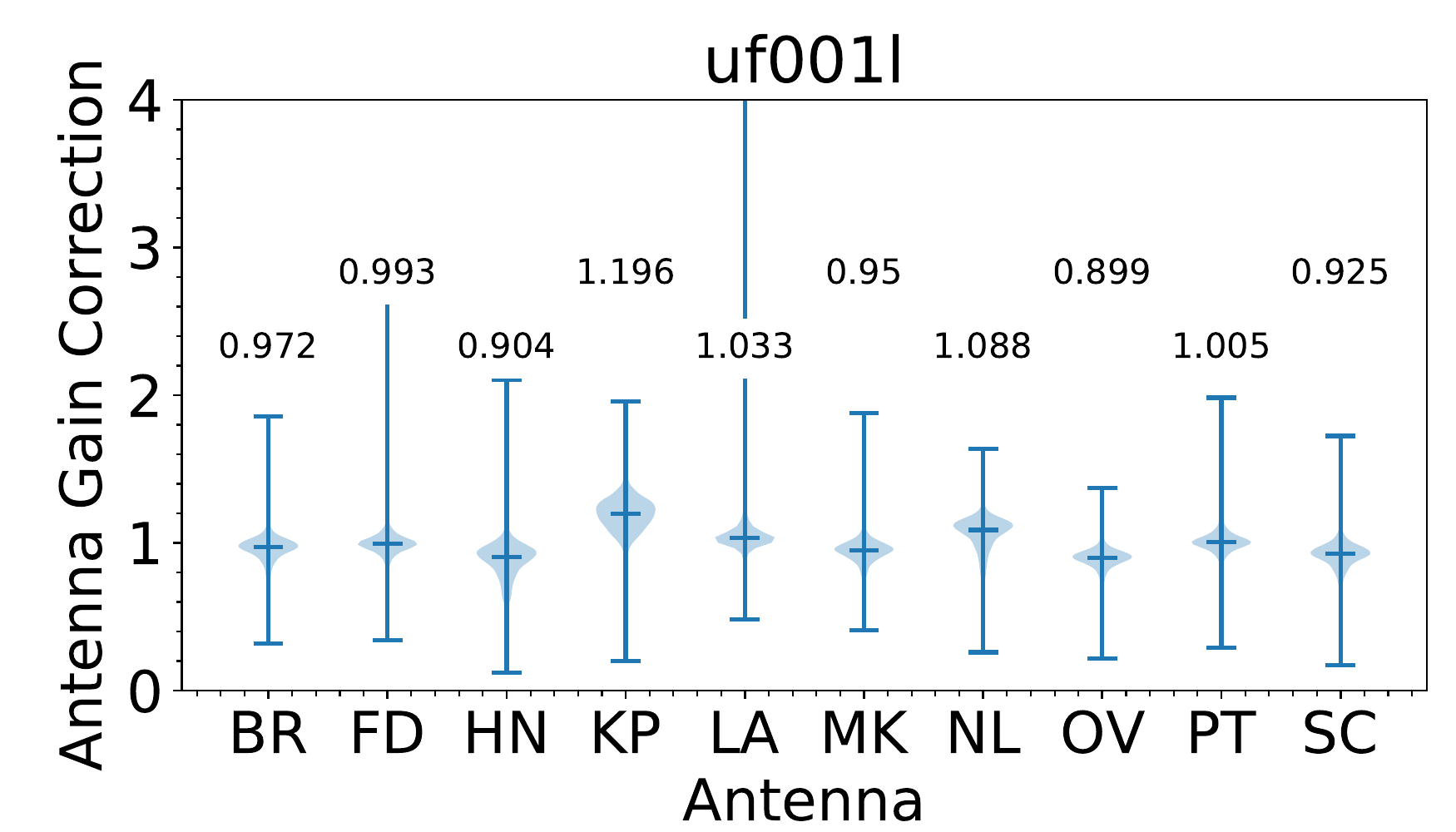}{0.48\linewidth}{(l)}}
    \caption{Violin plots showing the distribution and span of the gain corrections calculated from self-calibration for each antenna for each scan at 8.7 GHz. The median value of the gain corrections is indicated by the horizontal line in the center, and shown as text in each plot. }
    \label{fig:X_BAND_GAINCAL_VIOLIN_PLOTS_2}
\end{figure*}
\begin{figure*}[h]
\figurenum{11}
    \gridline{\fig{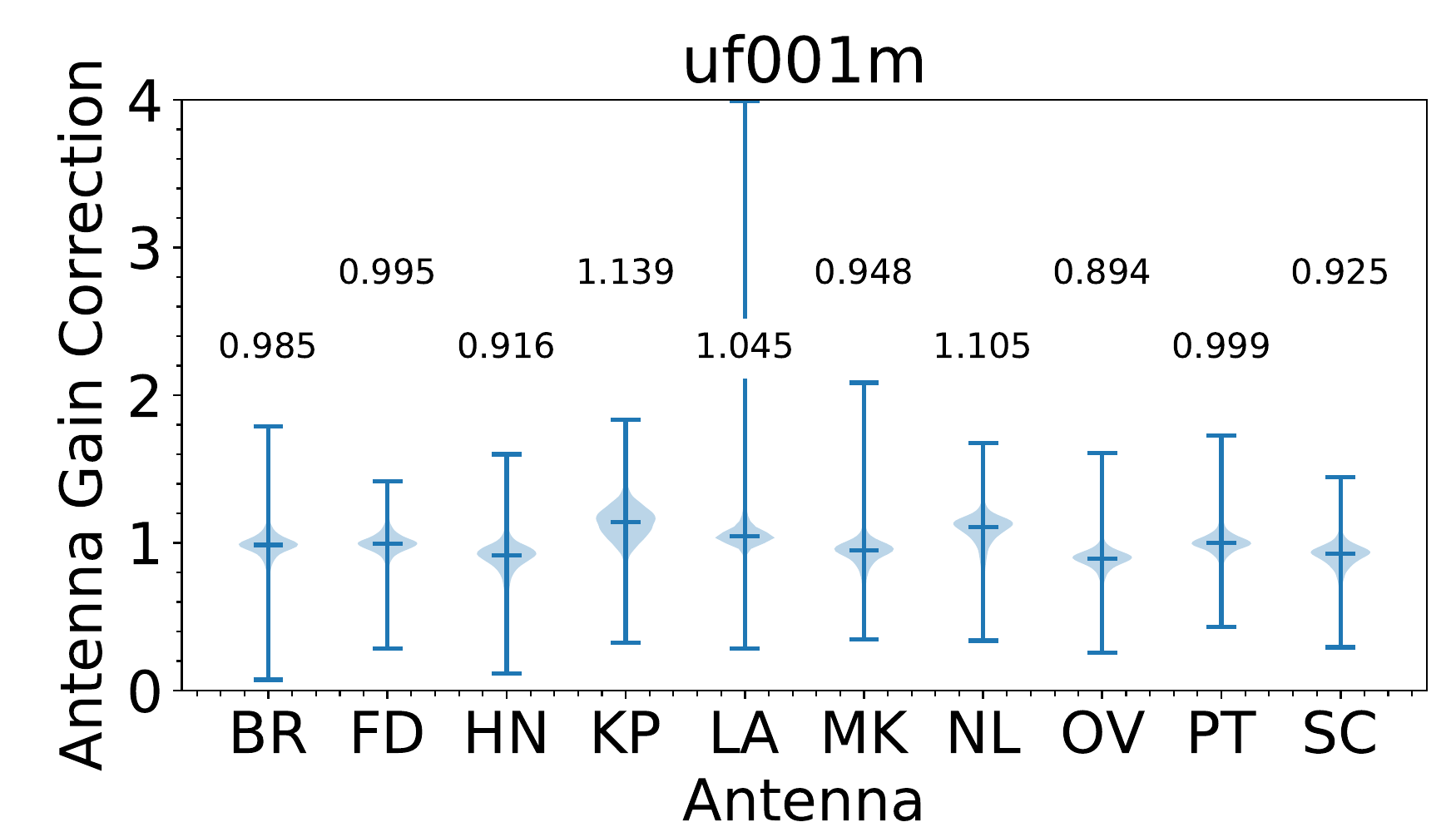}{0.48\linewidth}{(m)}
              \fig{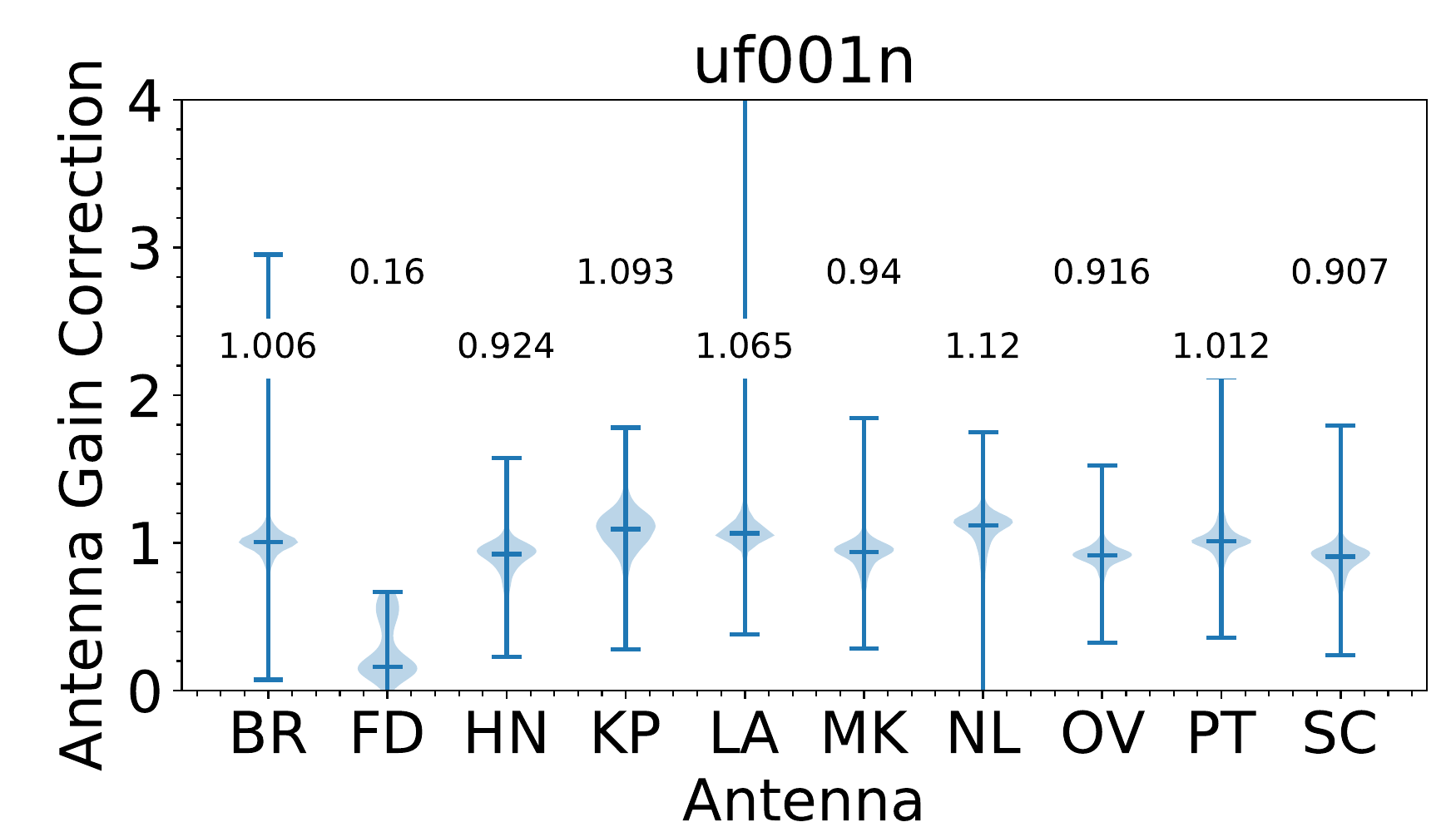}{0.48\linewidth}{(n)}}
    \gridline{\fig{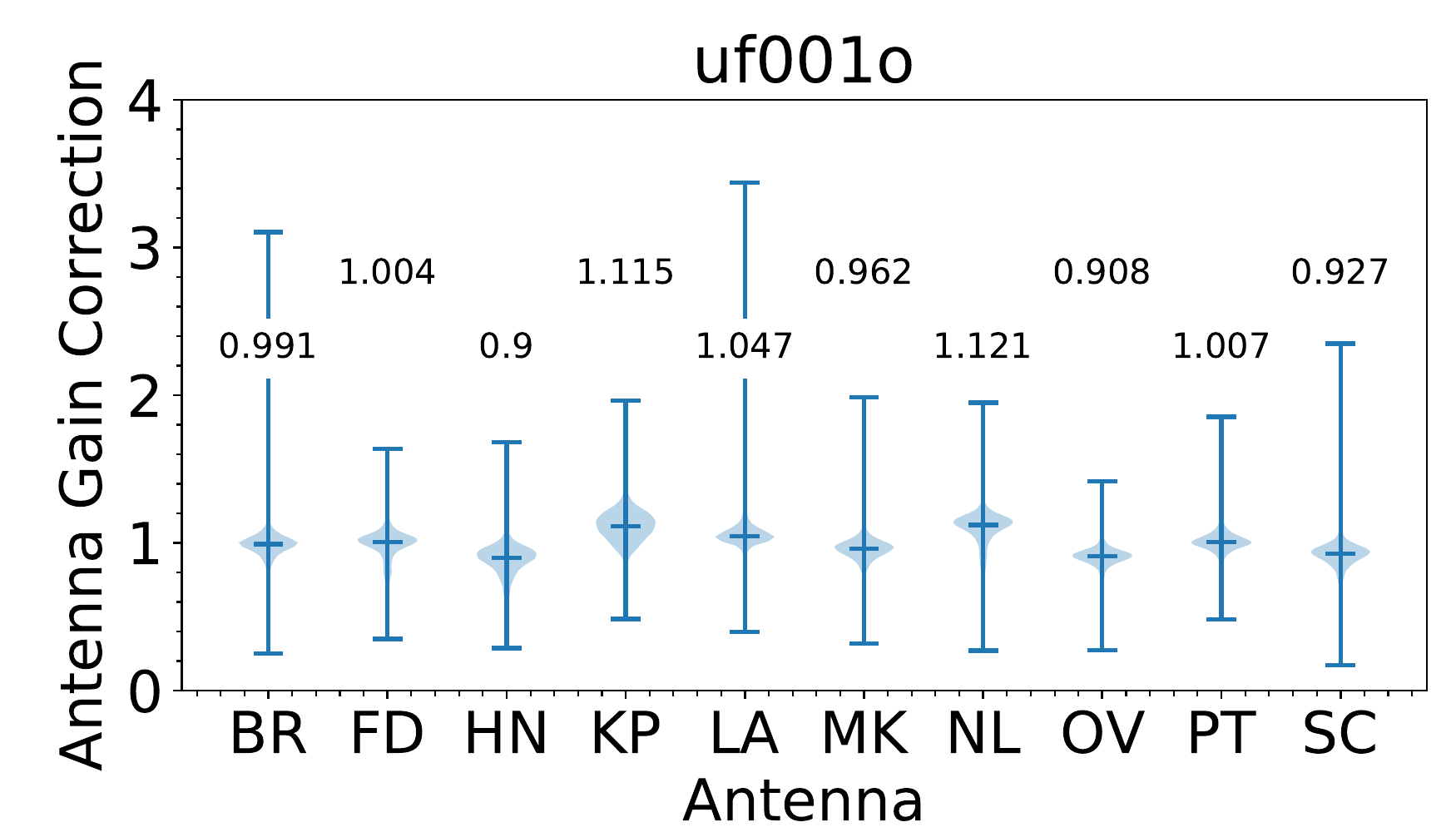}{0.48\linewidth}{(o)}
              \fig{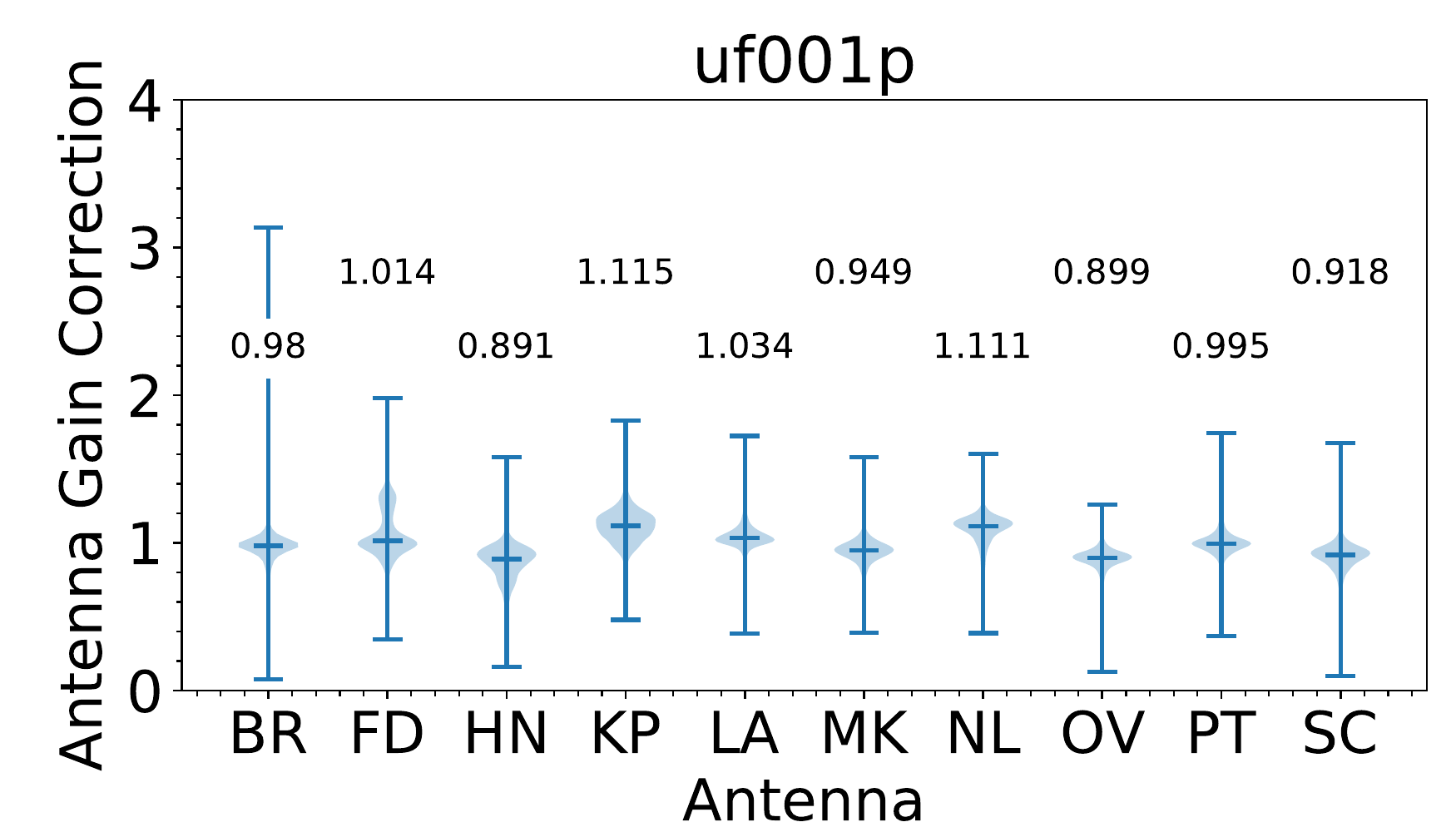}{0.48\linewidth}{(p)}}
    \gridline{\fig{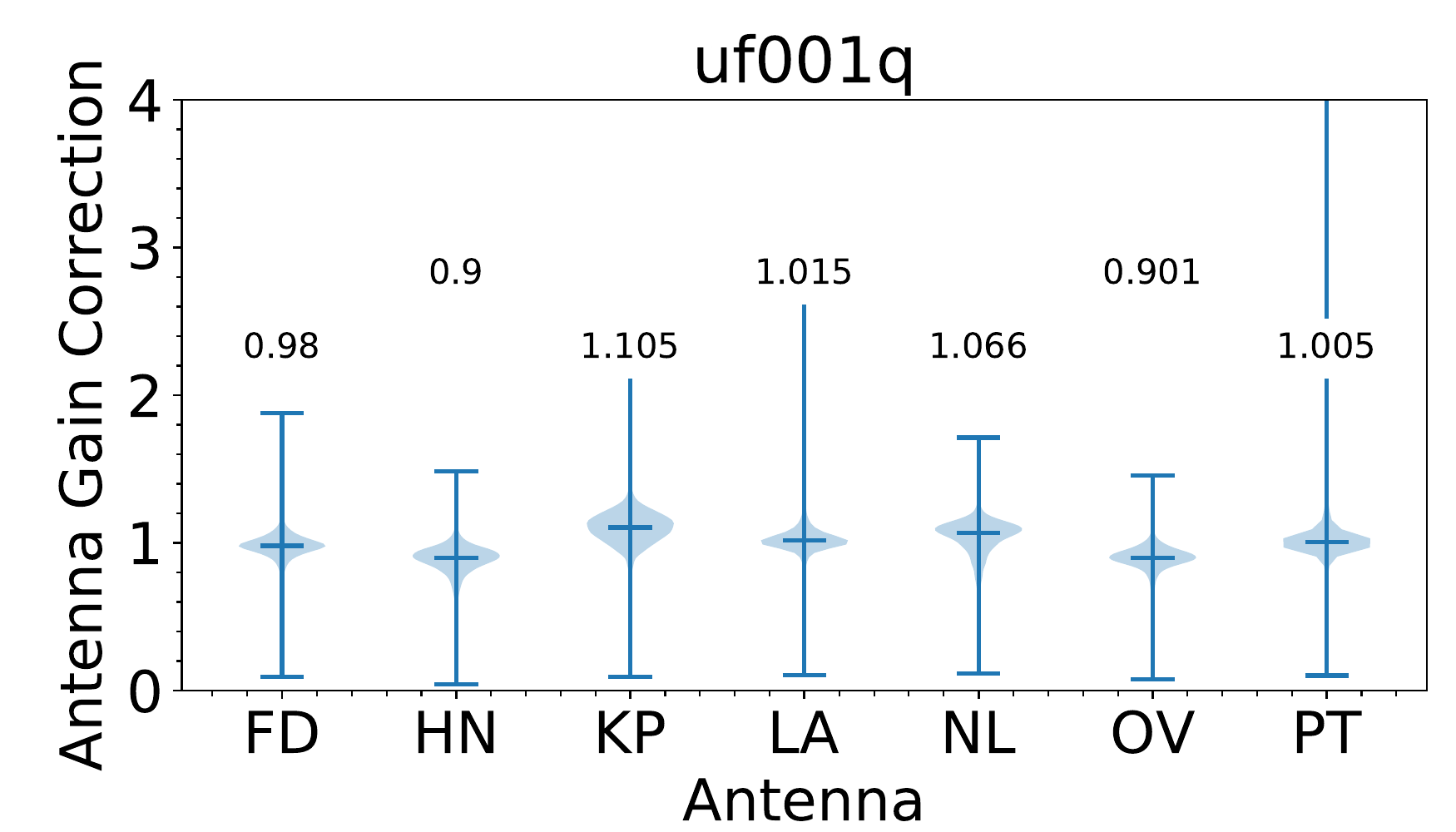}{0.48\linewidth}{(q)}
              \fig{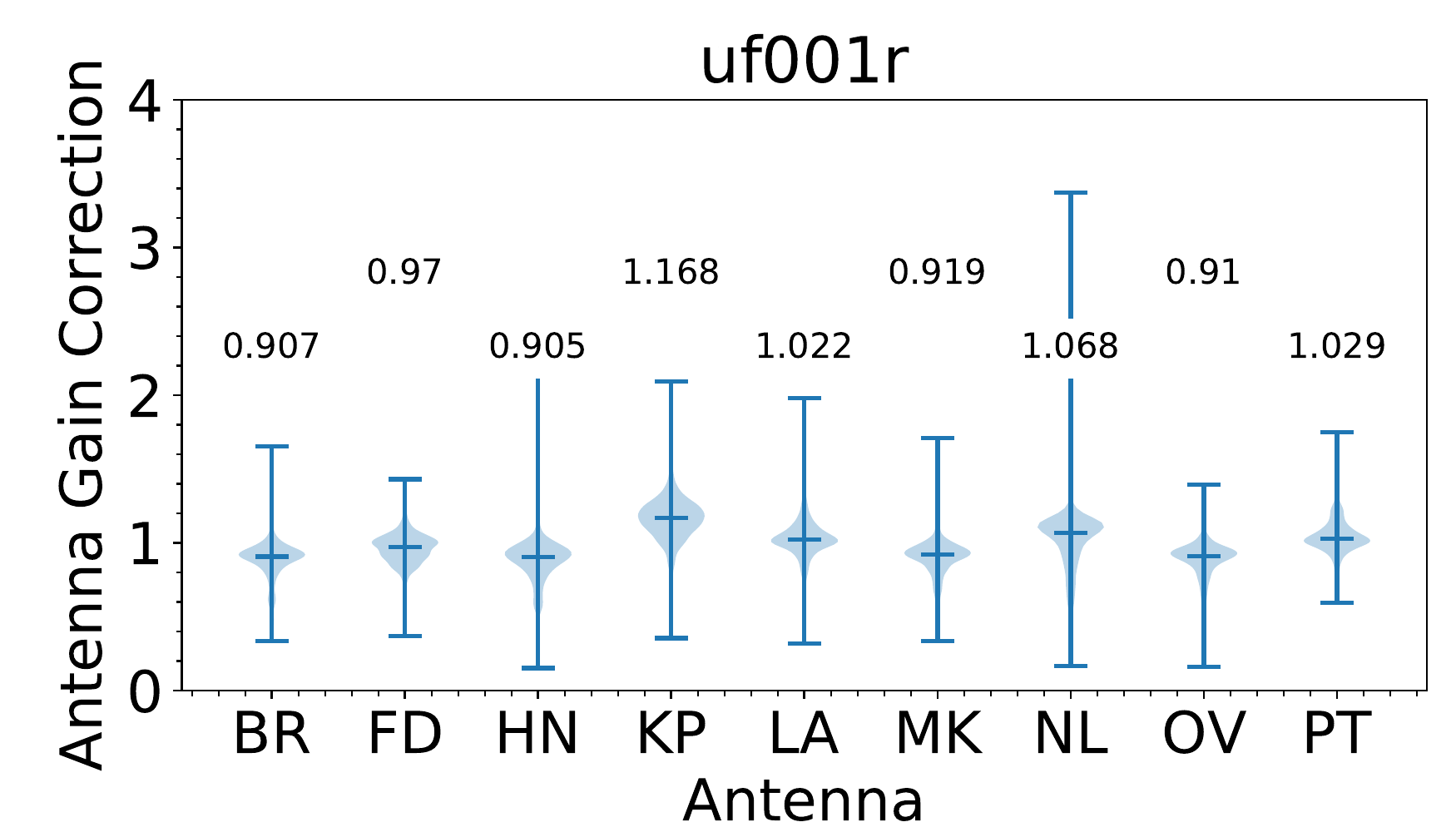}{0.48\linewidth}{(r)}}
    \caption{Violin plots showing the distribution and span of the gain corrections calculated from self-calibration for each antenna for each scan at 8.7 GHz. The median value of the gain corrections is indicated by the horizontal line in the center, and shown as text in each plot. }
    \label{fig:X_BAND_GAINCAL_VIOLIN_PLOTS_3}
\end{figure*}
\begin{figure*}[h]
\figurenum{11}
    \gridline{\fig{uf001q_x_violinplot.pdf}{0.48\linewidth}{(s)}
              \fig{uf001r_x_violinplot.pdf}{0.48\linewidth}{(t)}}
    \caption{Violin plots showing the distribution and span of the gain corrections calculated from self-calibration for each antenna for each scan at 8.7 GHz. The median value of the gain corrections is indicated by the horizontal line in the center, and shown as text in each plot. }
    \label{fig:X_BAND_GAINCAL_VIOLIN_PLOTS_4}
\end{figure*}

\begin{figure*}[h]
\figurenum{12}
    \centering
    \gridline{\fig{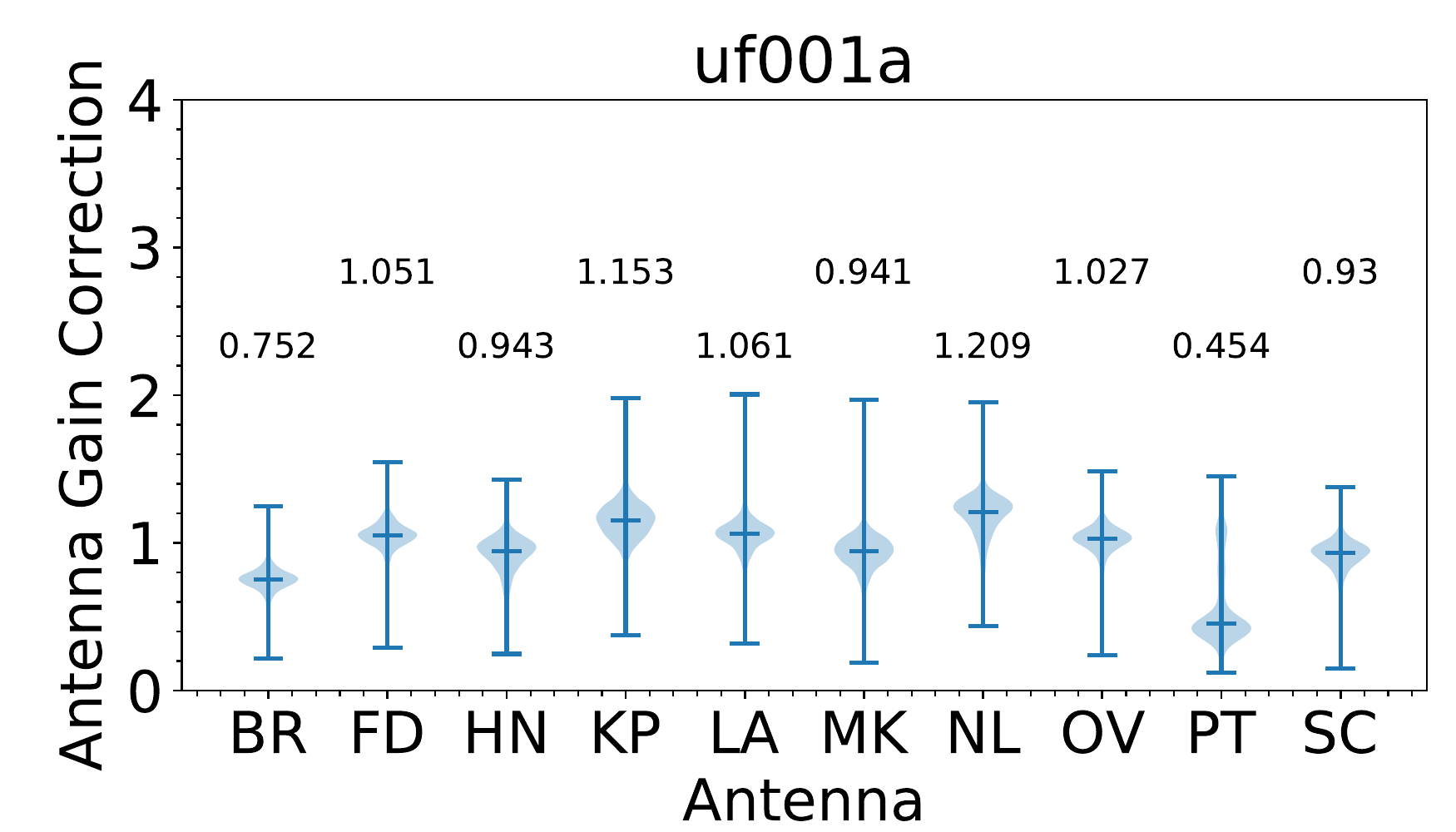}{0.48\linewidth}{(a)}
              \fig{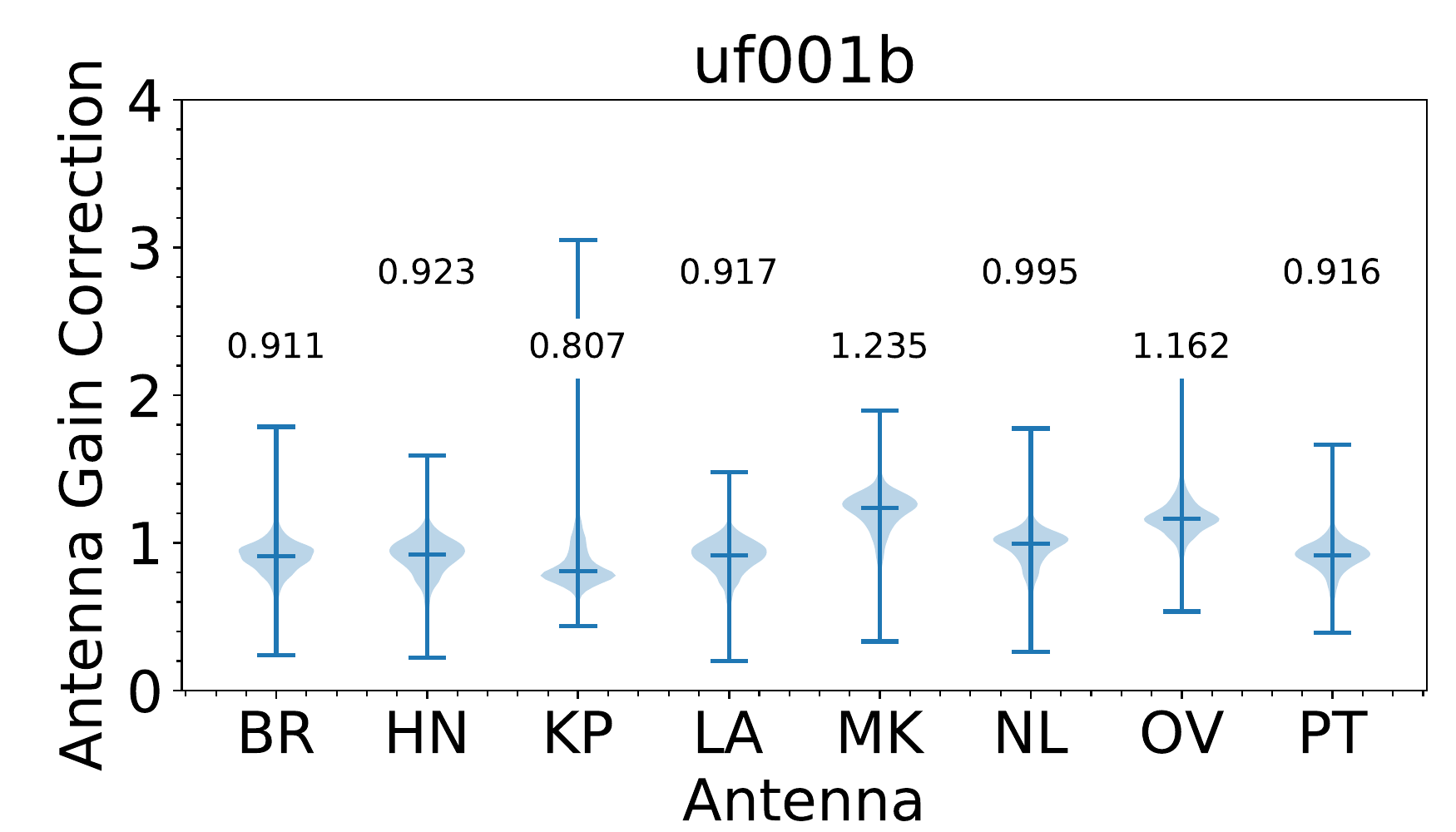}{0.48\linewidth}{(b)}}
    \gridline{\fig{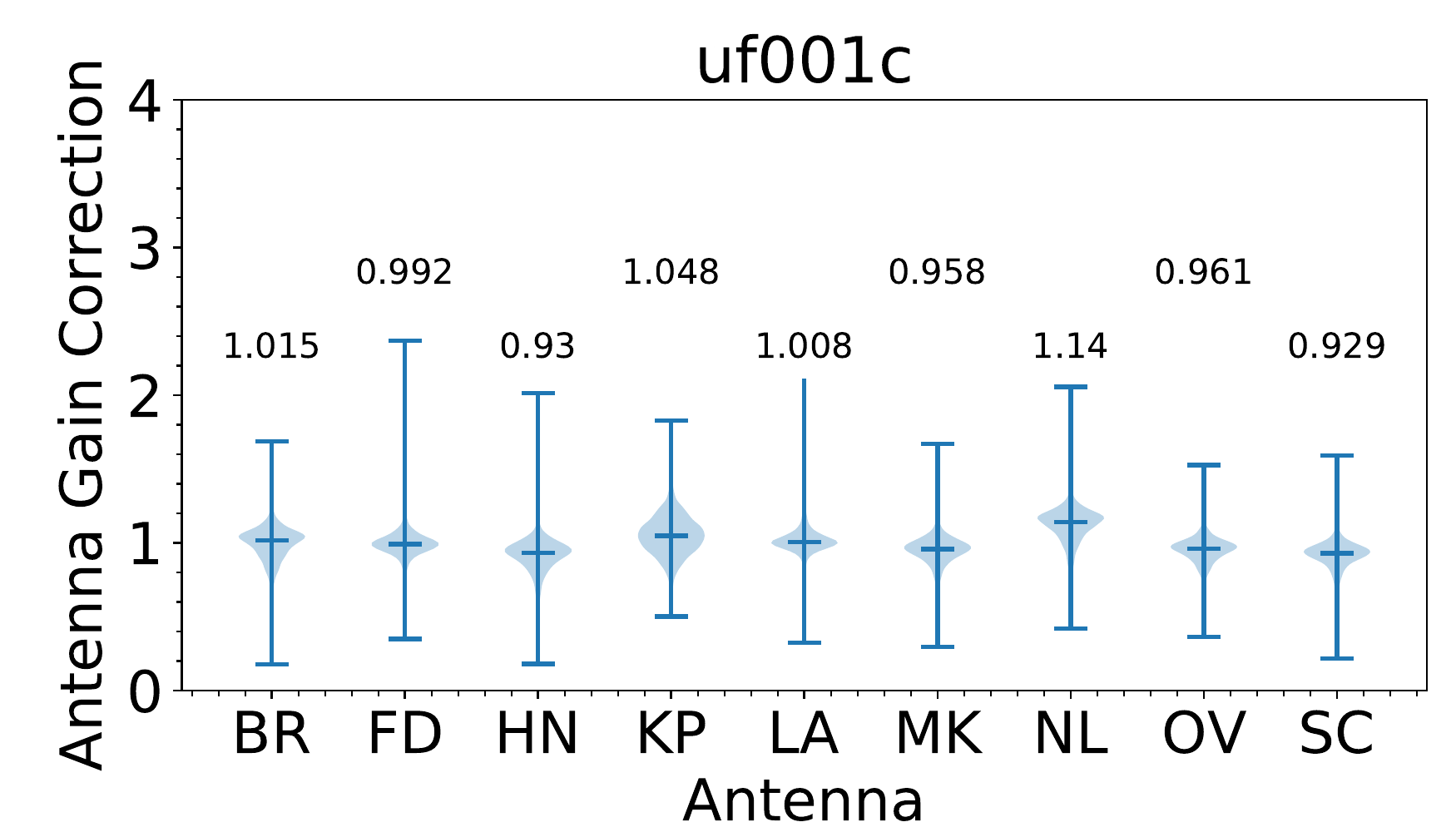}{0.48\linewidth}{(c)}
              \fig{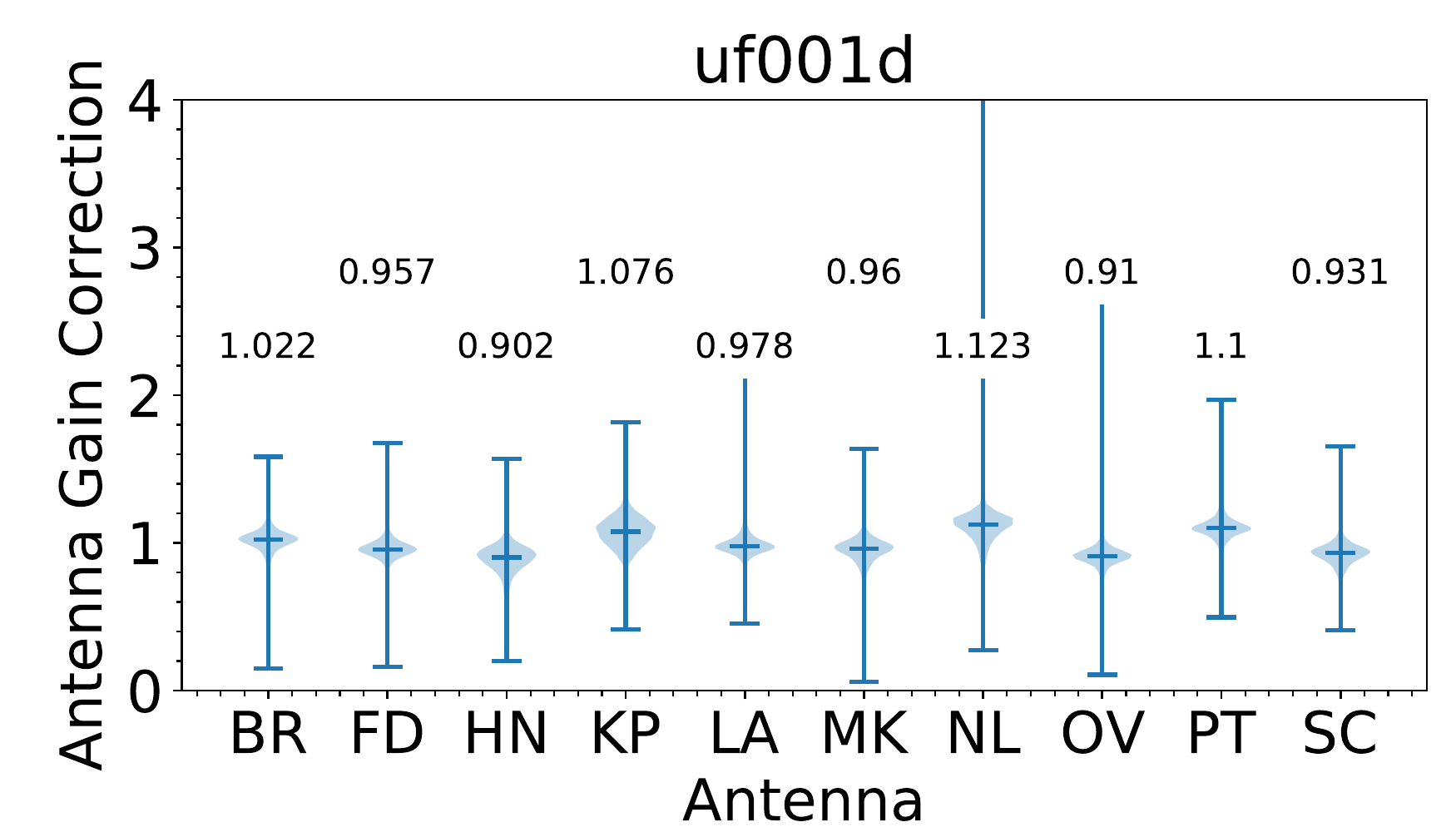}{0.48\linewidth}{(d)}}
    \gridline{\fig{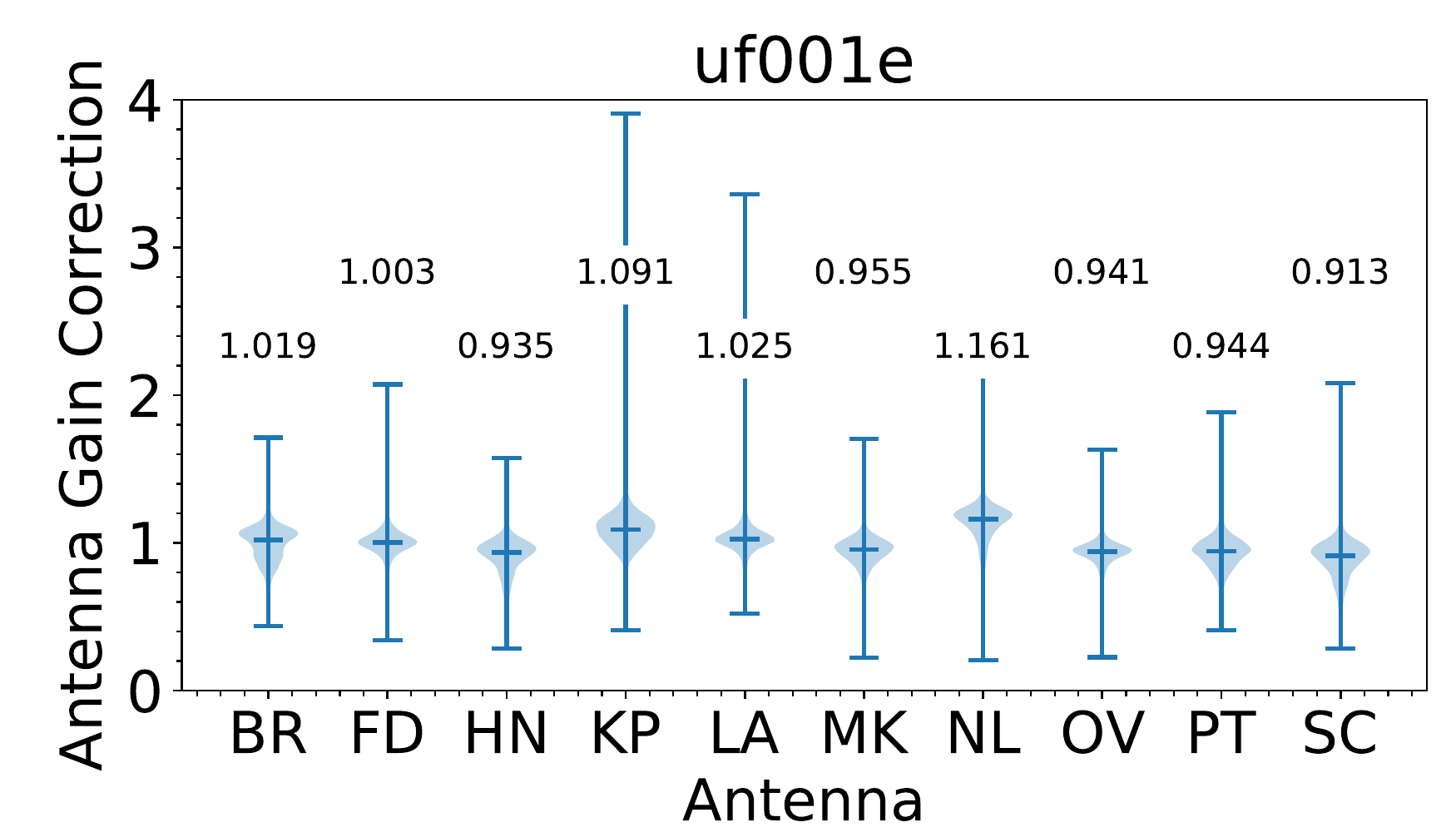}{0.48\linewidth}{(e)}
              \fig{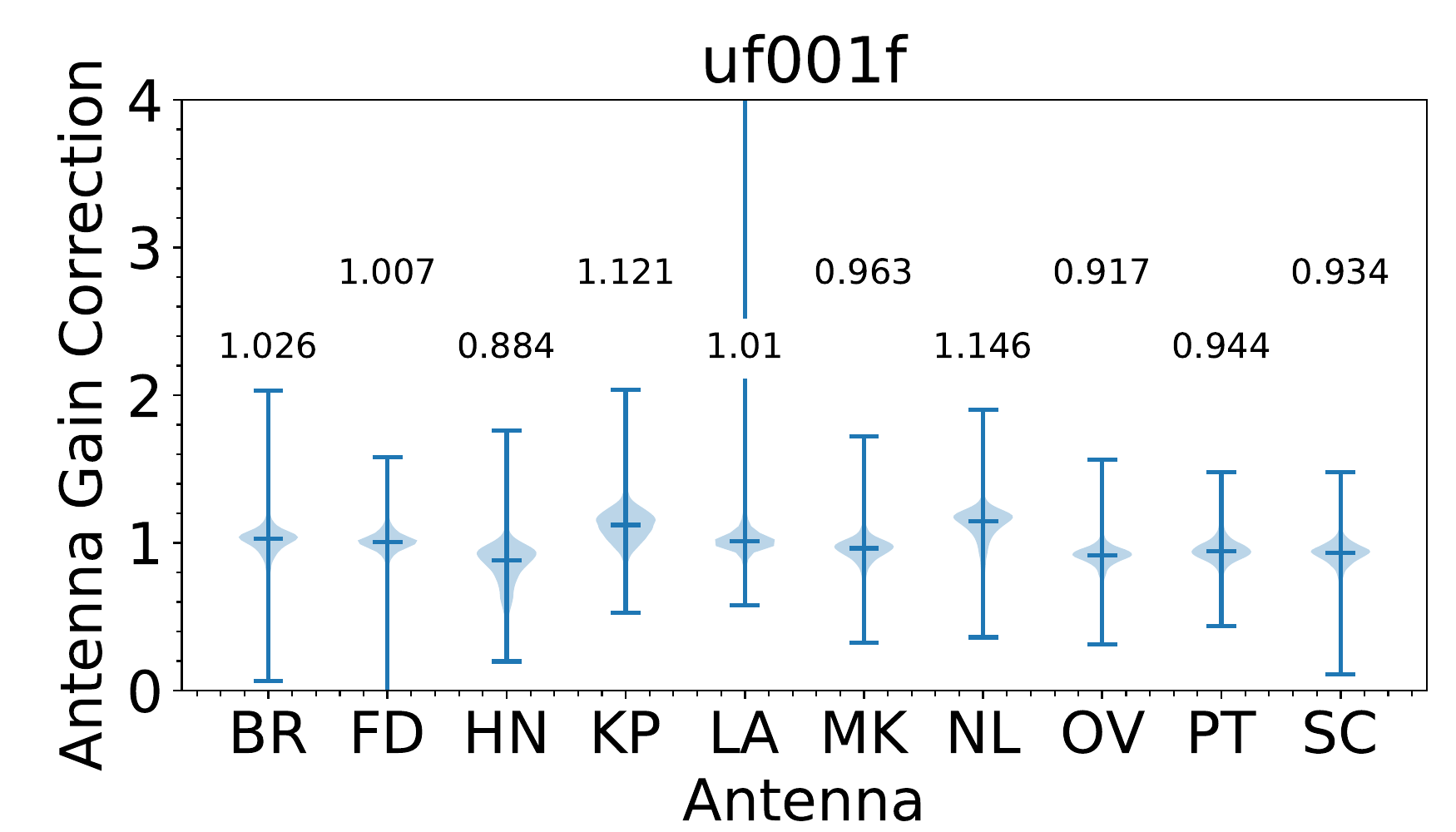}{0.48\linewidth}{(f)}}
    \caption{Violin plots showing the distribution and span of the gain corrections calculated from self-calibration for each antenna for each scan at 2.3 GHz. The median value of the gain corrections is indicated by the horizontal line in the center, and shown as text in each plot. }
    \label{fig:S_BAND_GAINCAL_VIOLIN_PLOTS}
\end{figure*}
\begin{figure*}[h]
\figurenum{12}
    \gridline{\fig{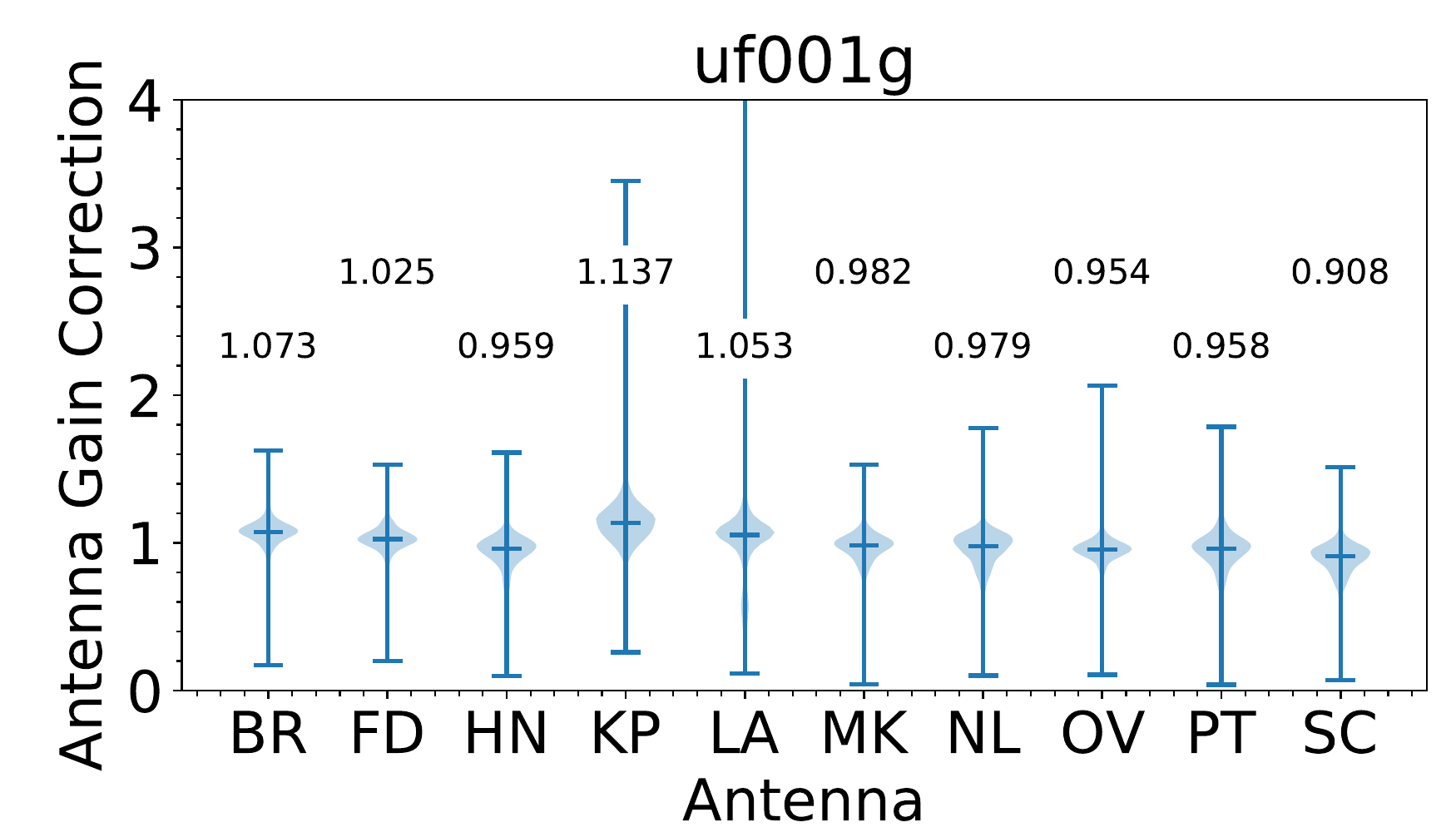}{0.48\linewidth}{(g)}
              \fig{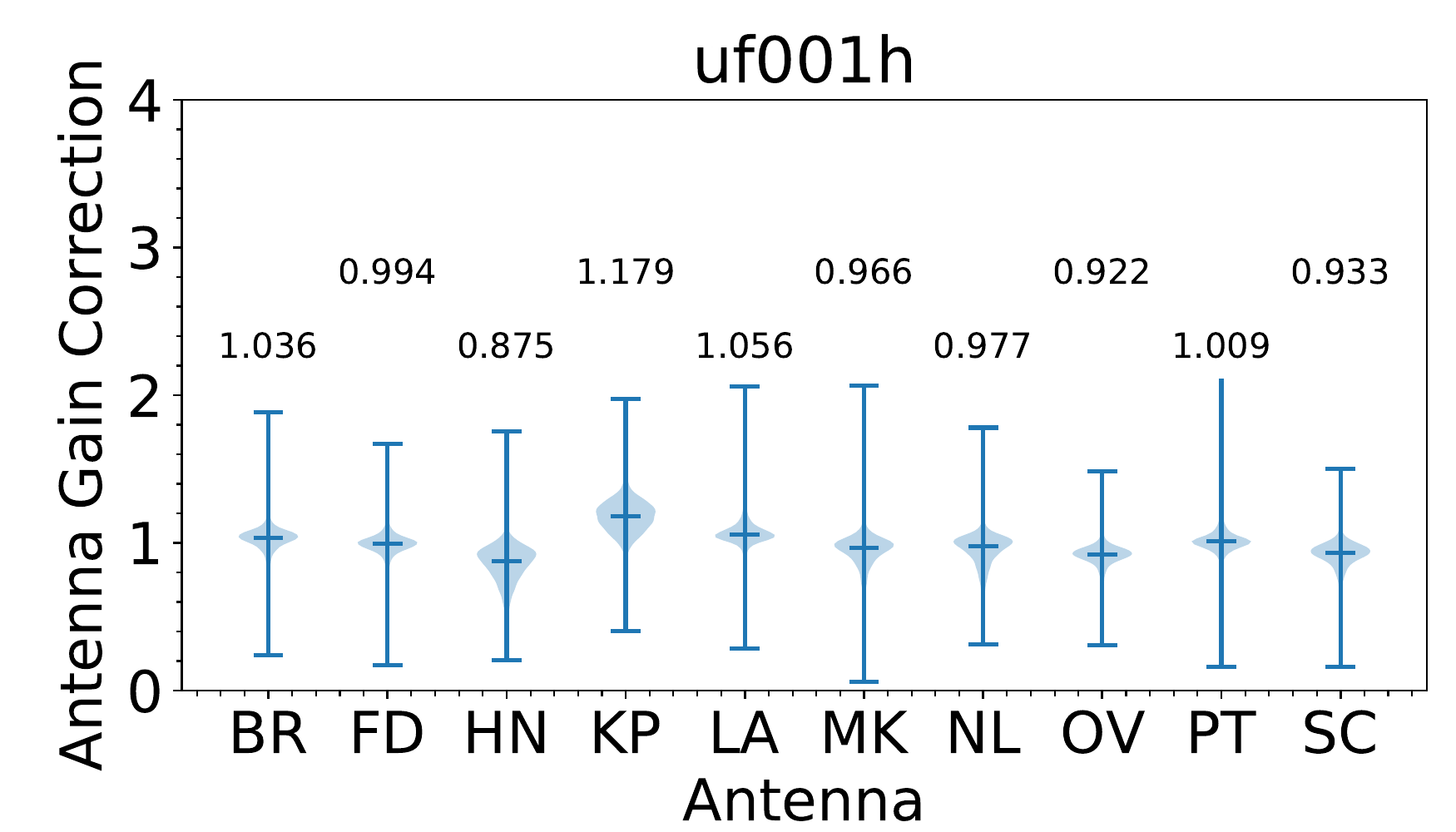}{0.48\linewidth}{(h)}}
    \gridline{\fig{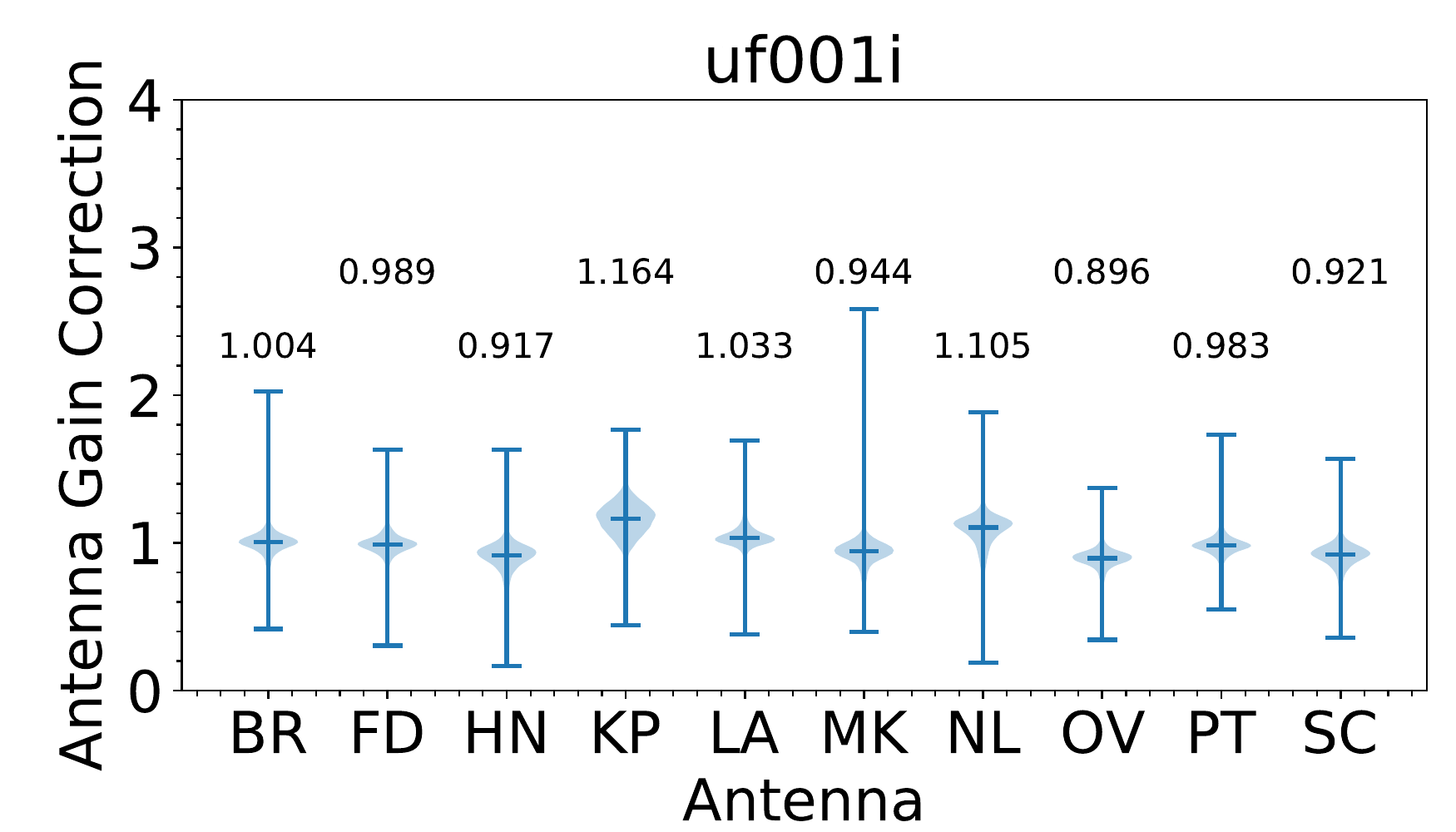}{0.48\linewidth}{(i)}
              \fig{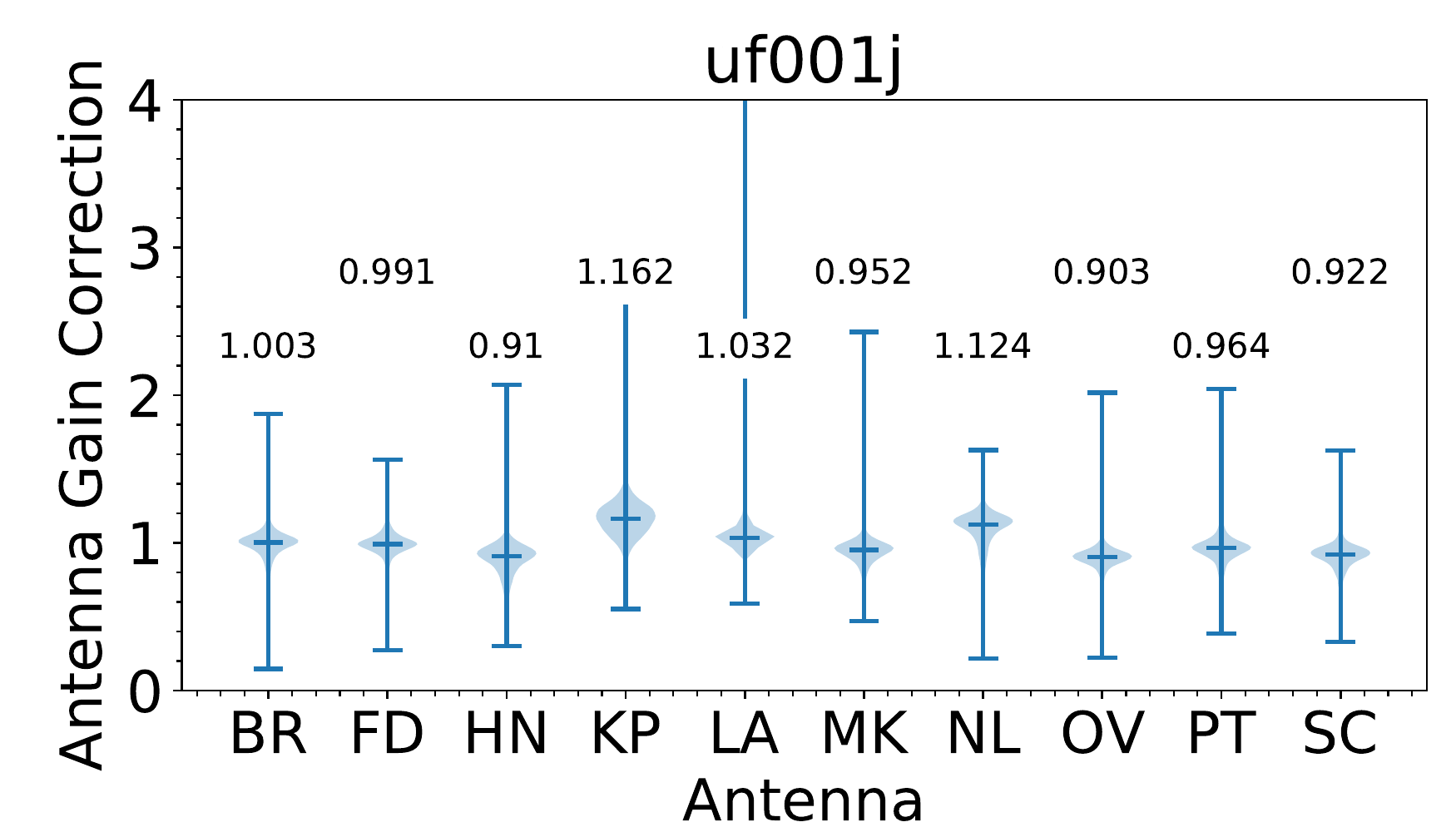}{0.48\linewidth}{(j)}}
    \gridline{\fig{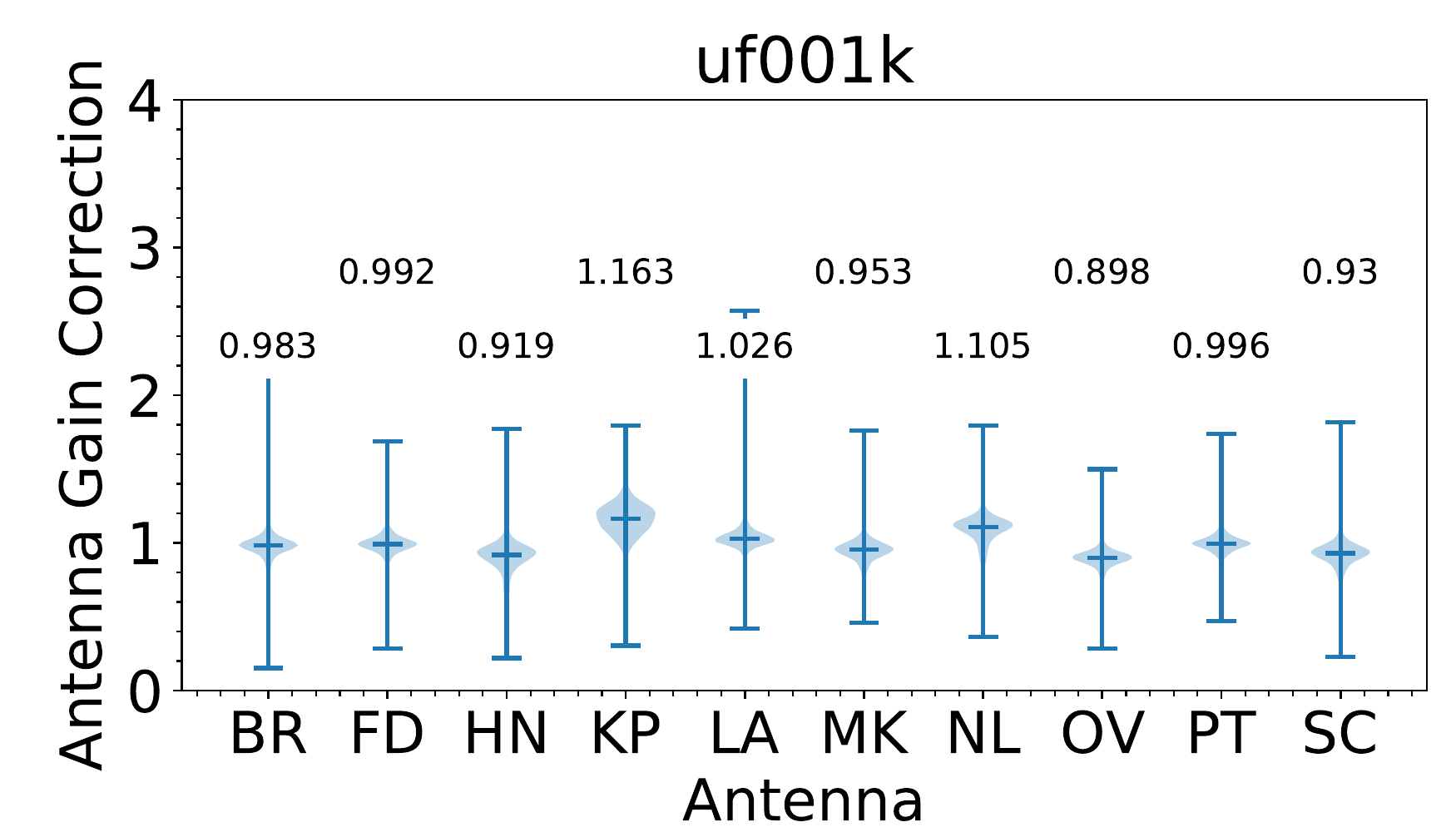}{0.48\linewidth}{(k)}
              \fig{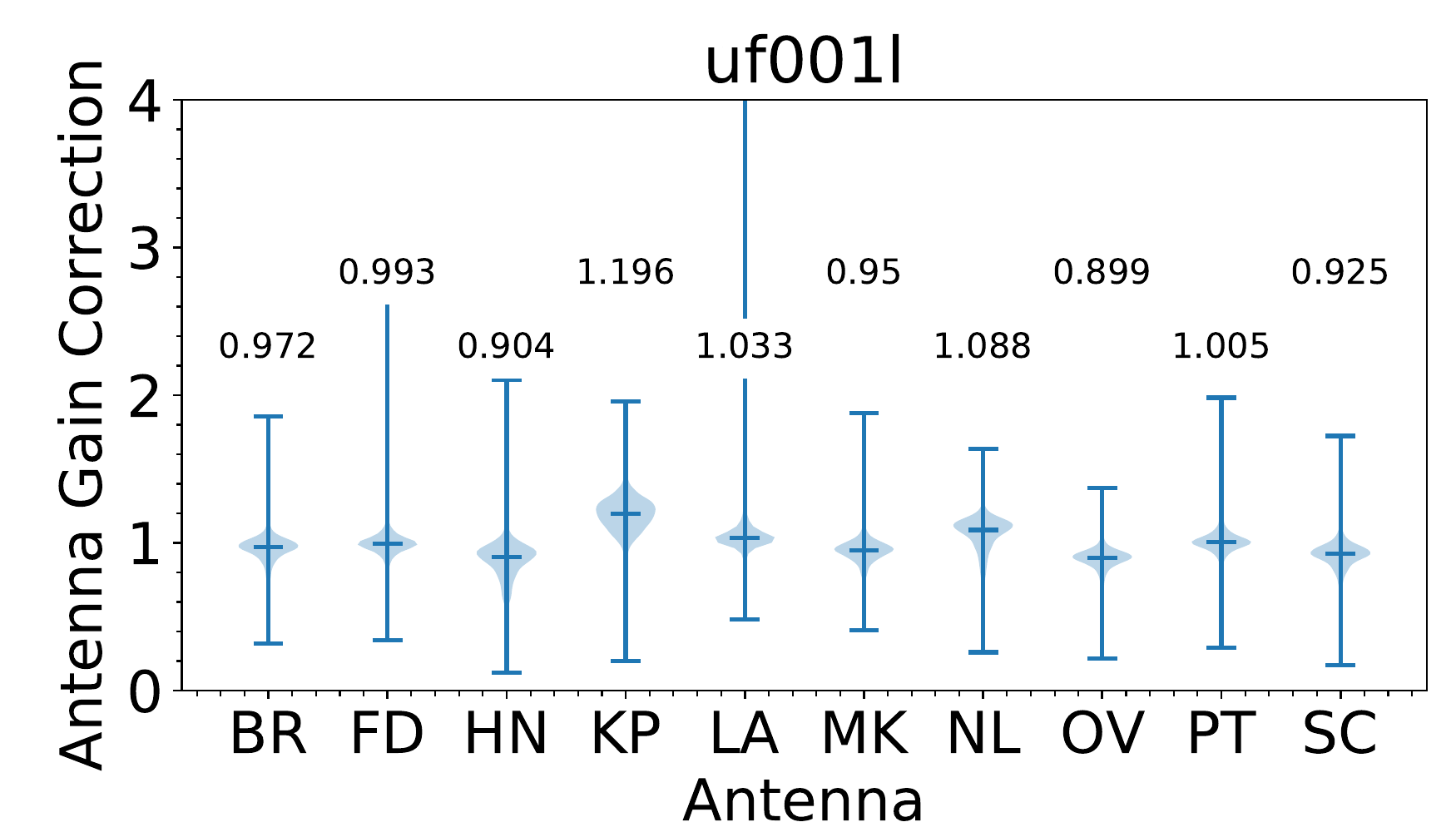}{0.48\linewidth}{(l)}}
    \caption{Violin plots showing the distribution and span of the gain corrections calculated from self-calibration for each antenna for each scan at 2.3 GHz. The median value of the gain corrections is indicated by the horizontal line in the center, and shown as text in each plot. }
    \label{fig:S_BAND_GAINCAL_VIOLIN_PLOTS_2}
\end{figure*}
\begin{figure*}[h]
\figurenum{12}
    \gridline{\fig{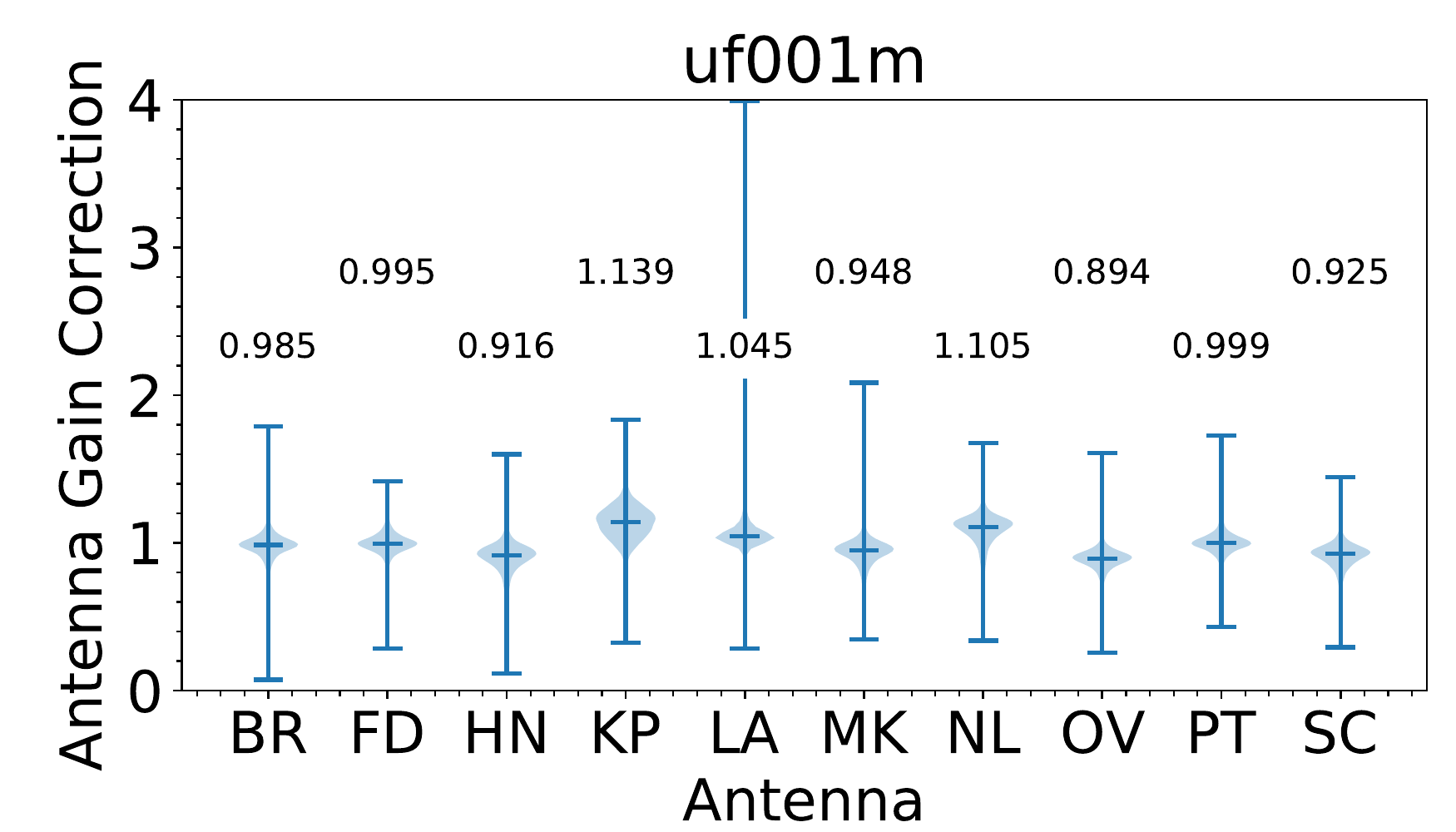}{0.48\linewidth}{(m)}
              \fig{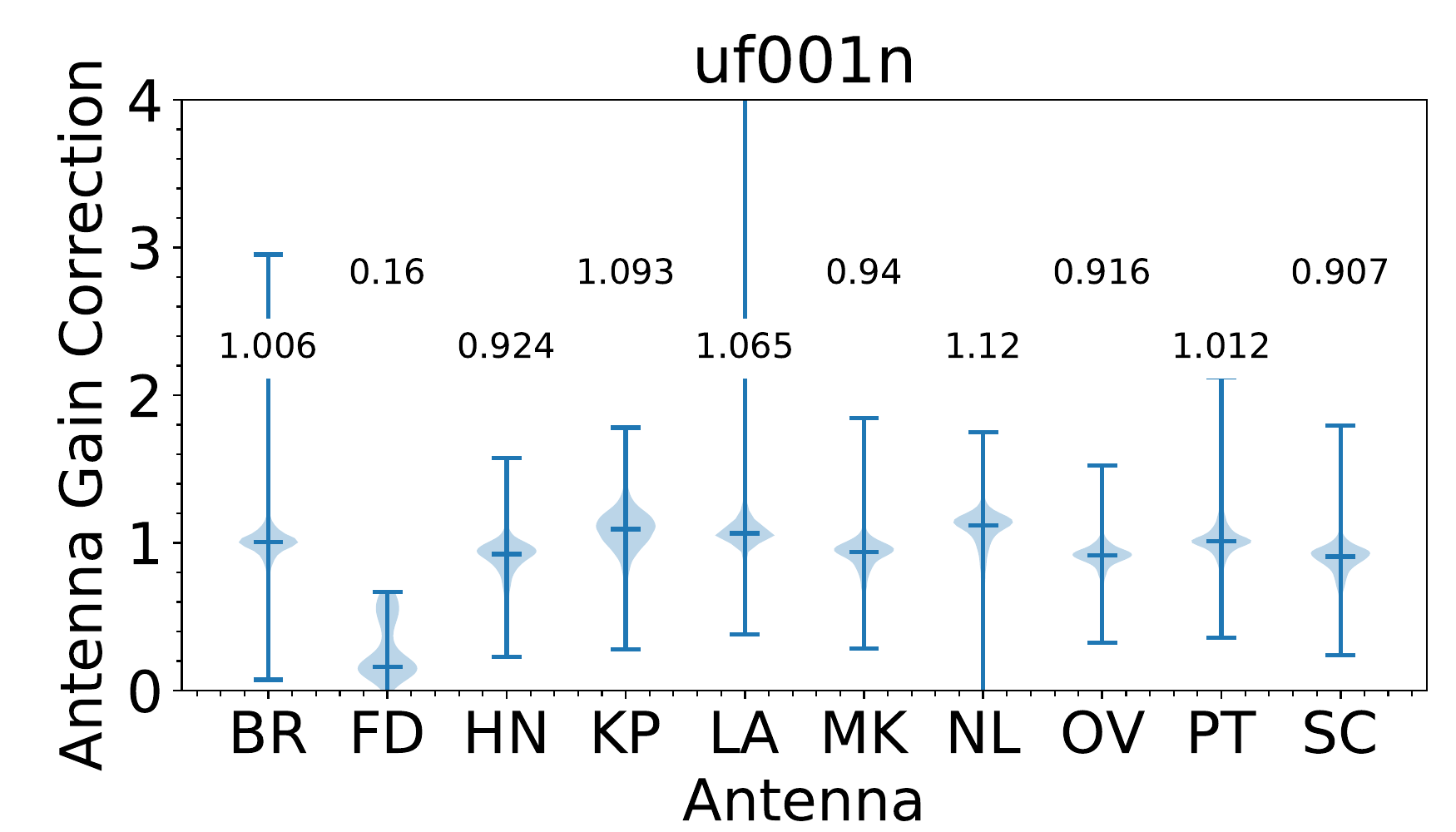}{0.48\linewidth}{(n)}}
    \gridline{\fig{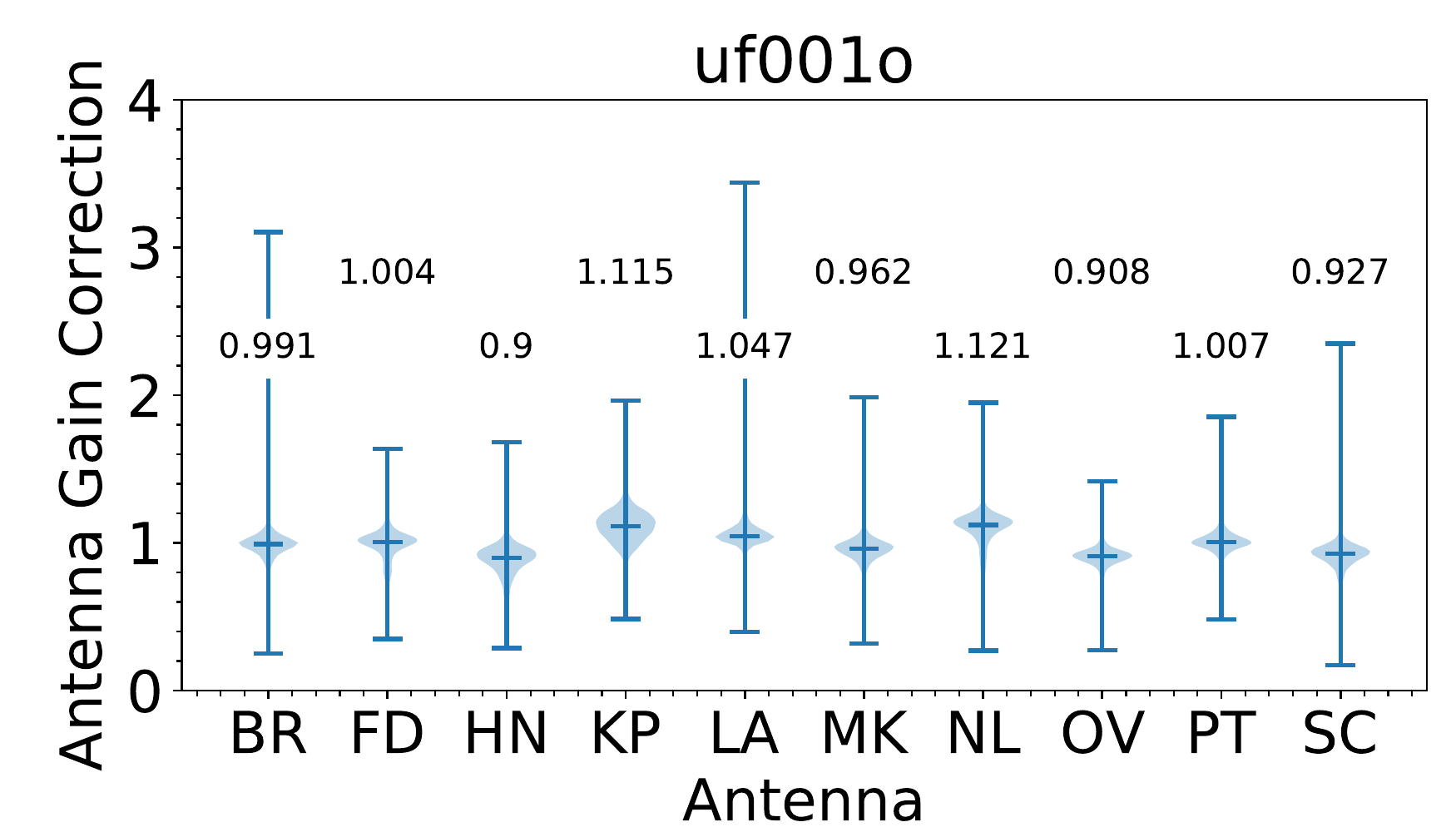}{0.48\linewidth}{(o)}
              \fig{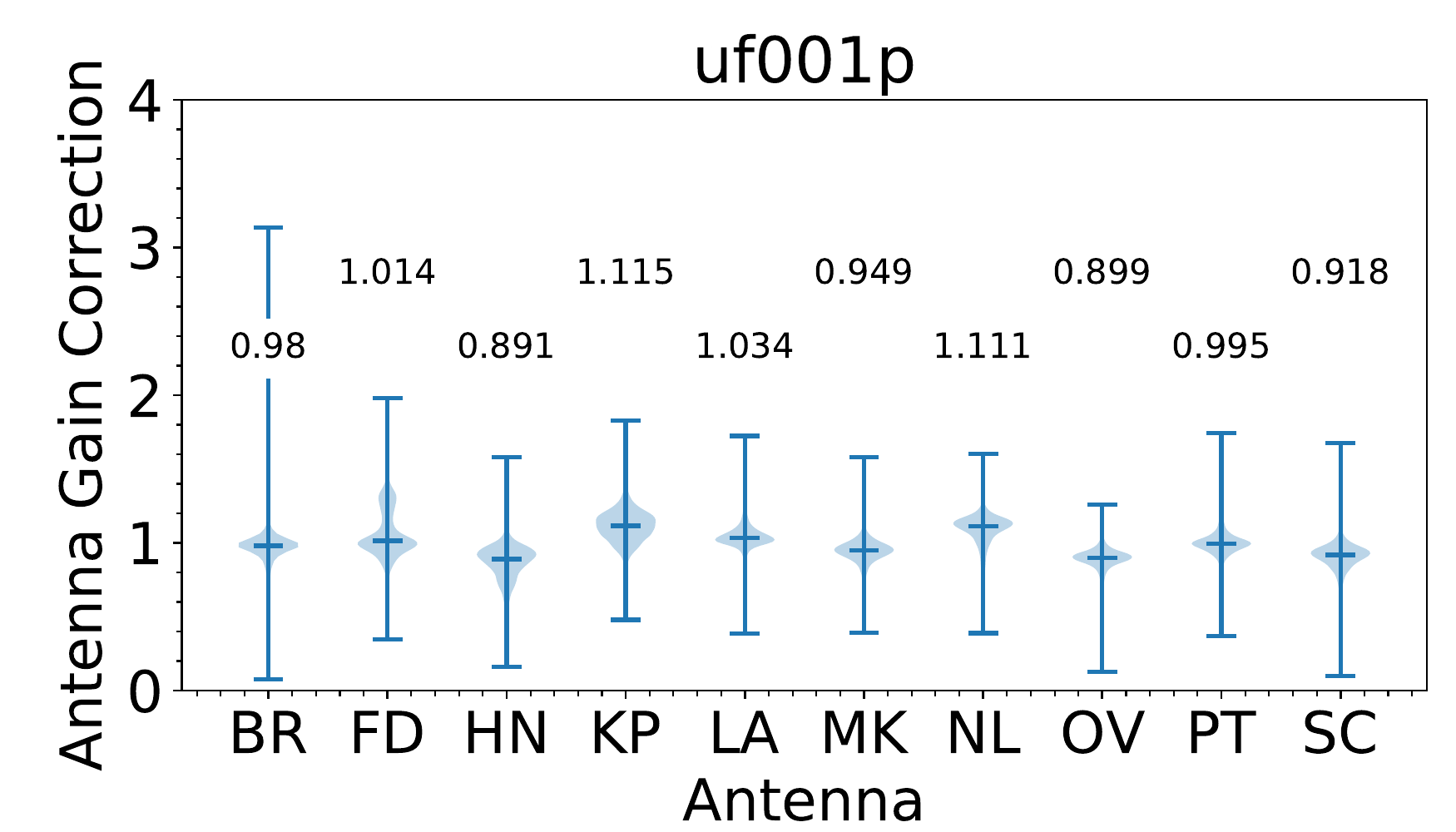}{0.48\linewidth}{(p)}}
    \gridline{\fig{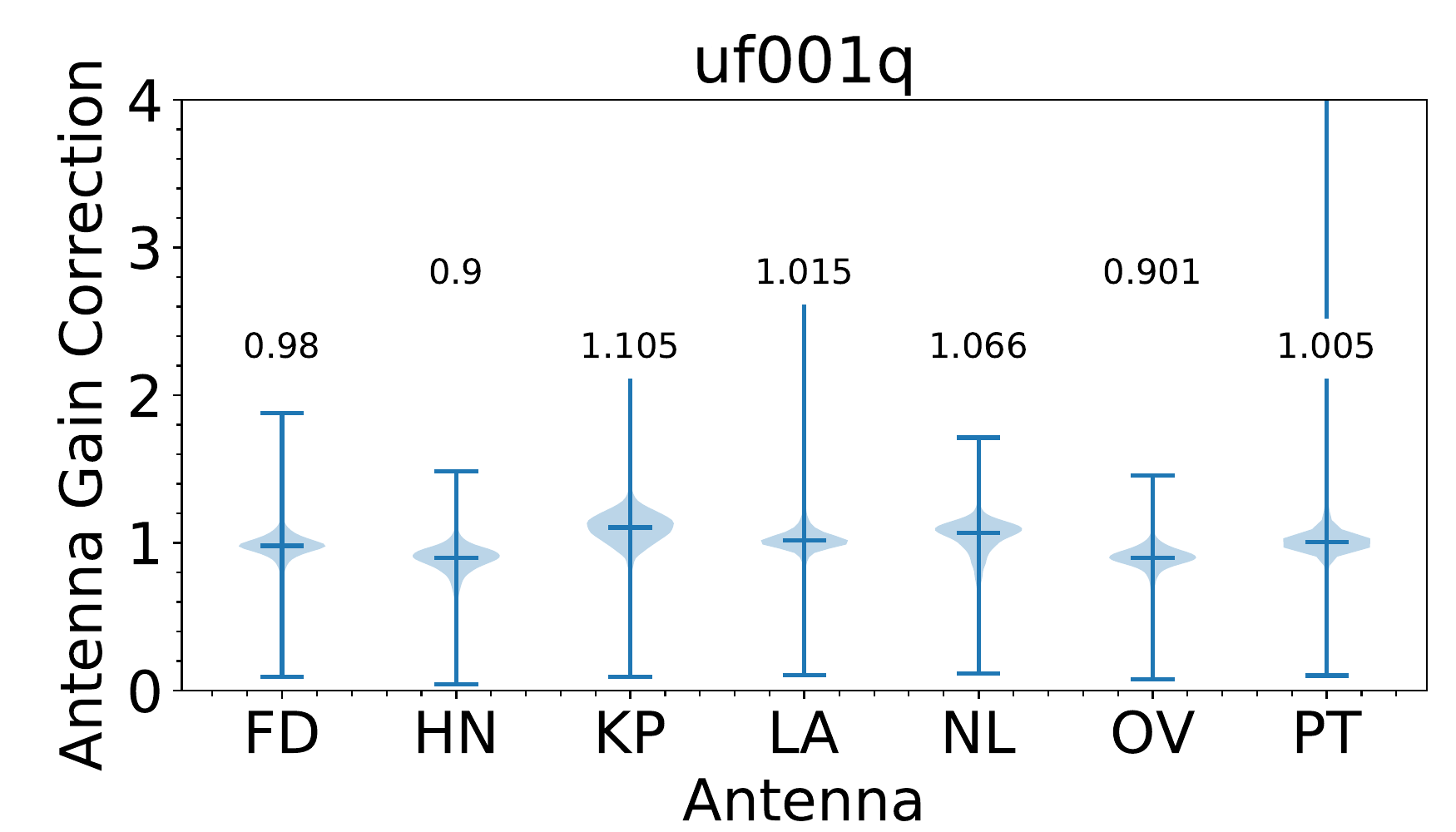}{0.48\linewidth}{(q)}
              \fig{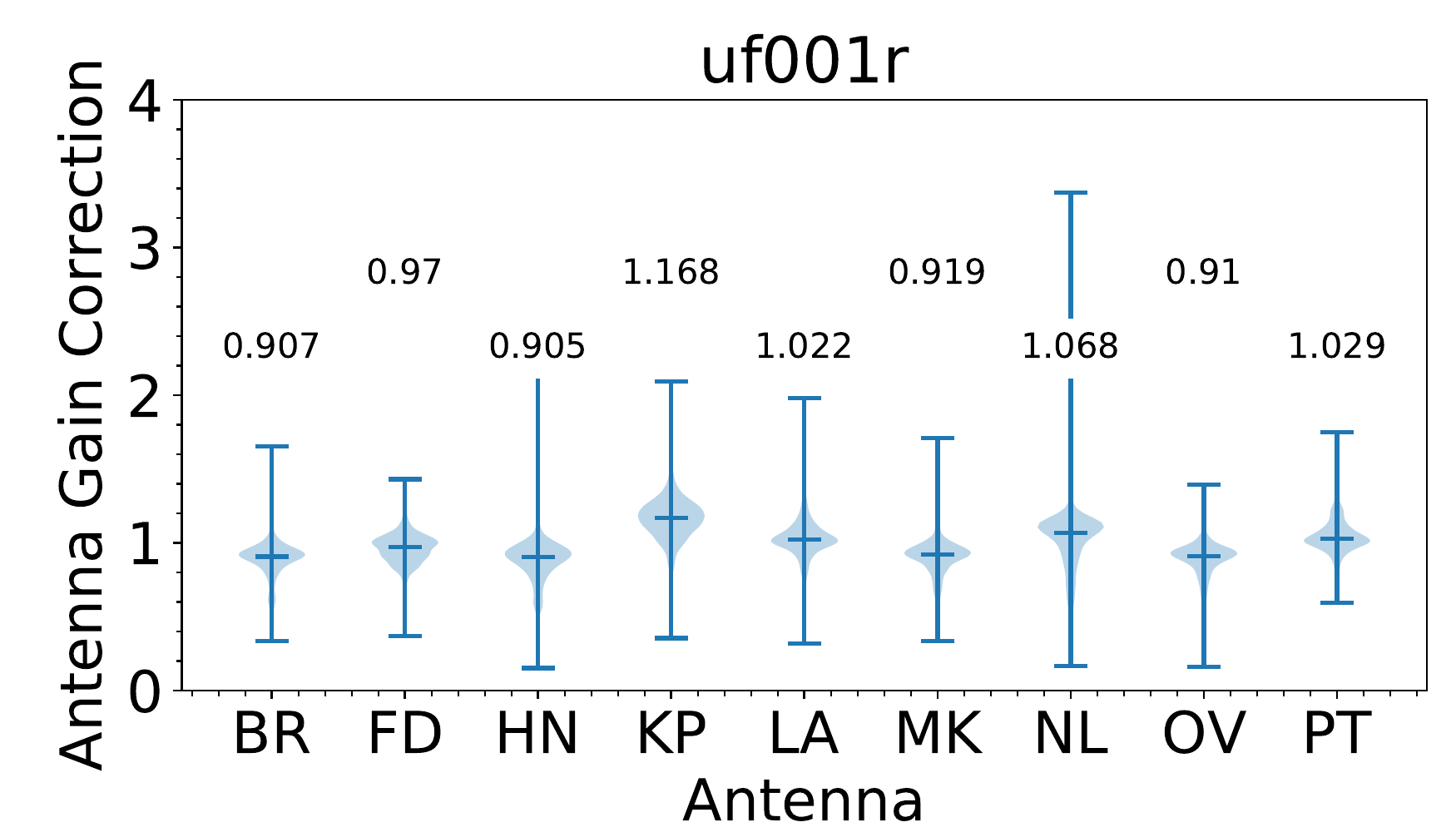}{0.48\linewidth}{(r)}}
    \caption{Violin plots showing the distribution and span of the gain corrections calculated from self-calibration for each antenna for each scan at 2.3 GHz. The median value of the gain corrections is indicated by the horizontal line in the center, and shown as text in each plot. }
    \label{fig:S_BAND_GAINCAL_VIOLIN_PLOTS_3}
\end{figure*}
\begin{figure*}[h]
\figurenum{12}
    \gridline{\fig{uf001q_s_violinplot.pdf}{0.48\linewidth}{(s)}
              \fig{uf001r_s_violinplot.pdf}{0.48\linewidth}{(t)}}
    \caption{Violin plots showing the distribution and span of the gain corrections calculated from self-calibration for each antenna for each scan at 2.3 GHz. The median value of the gain corrections is indicated by the horizontal line in the center, and shown as text in each plot. }
    \label{fig:S_BAND_GAINCAL_VIOLIN_PLOTS_4}
\end{figure*}

\bibliography{library}

\end{document}